\documentclass[preprint,flushrt]{aastex}

\usepackage[utf8]{inputenc}
\usepackage{amssymb}
\usepackage{amsmath}
\usepackage{graphicx}
\usepackage{color}
\usepackage{natbib}

\bibpunct{(}{)}{;}{a}{}{,}

\newcommand{\mat}[1]{\boldsymbol{#1}}
\renewcommand{\vec}[1]{\boldsymbol{#1}}

\newcommand{\cm}{\checkmark}

\shorttitle{ISN He from IBEX: Uncertainties and Backgrounds}
\shortauthors{Swaczyna et al.}
\keywords{instrumentation: detectors -- ISM: atoms -- ISM: kinematics and dynamics -- methods: data analysis -- methods: statistical -- Sun: heliosphere}

\begin{document}

\title{Interstellar Neutral Helium in the Heliosphere from \emph{IBEX} Observations. \\
I. Uncertainties and Backgrounds in the Data\\ and Parameter Determination Method}

\author{
P.~Swaczyna\altaffilmark{1},
M.~Bzowski\altaffilmark{1}, 
M.~A.~Kubiak\altaffilmark{1}, 
J.~M.~Sok{\'o}{\l}\altaffilmark{1}, 
S.~A.~Fuselier\altaffilmark{2,3}, 
D.~Heirtzler\altaffilmark{4}, 
H.~Kucharek\altaffilmark{4}, 
T.~W.~Leonard\altaffilmark{4}, 
D.~J.~McComas\altaffilmark{2,3}, 
E.~M{\"o}bius\altaffilmark{4}, 
N.~A.~Schwadron\altaffilmark{4,2}
}
\email{pswaczyna@cbk.waw.pl}

\altaffiltext{1}{Space Research Centre of the Polish Academy of Sciences, Warsaw, Poland}
\altaffiltext{2}{Southwest Research Institute, San Antonio, TX, USA}
\altaffiltext{3}{University of Texas at San Antonio, San Antonio, TX, USA}
\altaffiltext{4}{Space Science Center and Department of Physics, University of New Hampshire, Durham, NH, USA}

\begin{abstract}
This paper is one of three companion papers presenting the results of our in-depth analysis of the interstellar neutral helium (ISN He) observations carried out using the IBEX-Lo during the first six \emph{Interstellar Boundary Explorer} (\emph{IBEX}) observation seasons. We derive corrections for losses due to the limited throughput of the interface buffer and determine the \emph{IBEX} spin-axis pointing. We develop an uncertainty system for the data, taking into account the resulting correlations between the data points. This system includes uncertainties due to Poisson statistics, background, spin-axis determination, systematic deviation of the boresight from the prescribed position, correction for the interface buffer losses, and the expected Warm Breeze (WB) signal. Subsequently, we analyze the data from 2009 to examine the role of various components of the uncertainty system. We show that the ISN He flow parameters are in good agreement with the values obtained by the original analysis. We identify the WB as the principal contributor to the global $\chi^2$ values in previous analyses. Other uncertainties have a much milder role and their contributions are comparable to each other. The application of this uncertainty system reduced the minimum $\chi^2$ value 4-fold. The obtained $\chi^2$ value, still exceeding the expected value, suggests that either the uncertainty system may still be incomplete or the adopted physical model lacks a potentially important element, which is likely an imperfect determination of the WB parameters. The derived corrections and uncertainty system are used in the accompanying paper by Bzowski et al. in an analysis of the data from six seasons.
\end{abstract}

\section{INTRODUCTION}
\label{introduction}

Measurements of the temperature and inflow velocity vector of interstellar neutral (ISN) gas are an important source of information on the interstellar medium ahead of the heliosphere and are important for the heliosphere itself. Measurements of interstellar helium are critical because helium has a high ionization potential and is the second most abundant species. Since it is insensitive to solar radiation pressure and weakly ionized inside the heliosphere, it is plentiful at 1~AU from the Sun, and thus can be measured by the \emph{Interstellar Boundary Explorer} (\emph{IBEX}) with high statistics, and the measurements are relatively easy to interpret using theoretical models. The high statistics potentially facilitate studies of possibly subtle effects, including departures from the fully thermalized (equilibrium) state of the parent population in the local interstellar medium, the presence of additional populations, etc. The history of both direct and indirect interstellar neutral helium (ISN He) observations is available in \citet{frisch_etal:13a,frisch_etal:15a}.

\emph{IBEX} \citep{mccomas_etal:09a} has been measuring the ISN He flow at Earth's orbit since 2009.  ISN He is observed by the \emph{IBEX}-Lo instrument \citep{fuselier_etal:09b, mobius_etal:09b, wurz_etal:09a}. An overview of the ``first light'' observations was presented by \citet{mobius_etal:09a}. The first comprehensive analyses of ISN He observed by \emph{IBEX} in 2009 and 2010 \citep{bzowski_etal:12a,mccomas_etal:12b,mobius_etal:12a} showed that the ISN He speed, longitude, and temperature are tightly correlated and form a narrow ``tube'' in this four-dimensional (4D) parameter space. The statistically most likely solution seemed to differ by $\sim$4$^{\circ}$ from the direction previously suggested by \citet{witte:04} based on \emph{Ulysses} observations. As a result of the parameter correlation, the speed was also lower by $\sim$3.5~km~s$^{-1}$, and the temperature was in agreement. The longitude obtained from \emph{Ulysses} was within the uncertainty range obtained from the \emph{IBEX} observations. However, due to the parameter correlation, the adoption of the longitude from \emph{Ulysses} required a much higher temperature than was obtained from \emph{Ulysses}, namely, $\sim$8500~K, in contrast to 6300~K originally published by \citet{witte:04}.

Some researchers challenged the findings from \citet{bzowski_etal:12a}, \citet{mccomas_etal:12b}, and \citet{mobius_etal:12a} and suggested that due to alleged errors in the data treatment by the \emph{IBEX} team, in particular, to neglecting the dead time effect in the instrument, as well as other unspecified deficiencies in the analysis and modeling, the ISN He flow vector and temperature obtained from \emph{IBEX} were not credible and thus the \emph{Ulysses} results were the correct ones \citep{lallement_bertaux:14a}. The errors in those suggestions have been pointed out by \citet{frisch_etal:15a} and \citet{mobius_etal:15a}, as well as several subsequent independent studies. \citet{bzowski_etal:14a} and \citet{wood_etal:15a} reanalyzed the GAS data, including the portion of the data from 2007 from the last \emph{Ulysses} orbit which had not previously been analyzed. The results from these studies, which markedly agree with each other, suggested that the original flow vector obtained by \citet{witte:04} is supported within the uncertainties, but that the temperatures originally reported by \citet{witte:04} was too low: \citet{bzowski_etal:14a} and \citet{wood_etal:15a} found ISN He temperature equal to $7500 \pm 1500$~K and $7260 \pm 270$~K, respectively.
\citet{mccomas_etal:15a} and \citet{leonard_etal:15a} analyzed data taken by \emph{IBEX} in 2012--2014 and found that the parameter ``tube'' suggested by \citet{mccomas_etal:12b} is still correct and suggested a new explanation including a flow longitude and speed similar to \emph{Ulysses}, but a temperature which was markedly higher than that found by \citet{witte:04}.

This paper is one of three complementary studies presenting the results of an in-depth analysis of the ISN He observations carried out using \emph{IBEX}-Lo during the first six \emph{IBEX} observation seasons. We develop a sophisticated uncertainty system for the \emph{IBEX}-Lo data which takes into account all of the known sources of error and uncertainty and the resulting correlations between the data points. We also develop an ISN He parameter fitting method that utilizes this uncertainty system, and we present corrections for losses due to the limited throughput of the interface buffer, mistaken by \citet{lallement_bertaux:14a} for the sensor dead time effect. We also calculate the \emph{IBEX} spin-axis pointing with a very high accuracy. With those elements in hand, we analyze in greater detail data from one of the seasons used in the original analysis by \citet{bzowski_etal:12a} and \citet{mobius_etal:12a}, and illustrate the role of various uncertainties in the data fitting process to determine the most important ones. The ISN He flow model required in the analysis is presented in the companion paper by \citet{sokol_etal:15b} and the results of the global analysis of ISN He observation and their implications are provided by \citet{bzowski_etal:15a}. These papers are part of a Special Issue of the \emph{Astrophysical Journal Supplement Series} that addresses many important ISN observations from the \emph{IBEX} mission---see \citet{mccomas_etal:15b} for an overview of the entire Special Issue.
 
\section{STRATEGY OF ISN OBSERVATIONS BY \emph{IBEX}}
\label{strategy}
\emph{IBEX} was launched into a highly elliptical orbit around Earth to maximize the observation time outside Earth's magnetosphere. \emph{IBEX} is spinning around its Z-axis in the spacecraft coordinate system at about 4.17 rpm. The orientation of the Z-axis is periodically changed to approximately follow the Sun. Originally, the orientation of the spin-axis was modified after the perigee in each orbit \citep{scherrer_etal:09a}. After orbit 127 in 2011 June, \emph{IBEX} was placed into a long-term stable lunar-resonance orbit \citep{mccomas_etal:11a} and since then the \emph{IBEX} spin-axis has been re-oriented twice per orbit, near apogee and perigee. This new reorientation scheme divides each orbit into two arcs, denoted as a and b, with the spin-axis fixed in different directions in each arc. The \emph{IBEX}-Lo boresight points toward the negative Y-axis in the spacecraft frame \citep{scherrer_etal:09a}, and thus it views a great circle on the sky perpendicular to the spin-axis. 

\emph{IBEX}-Lo \citep{fuselier_etal:09b} is a single-pixel neutral atom camera: it consists of a time-of-flight (TOF) ion mass spectrometer with an entrance system composed of a collimator to restrict the field of view (FOV) and a conversion surface to turn the incoming neutral atoms into negative ions which are electrostatically analyzed and then detected. The \emph{IBEX}-Lo collimator is composed of four quadrants: three of low resolution, with FWHM FOV 6.5$^\circ$, and one of high resolution, with FWHM FOV 3.2$^\circ$. Neutral helium atoms, having passed through the collimator, hit a specially prepared conversion surface \citep{wurz_etal:06a} and sputter H$^-$ ions, which are detected by the \emph{IBEX}-Lo detector \citep{wieser_etal:07a}. \emph{IBEX}-Lo observes in eight logarithmically spaced energy bands, called energy steps, which are sequentially cycled during the observations, providing a coverage in energies from $\sim$10~eV to $\sim$2~keV. ISN He is measured mostly in energy steps 1 through 3. In this paper, as well as in the accompanying paper by \citet{bzowski_etal:15a} and as in the previous analyses of ISN He data from \emph{IBEX}-Lo \citep{bzowski_etal:12a, mobius_etal:12a, leonard_etal:15a, mccomas_etal:15a}, we use data from energy step 2 with the center energy 27~eV and energy resolution $\Delta E/E=0.7$ \citep{fuselier_etal:09b}. The measured H$^-$ ions are sputtered from the conversion surface by impacting ISN He atoms, and thus the energy information of the parent He atom is lost.

The ISN helium fluxes are observed when the spacecraft is in the portions of Earth's orbit where the bulk velocity of the ISN He is almost tangent to Earth's trajectory \citep{mobius_etal:09b}. This happens twice per year, in the spring and fall. In spring, the bulk velocity of ISN He is antiparallel to the velocity of the Earth and the relative velocity is high enough to sputter atoms from the \emph{IBEX}-Lo conversion surface \citep{mobius_etal:09a}. During fall, the situation is reversed and the energy is small (for a discussion of the ISN He gas observability during the fall season, see \citealt{galli_etal:15a} and \citealt{sokol_etal:15a}). Consequently, only the spring season is actually suitable for ISN He analysis. 

\subsection{Spacecraft Orientation}
\label{strategy:orientation}
The original determination of the \emph{IBEX} spin-axis pointing was presented by \citet{hlond_etal:12a} based on data from the spacecraft attitude control system \citep[ACS;][]{scherrer_etal:09a} and Star Sensor observations \citep{fuselier_etal:09b}. The latter observations were used to confirm the credibility and precision of the ACS system. These authors found that the ACS system with post-processing on the ground routinely achieved an accuracy of the spin-axis determination within $\sim$0.2$^{\circ}$. However, \citet{hlond_etal:12a} found that because of the complex patterns of stars on a relatively bright extended background in the FOV, Star Sensor data usually could not be used to determine the spin-axis orientation independently of ACS during the spring interstellar gas observation campaign. Therefore, during this portion of year, only the ACS system could be used. The spin-axis pointings used by \citet{bzowski_etal:12a} and \citet{mobius_etal:12a} in their analyses of neutral interstellar He inflow were obtained from the automatic scripts described below operating on ACS data.  

ACS consists of various sensors, including the Star Tracker, which measures the spacecraft attitude by referencing stars in the FOV. Information from the Star Tracker is given in the form of quaternions that describe the orientation of the \emph{IBEX} coordinate system with respect to the Earth equatorial reference system. To determine the spin-axis orientation needed for science data analysis, automated scripts operating at the \emph{IBEX Science Operations Center} \citep[ISOC;][]{schwadron_etal:09a} process those quaternions on an orbit by orbit (since the orbit perigee rise, on an arc by arc) basis.

If a bright celestial body is in the Star Tracker FOV, e.g., the Moon, then the determined quaternions are of poor quality and should not be used in the determination of the spin-axis pointing. The automated scripts generally reject the part of the data where this problem occurs during an orbit, but not just after the spacecraft reorientation. In this case, the obtained value could differ at most by 0.2$^\circ$. In Section \ref{uncertainties:orientation}, we discuss the influence of the spacecraft orientation on the obtained signal. We found that the uncertainty of the spin-axis pointing is an important source of uncertainty in the data, especially in the portion of the data with the highest statistics.

We reanalyzed the Star Tracker quaternions and determined the spin-axis pointing again, carefully omitting the affected part of the data. The algorithm we developed is independent from that previously used. We determined the spin-axis orientation for 368 orbits or arcs. In about 90\% cases, the values obtained in this analysis differ by less than 0.01$^\circ$ from the values obtained using the original algorithm. Unfortunately, the remaining 10\% occur mostly in orbits used in the analysis of ISN He. In Table~\ref{strategy:orientation:table}, we present the previously used ISOC values and the values obtained in this analysis of the spin-axis pointing for orbits used in the analysis of ISN He in season 2009. The angular distances between the previous values and those used in this analysis are larger than 0.01$^\circ$ only for orbits 14 and 16. However, the old and new coordinates are safely within the reported uncertainty of the original determination. In the ISN analysis, we assume that the present uncertainty of the spin-axis pointing is approximately 0.01$^\circ$ (see Section~\ref{uncertainties:orientation}). The \emph{IBEX} spin-axis pointings for ISN seasons are provided in the \emph{IBEX} data release (Section~\ref{datarelease}).

\begin{deluxetable}{crrrrr}
 \tablewidth{0pt}
 \tablecolumns{6}
 \tablecaption{\emph{IBEX} Spin-axis Pointing in ISN Season 2009 \label{strategy:orientation:table}}
 \tablehead{
    \colhead{Orbit \#} 					& 
    \multicolumn{2}{c}{ISOC}			&
    \multicolumn{2}{c}{This Analysis}	&
    \colhead{Difference} 				\\
    \colhead{} 							& 
    \colhead{R.A. ($^\circ$)}				&
    \colhead{decl. ($^\circ$)}			&
    \colhead{R.A. ($^\circ$)}				&
    \colhead{decl. ($^\circ$)}			&
    \colhead{($^\circ$)} 			
 }
 
 \startdata
13 & 291.103 &   -21.367 &  291.100 &     -21.371 &    0.0048\\
14 & 299.404 &   -20.018 &  299.456 &     -20.080 &    0.0655\\
15 & 307.070 &   -18.351 &  307.070 &     -18.348 &    0.0030\\ 
16 & 315.073 &   -16.298 &  315.261 &     -16.281 &    0.1581\\
17 & 322.958 &   -13.940 &  322.956 &     -13.941 &    0.0013\\
18 & 330.488 &   -11.300 &  330.486 &     -11.304 &    0.0038\\
19 & 337.728 &   -8.553  &  337.725 &     -8.554  &    0.0021\\
 \enddata
 
\end{deluxetable}

\subsection{Interstellar He Data Selection}
\label{strategy:dataselection}
The goal of the data analysis reported in the present series of papers is to determine the temperature and bulk velocity vector of the ISN He inflowing into the heliosphere from the interstellar medium. We assume that the  undisturbed parent distribution function of this gas is given by the Maxwell--Boltzmann function (for the adopted definition, see Equation (1) in \citet{sokol_etal:15a}). The analysis focuses on this component of the neutral He observed by \emph{IBEX}-Lo, and the data selection keeps this goal in mind. We extract a data subset with minimized contributions from other sources, sucha as the Warm Breeze (WB) discovered by \citet{kubiak_etal:14a}, to avoid possible bias.

In this paper, we show how different uncertainty sources impact the fitted parameters of the parent ISN He population and their uncertainties. We develop an uncertainty system that takes into account various correlations introduced to the data during the measurement process and we describe the process of ISN He parameter searching. For this task, we use a subset of the data limited to the 2009 season so that our analysis is directly comparable to the results presented by \citet{bzowski_etal:12a} and \citet{mobius_etal:12a}. The final data selection and analysis of the full six-year data set are presented by \citet{bzowski_etal:15a}.

The data used for the ISN He flow analysis were collected during the spring observation season, when \emph{IBEX} was located between $115^{\circ}$ and $160^{\circ}$ ecliptic longitude. The orbits with the highest count rate in the peak have ecliptic longitude close to $131^{\circ}$ \citep{mobius_etal:12a, mobius_etal:15a}. The data selection used now is similar to the data selection used by \citet{bzowski_etal:12a}, but with the insights gathered in this paper taken into account. Specifically, we now know that data collected at ecliptic longitudes less than $\sim$115$^{\circ}$ have a large contribution from the WB. At the other end of this range, the limit of the ecliptic longitude of 160$^\circ$ is adopted to avoid a contribution to the signal from the interstellar H population, which is known to exist in this region of Earth's orbit \citep{saul_etal:12a, saul_etal:13a, schwadron_etal:13a}. Both the WB and ISN H could bias the results if we included this part of the data in our analysis. Based on these criteria, for our analysis we only selected orbits 13 through 19. 

The \emph{IBEX}-Lo data are being collected in 60 time slots per spin-period, which is equivalent to 60 spin-angle bins with a width of $6^{\circ}$. This data product is referred to as the Histogram-Binned (HB) data. Another data product is the Direct Events data, which are not used in our analysis and will not be discussed here. The HB data set does not contain any additional information on the time, except for the start time of the histogram accumulation, the number of the spin-angle bin, and the number of \emph{IBEX} spins included.

Generally, science data are taken throughout the High Altitude Science Operations (HASO) phase of \emph{IBEX} orbits. However, data from some time intervals turn out to be unsuitable for analysis and are rejected. The criteria and the procedure used for data selection are described by \citet{mobius_etal:12a}, \citet{mobius_etal:15b}, and \citet{leonard_etal:15a}. Generally speaking, what is rejected are those intervals when the spacecraft was inside the magnetosphere, when the Moon or Earth were in or near the FOV, when temporary spin-pulse/spin-period synchronization glitches occurred, when high background appeared, etc. The selection process results in a set of ISN He ``good time'' intervals for each orbit/arc, from which the data are analyzed. During this data selection step, no spin-angle filtering is performed.

The data from the good time intervals are subdivided into blocks corresponding to 512 spacecraft spins, i.e., to time intervals approximately two hours long. This subdivision is needed to correct for data losses due to reasons discussed in Section \ref{uncertainties:throughput}. Once the correction is performed separately for each of the 512-spin subintervals, the resulting corrected count rates are re-assembled into orbit-averaged count rates. Note that the presence of the inevitable gaps between the good time intervals implies that the averaging is not done over the entire HASO time \citep[see][Figure 1]{bzowski_etal:15a}. 

The final step of data point selection after correcting for throughput losses (see Section~\ref{uncertainties:throughput}) and reassembling the corrected data into good time averages is rejecting the spin-angle bins, which may be contaminated by background or unidentified populations of neutral atoms, and thus may bias the results. For this selection, we use the criterion based on the observation by \citet{bzowski_etal:12a} that the simulated signal $F(\psi)$ from the ISN He primary population approximately follows the Gaussian function in spin-angle: $F(\psi) = f_0 \exp[-\left(\psi - \psi_0\right)^2/(2 \omega)^2]$. 
Based on this insight, we decided to select the data to fit the Gaussian shape so that the absolute value of the residual for each selected data point does not exceed two times the point uncertainty. The filtering is done individually for each orbit. In other words, we require that the data fit a Gaussian function, but we do not impose any additional criteria on the Gaussian function parameters. This criterion is adopted to optimize the filtering of the signal against potential additional non-ISN components. For each orbit, we select the maximum number of the spin-angle bins with the highest rates that fulfill this criterion. Admittedly, this criterion is heuristic. It follows from statistics that it may happen that we reject valid data points. 
However, this potential rejection of valid data only occurs in the wings of the adopted angular distribution because application of our criterion did not result in the rejection of any points with high count rates, and the resulting spin-angle coverages for individual orbits are contiguous and include at least spin-angle bins in the range 252$^\circ$--282$^\circ$. The included range covers the core of the Gaussian shape within at least $\pm 1.6\omega$ from the peak. We estimate that with the normal distribution of residuals, only $\sim$5\% of valid data would be left out. For a total number of points in the sample of $\sim$50 this process is expected to reject only about 3 valid points, and so the described procedure is not likely to bias our fitting results. On the other hand, using a well-defined criterion enables an objective rejection of the portions of the signal that still may be significantly contaminated by background or unknown additional populations. Applying this criterion results in the adoption of data points with a contrast sufficiently high contrast to not bias our results, and which are simultaneously well above the background. This procedure leads to a bin selection similar to that originally used by \citet{bzowski_etal:12a}.

Since the criterion presented above is arbitrary, we also performed tests with alternative criteria. These included the requirement that all of the adopted count rates were not less than $B$\% of the maximum count rate for a given season, where $B$ was varied from 2\% up to 10\%. A discussion of the impact of our selection of the data points is presented in Section~\ref{results:selection}.

\section{SOURCES OF UNCERTAINTIES}
\label{uncertainties}
Uncertainties in a series of experimental measurements generally belong to two classes: statistical (random) and systematic. In counting experiments, statistical uncertainties include the Poisson statistical scatter, and thus are uncorrelated within the series of measurements. Systematic uncertainties include errors and uncertainties that affect all members of a data set or some groups within the set because they have a common origin and thus are not independent. The lack of independence of a portion of uncertainties in data points must be taken into account in the analysis, especially if the correlations are strong. In the previous analyses of interstellar He \citep{bzowski_etal:12a, mccomas_etal:12b, mccomas_etal:15a, mobius_etal:12a, kubiak_etal:14a, leonard_etal:15a}, those correlations were assumed to be insignificant, and thus were neglected.

The uncertainty system in the case of systematic, correlated uncertainties is commonly described by a covariance matrix \citep[e.g.][]{pdg:2014} in which out-of-diagonal terms describe the relations between uncertainties for different point pairs. In this paper, we present a procedure to reduce the available ISN data from \emph{IBEX} and prepare an adequate covariance matrix of the reduced data. The model used to calculate the expected flux value to compare with the data \citep{bzowski_etal:15a} carefully takes into account all known measurement effects that affect the measured flux, but it also uses some parameters that are only approximately known. The uncertainties related to the measurement process introduce correlated uncertainties in the data, which are described in this section. In addition to the uncertainties related to the measurement process, there are also uncorrelated statistical uncertainties that are always present due to the nature of the measured quantities. 

The raw data that serve as the starting point in the analysis are the numbers of counts recorded in the 6$^\circ$ spin-angle bins over time blocks corresponding to 512 \emph{IBEX} spins, chosen through the data selection process described in Section \ref{strategy:dataselection}. Splitting the data into these time blocks is crucial for the determination of the throughput correction (TC). After the correction factors have been obtained, we sum over all of the time blocks. The duration of the total observation time for a given bin is equal to the sum of the durations of the good times divided by the number of bins in spin-angle (60) and the number of observed energy channels (8). For a given orbit/arc, all of the adopted bins have identical observation times. We denote the number of counts in each bin as $d_{i}$. For simplicity we assume that the index $i$ describes both the orbit number and the 6$^\circ$ bin on this orbit. We also define a function $\mathcal{O}(i)$ that returns the orbit number for index $i$ (effectively, a look-up table). With these definitions, the mean count rate $c_i$ for bin $i$ can be obtained from the following formula:
\begin{equation}
 c_i=\frac{d_{i}\gamma_{i} }{t_i} - b_{i}-w_{i}\, , \label{eq:countrate}
\end{equation}
where $\gamma_{i}$ is the so-called TC factor, $t_i$ is the duration of the bin observation, and $b_i$ and $w_i$ are the counts from  the background and WB, respectively, expected in bin $i$. The count rates obtained for the selected data points (Section~\ref{strategy:dataselection}) form a set which we use to compare the measurements with results of the Warsaw Test Particle Model (WTPM, Section~\ref{application}).

Below, we describe the identified sources of uncertainties. First, we define the uncertainties related to the statistical aspect of the recorded count numbers (Section~\ref{uncertainties:counts}). Subsequently, we discuss the limited throughput of the \emph{IBEX}-Lo interface buffer due to the high rate of measured electrons (Section~\ref{uncertainties:throughput}). Another source of uncertainty could potentially be related to the WB and the estimated background level. These uncertainties are directly related to the quantities in Equation~\eqref{eq:countrate}. 

A different source of correlated uncertainties is the geometry of the observations. The index $i$ could be translated into the collimator boresight position, which is needed to perform the simulations. This position is known with some uncertainty due to the uncertainty of the spacecraft orientation in each orbit and the orientation of the \emph{IBEX}-Lo boresight with respect to the spin-axis of the spacecraft. We take these two sources into account as described in Sections~\ref{uncertainties:orientation} and \ref{uncertainties:boresight}. 

\subsection{Number of Counts}
\label{uncertainties:counts}
The data stored in histogram bins (HB, Section~\ref{strategy:dataselection}) are built up from individual events, qualified as valid counts of the H$^-$ ions that are sputtered off the \emph{IBEX}-Lo conversion surface by the incoming He atoms. Each measured event is independent of the other measured events. Consequently, the observed count numbers follow the Poisson distribution, and the related Poisson uncertainty of the number of counts is given by
\begin{equation}
 \delta d_{i}=\sqrt{d_{i}}\, . \label{eq:unc:poiss}
\end{equation}

\subsection{Limited Throughput of the IBEX-Lo Interface Buffer}
\label{uncertainties:throughput}

All of the events observed by \emph{IBEX}-Lo that have generated at least one valid TOF measurement are transmitted via an interface buffer to the Central Electronics Unit (CEU). The interface buffer has a finite capacity, which is controlled by the finite transfer time between the interface buffer and the CEU. The interface buffer is laid out as a double buffer, with two slots that can temporarily hold up to two consecutive events. Once the CEU has accepted the most recent event, it checks the interface buffer again and transfers the next event in line, provided that the interface buffer is not empty. Only if another event appears when both slots of the interface buffer are occupied is such an event lost.

The \emph{IBEX}-Lo interface buffer was designed to safely handle valid event rates up to about 100 events per second without noticeable losses, which is more than adequate for heliospheric ENAs, including the peak coincidence rates when observing the ISN He flow. The TOF system itself is capable of handling $> 50,000$ events per second, and thus does not contribute to any suppression at all \citep{fuselier_etal:09b}.

After launch, the negative voltage on the collimator could not be raised to the originally planned value to reject ambient electrons \citep{mobius_etal:12a}. Therefore, an excessive amount of electrons reach the detector system and generate a valid TOF coincidence in the delay line of the Stop anode \citep{fuselier_etal:09b}. They are subsequently identified as electrons and eliminated by the CEU. The total event rate transmitted to the interface buffer (including both electrons and H$^-$ ions) approach $\sim$300--500 events per second in energy step 2, used in this analysis. This resulted in a fractional event loss due to the overflow of the interface buffer. Data lost during the interface buffer overflow can be estimated statistically and corrected for. Data from the ISN seasons starting with 2013 do not require further corrections because a change in the \emph{IBEX}-Lo TOF logic \citep{mobius_etal:15a} was implemented in mid-2012. This logic now requires a valid TOF between the two carbon foils of the TOF system (valid TOF2), and thus rejects almost all electron events before transmission to the interface buffer. We have verified that after this change, the suppression of the ISN He event rate is negligible ($<0.002$ in the ISN flow peak). However, for 2009 through 2012, a small correction is needed for losses in the interface buffer.

\begin{figure*}[ht!]
 \includegraphics[width=.24\textwidth]{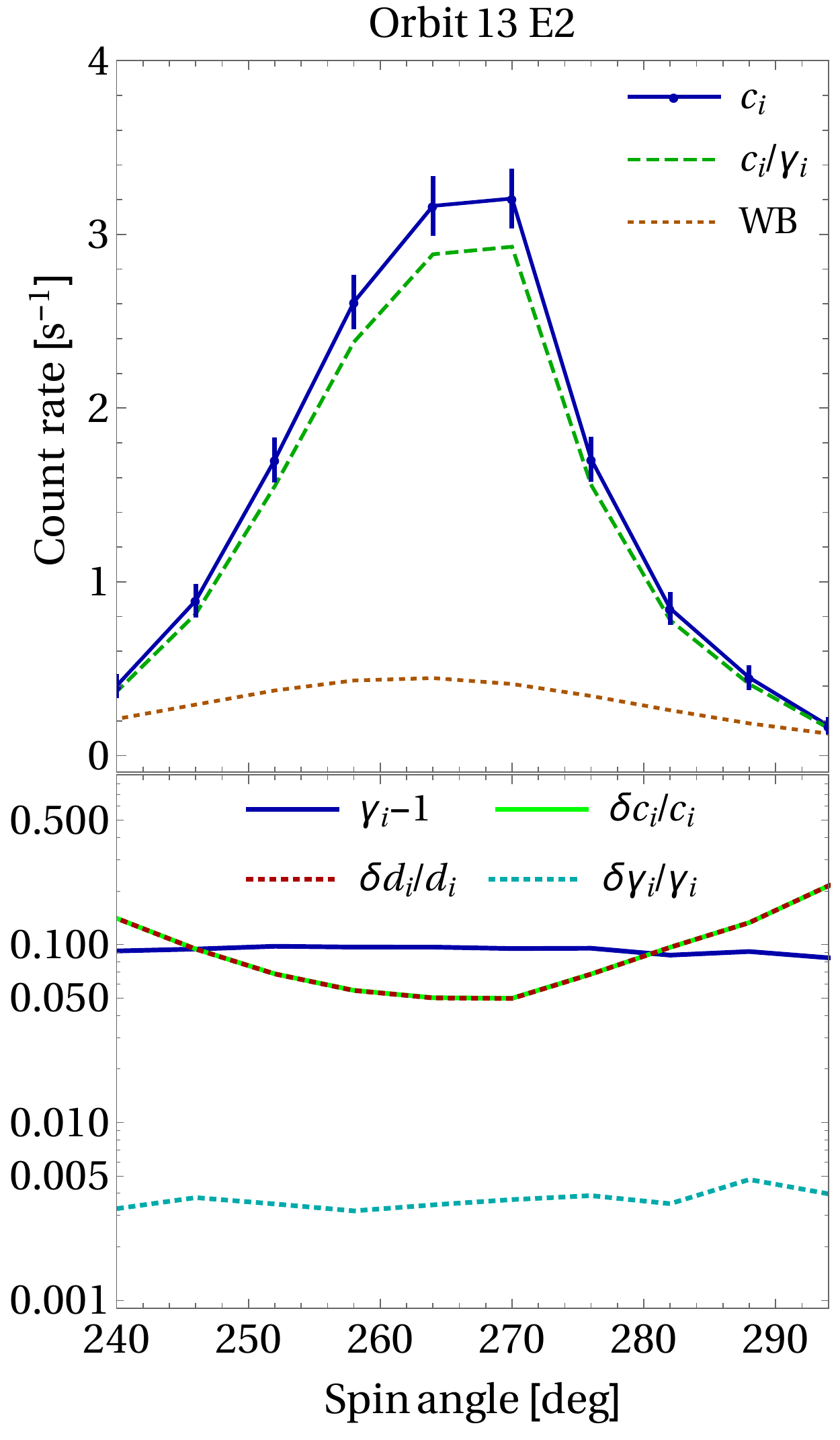}
 \includegraphics[width=.24\textwidth]{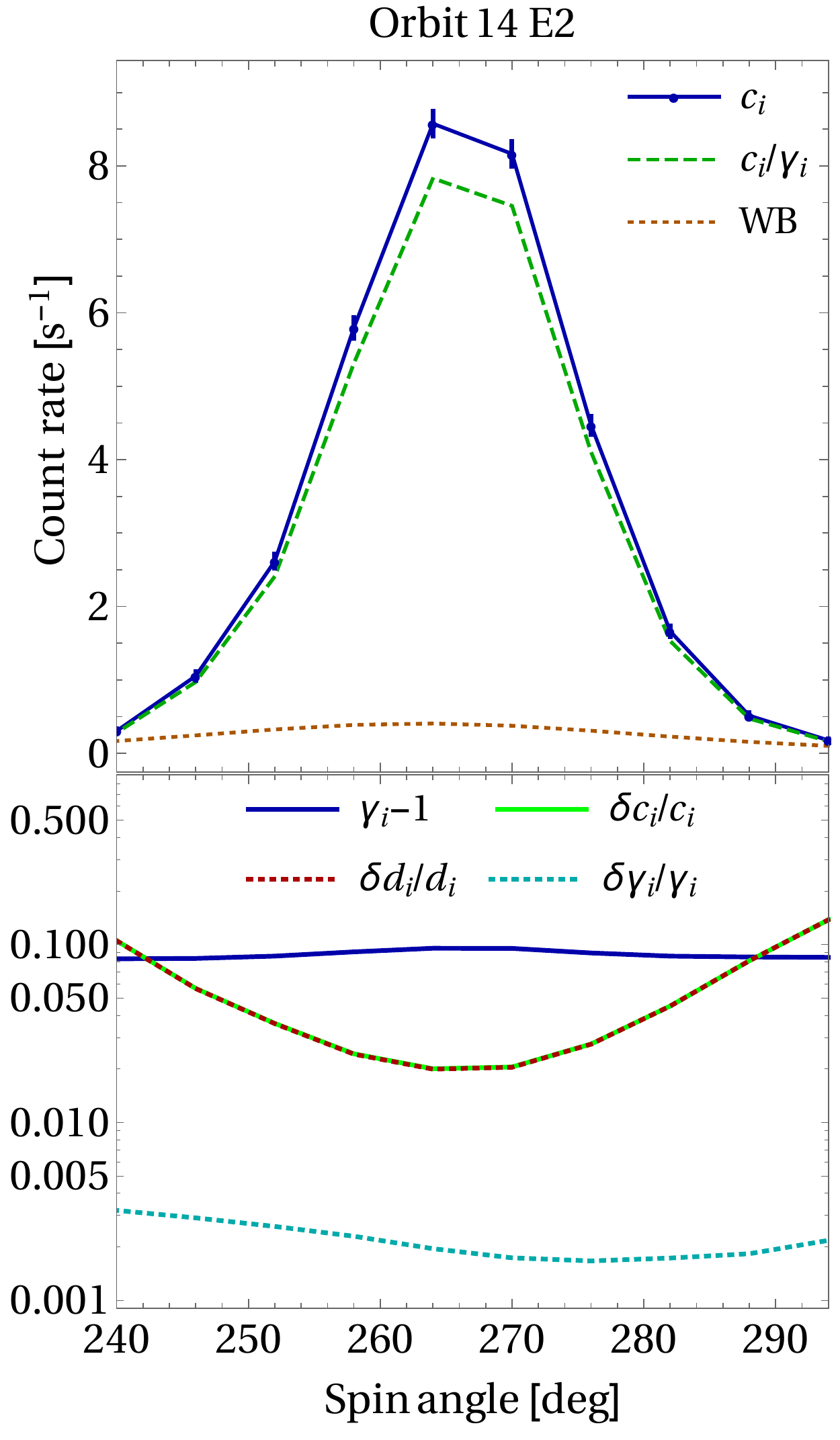}
 \includegraphics[width=.24\textwidth]{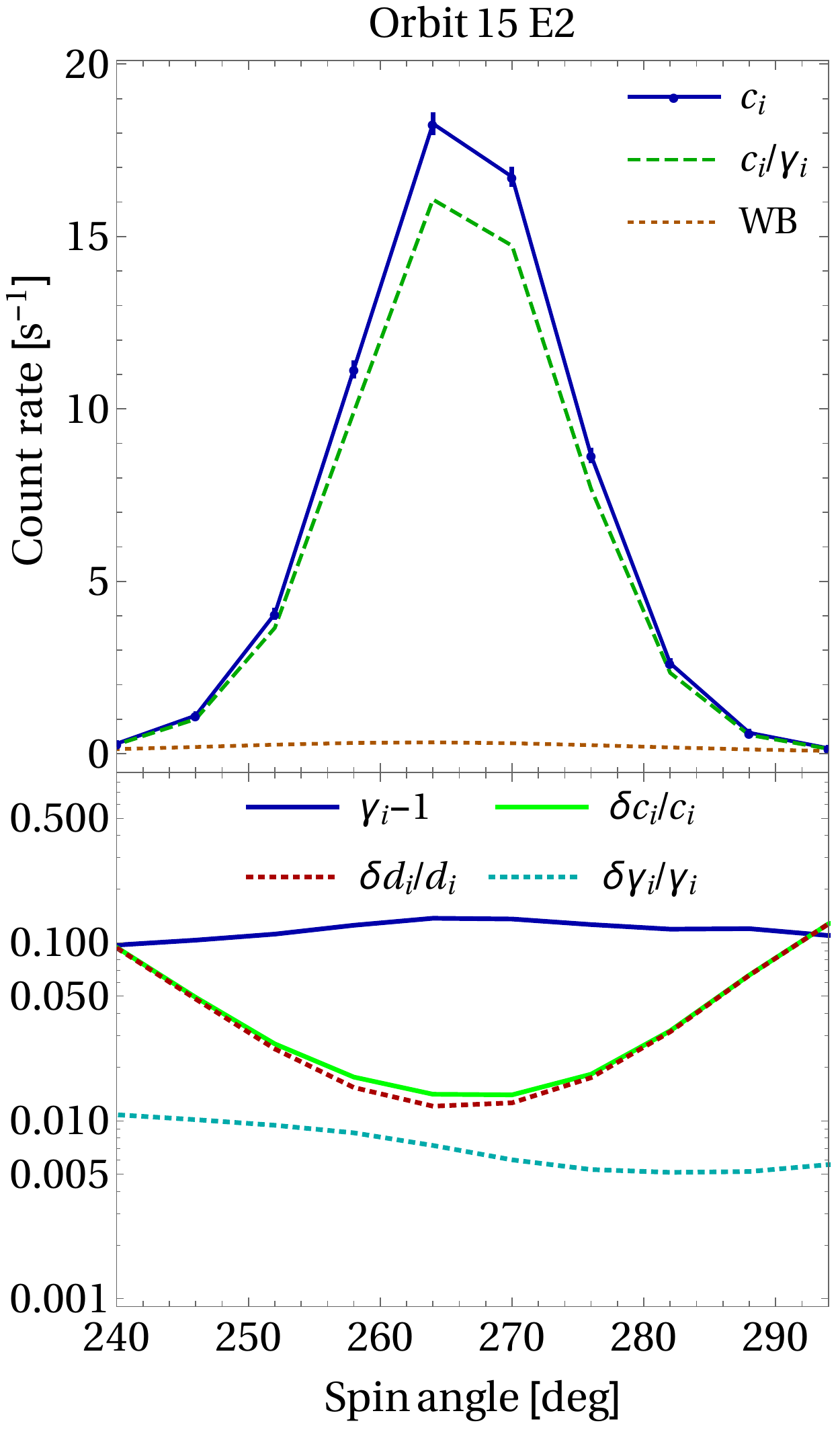}
 \includegraphics[width=.24\textwidth]{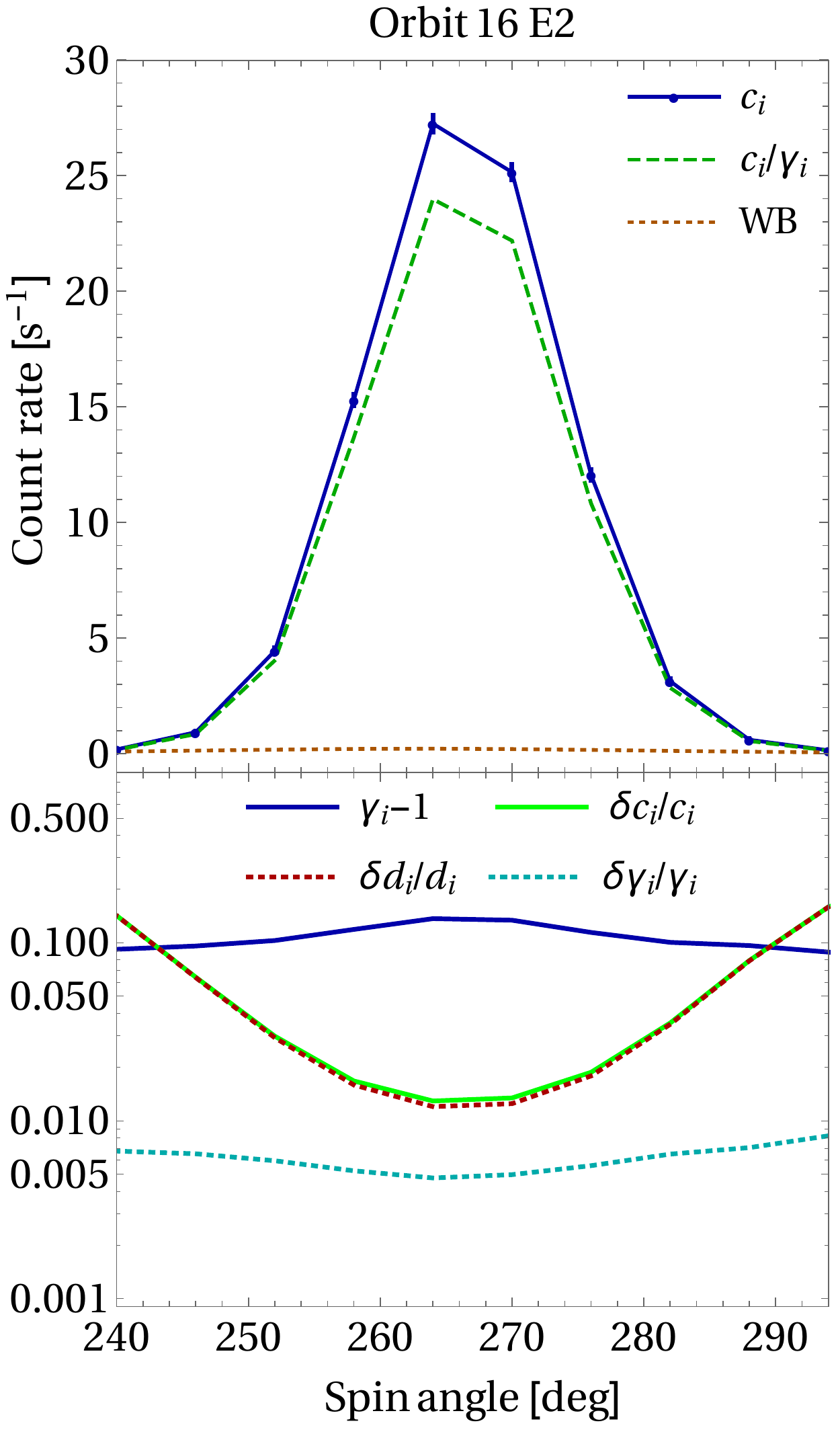}
 
 \hspace{1cm}
 
 \includegraphics[width=.24\textwidth]{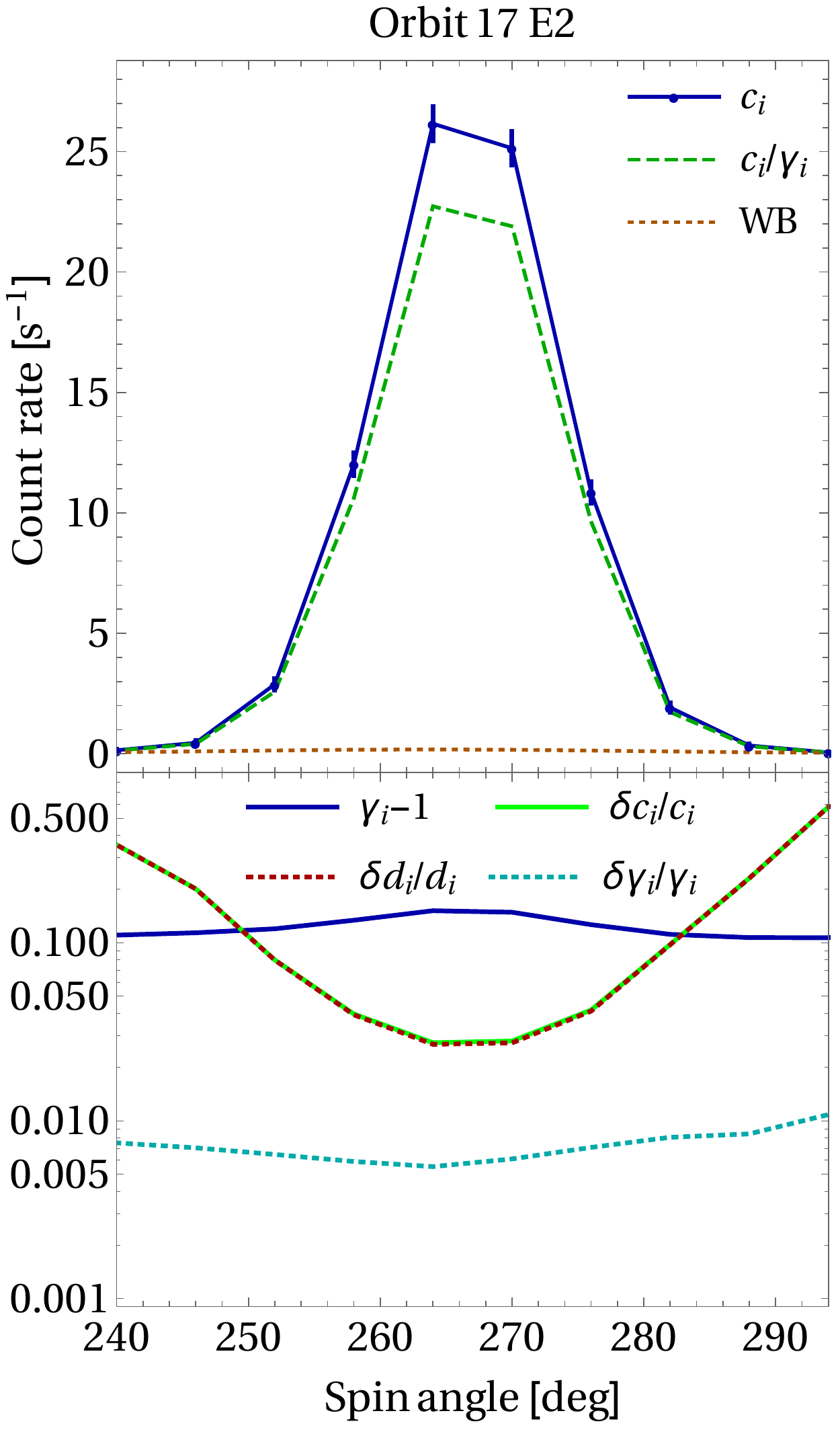}
 \includegraphics[width=.24\textwidth]{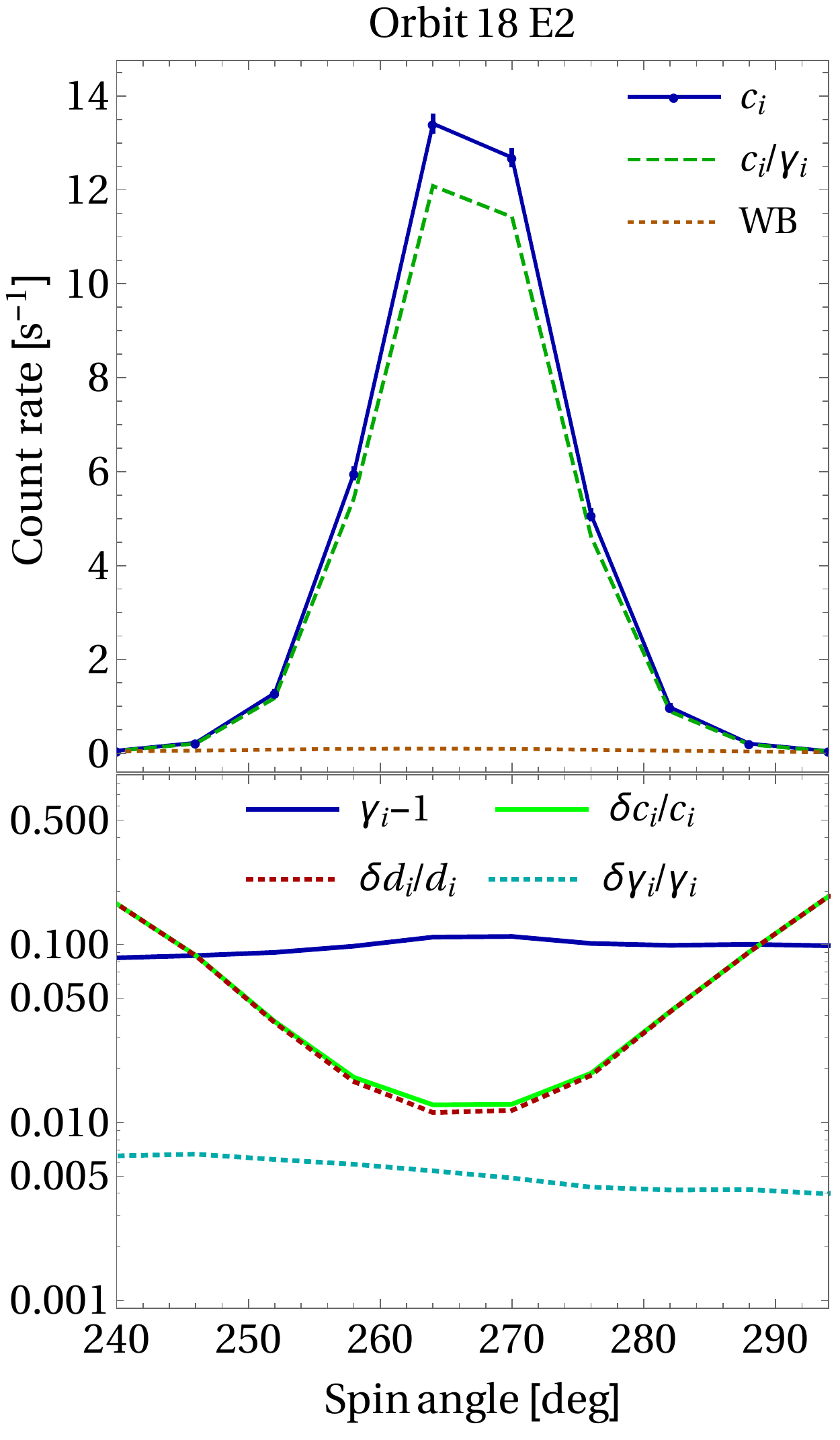}
 \includegraphics[width=.24\textwidth]{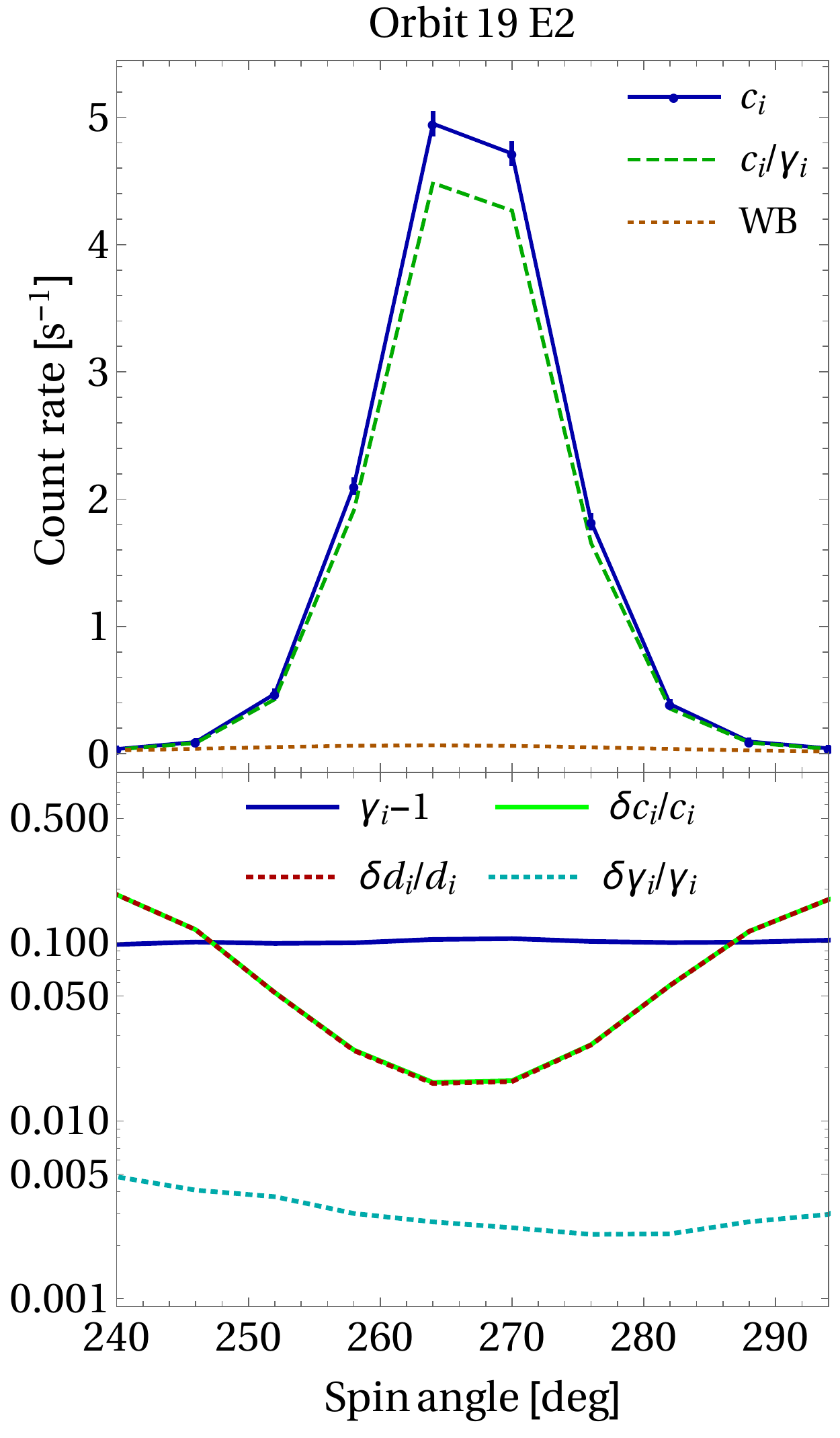}
 
 \caption{Upper part of each panel: bin- and good time-averaged count rates, measured in \emph{IBEX}-Lo energy step 2 during the 2009 ISN observation season in orbits 13--19 for the 6$^\circ$ bins from the spin-angle range from 240$^\circ$ to 294$^\circ$. The blue solid lines represent the count rate corrected for throughput reduction (with the Poisson and throughput correction uncertainties), and the green dashed lines the uncorrected rate. Lower part of each panel: comparison of the correction factor $\gamma_i-1$ (blue) with the uncertainty of the corrected count rate $\delta c_i/c_i$ (green), which includes the Poisson uncertainty $\delta d_i/d_i$ of accumulated counts $d_i$ (dashed red), and the uncertainty of the correction factor $\delta\gamma_i/\gamma_i$ (dashed cyan).}
 \label{unc:uncorr}
\end{figure*}

In Appendix~\ref{throughput}, we derive an analytical model of the interface buffer to statistically predict these losses. We also describe in detail how we estimate the rate of the different types of events transmitted to the CEU. With this model, we can calculate an appropriate TC factor and its uncertainty for each data point. The effective TC factor $\gamma_i$ for bin $i$ is given by Equation~\eqref{eq:lambda}, and its uncertainty $\delta\gamma_i$ by Equation~\eqref{eq:dlambda}. In Figure~\ref{unc:uncorr}, we plot the uncorrected and corrected count rates of sputtered H$^-$ from ISN He measured by \emph{IBEX}-Lo in orbits 13--19. We also show the Poisson uncertainties of the collected data for comparison with the uncertainties of the TC factor. 

During these ISN seasons, the magnitude of the TC substantially exceeds the Poisson uncertainty of the data for all ISN data bins in the analysis, except for those with the lowest count rates. It is a relatively flat function of spin-angle. A typical value of this correction factor is 1.1--1.15, i.e., a correction by 10\%--15\%. However, for all orbits the TC is the largest for those spin-angle bins with the highest count rates, which have the lowest statistical uncertainties. The relative uncertainty of the correction varies between 0.5\% and 5\%, with typical values of about 1\% for the angle bins with the strongest signal, and thus is comparable to the Poisson uncertainties. Consequently, both the correction and its uncertainty must be taken into account in the error system of the analysis.

The bias in the results by \citet{bzowski_etal:12a} and \citet{mobius_etal:12a} that may have been caused by the throughput reduction is substantially lower than suspected by \citet{lallement_bertaux:14a}, and is even much lower than that given by \citet{mobius_etal:12a} as a conservative estimate of the effect which was not fully understood at the time. This has been discussed by \citet{mobius_etal:15a} and \citet{frisch_etal:15a}.

All counting systems with an inevitably finite event processing time suffer from event losses because of the stochastic distribution of the events over time. The losses occur when a new event appears while the measuring instrument is still busy processing the former event. The magnitude of the loss increases with the ratio of the processing time to the mean time between events, which is inversely proportional to the event rate. Using a double buffer mitigates the loss and modifies its dependence on the event rate. The conclusions drawn by \citet{lallement_bertaux:14a} are incorrect for two reasons. First, a double buffer in the \emph{IBEX}-Lo interface softens the loss as a function of the count rate, as we show in Appendix~\ref{throughput}. Second, and more important, the main factor that caused the throughput reduction was not the interstellar atom events but the events due to electrons, which dominated the event stream by an order of magnitude in the peak bins. Thus, the bias on the relation between the peak- and low count-rate data points, and thus on the angular ISN flow distribution, is even much lower. In essence, the relations between data points, which are critical for the determination of the ISN gas temperature and flow vector, are less affected.

\subsection{Background}
\label{uncertainties:background}
From the perspective of ISN observations, all of the events classified as hydrogen in the histogram bin data that are not due to ISN He atom impacts are effectively background. Generally, they are indistinguishable from those events due to ISN He and must be rejected by other methods as much as possible or considered in the uncertainty budget of the analysis. Potential sources of background in \emph{IBEX}-Lo measurements were presented by \citet{wurz_etal:09a}, and the actual background in \emph{IBEX}-Lo measurements was discussed by \citet{fuselier_etal:14a} and \citet{galli_etal:14a} for heliospheric studies and by \citet{kubiak_etal:14a} and \citet{galli_etal:15a} for ISN studies. These authors found that a baseline value for the background exists, independent of the observation season and the direction on the sky, with a count rate of $0.0089 \pm 0.001$~s$^{-1}$ in energy step 2 \citep{galli_etal:14a}. On top of this, a variable portion of the signal exists. It depends on the time and spin-angle, and may be due to remnant magnetospheric atoms, the distributed heliospheric signal, and perhaps other sources that were not rejected by the ``good time'' selection.

An assessment of the background level during the ISN observation portion of Earth's orbit is shown in Figure~3 in \citet{kubiak_etal:14a}. The background varies between $\sim$0.005 and $\sim$0.02 counts per second between orbits, while the maximum ISN signal count rate for orbit 16 (before TC) is $\sim$23~s$^{-1}$. The estimated background is $\sim$5$ \times 10^{-4}$ of the ISN peak rate. The lowest count rates in the data subset selected for ISN analysis are typically on a level of $\sim$2\% of the peak count rate \citep[see][Figure~3]{bzowski_etal:15a}, and thus they exceed the estimated background level by a factor of 30. Hence, the background ought to be subtracted from the signal when one is interested mostly in the low count-rate data bins. Here, we did subtract from the measured count rates the value of $0.0089\pm0.001\text{ s}^{-1}$, found for the background by \citet{galli_etal:14a}. The reported uncertainty is the uncertainty of the mean level of the background and does not concern the bin to bin variation. The statistical variation of the background, which is assumed to follow the Poisson distribution, is already included in the uncertainty of the number of counts.

\subsection{Warm Breeze}
\label{uncertainties:warmbreeze}

The WB discovered by \citet{kubiak_etal:14a} contributes a non-negligible portion to the observed signal. In most spin-angle bins used in the analysis of the ISN He, the signal from the WB is much smaller than the signal from ISN He. However, appropriate subtraction of this signal from the data is required before ISN He parameter fitting. The WB density at the 150 AU is $n_\text{WB}=0.07$ of the ISN He density there. The inflow parameters are $\lambda_\text{WB}=240.5^\circ$, $\beta_\text{WB}=11.9^\circ$, $T_\text{WB}=15\,000\text{ K}$, and $v_\text{WB}=11.3\text{ km s}^{-1}$. These parameters were determined with relatively large uncertainties, which we adopted as follows: $\delta n_\text{WB}=0.04$, $\delta\lambda_\text{WB}=10^\circ$, $\delta\beta_\text{WB}=7^\circ$, $\delta T_\text{WB}=7000\text{ K}$, and $\delta v_\text{WB}=4\text{ km s}^{-1}$. \citet{kubiak_etal:14a} do not present the correlation between the parameters, and thus we assume here that the covariance matrix is diagonal.

The count rates given by Equation~\eqref{eq:countrate} depend linearly on the contribution of the WB. We calculate the uncertainty of the WB contribution from simulations:
\begin{equation}
 \delta_\epsilon c_i=\frac{\partial c_i}{\partial \epsilon}\delta\epsilon=-\frac{\partial w_i}{\partial \epsilon}\delta\epsilon\approx \frac{w_i(-\delta\epsilon)-w_i(+\delta\epsilon)}{2}\, ,
 \label{eq:wbunc}
\end{equation}
where $\epsilon= n_\text{WB},\, \lambda_\text{WB},\, \beta_\text{WB},\,  T_\text{WB},\, v_\text{WB}$. 

\subsection{Spacecraft Orientation}
\label{uncertainties:orientation}
In Section~\ref{strategy:orientation}, we mentioned the reanalysis of the spacecraft orientation in the orbits used in the ISN analyses. We developed a new algorithm for determining the spin-axis pointing in equatorial coordinates, but both methods, the original and the new one, use the same data from Star Tracker. For most orbits not used in the ISN analysis, the pointing obtained using the new algorithm differs from the previous determination by about 0.005$^\circ$.  We assess the uncertainty of the spacecraft pointing as $\pm 0.01^\circ$ for all possible pairs of perpendicular directions.

The spin-axis pointing is one of the parameters used in the simulations to determine of the expected count rate for each set of the searched ISN He parameters in the WTPM. Propagating the spin-axis pointing uncertainties into the uncertainties of count rates $\delta c_i$ requires determining how the signal depends on the spin-axis pointing, which defines the great circle on the sky tracked by the \emph{IBEX}-Lo boresight. Determination of this uncertainty is not possible from the observations themselves because the pointing is constant for each orbit/arc (which we verified by analyzing the data from Star Tracker). Thus, we assess the changes related to the shift of the spin-axis pointing with auxiliary simulations of the ISN He and WB fluxes, which constitute almost the entire signal in the analyzed spin-angle bins. Specifically, we estimate the partial derivative of the count rate $c_i$ with respect to the direction of the spin-axis pointing by the finite difference of the simulated flux $s_i$:
\begin{equation}
 \frac{\delta_{\zeta}c_i}{c_i}=\frac{1}{c_i}\frac{\partial c_i}{\partial \zeta}\delta\zeta\approx \frac{1}{s_i}\frac{s_i(\zeta=+0.1^\circ)-s_i(\zeta=-0.1^\circ)}{0.2^\circ}\delta \zeta \, ,\label{eq:orientation}
\end{equation}
where $\zeta$ represents the deviation from the estimated spin-axis pointing, and $s_i$ is the appropriate simulated flux value in bin $i$. The spacecraft orientation is determined by two angles, and thus we need to choose two perpendicular directions to describe this uncertainty. For convenience, we take the directions of the declination ($\delta$) and right ascension ($\alpha$) of the spin-axis pointing. However, the final results do not depend on this choice.

The change with declination could be easily calculated by performing simulations with the spin-axis pointing shifted in declination by $\pm0.1^\circ$, whereas the shift in right ascension is $\pm (\cos \delta)^{-1}$ due to the properties of the equatorial coordinate system.\footnote{In fact, we evaluate Equation~\eqref{eq:orientation} for $\zeta=\tilde{\alpha}\equiv(\cos \delta)^{-1}\alpha$ with $\Delta\tilde{\alpha}=\pm0.1^\circ$.} We calculate the finite difference for larger deviations than the estimated uncertainty of the pointing to avoid a numerical bias in the result \citep[see][]{sokol_etal:15b}. The results obtained from Equation~\eqref{eq:orientation} are denoted as $\delta_\delta c_i$ and $\delta_\alpha c_i$ for the shifts of 0.01$^\circ$ in declination and right ascension, respectively.

\begin{figure*}[ht!]
 \includegraphics[width=.24\textwidth]{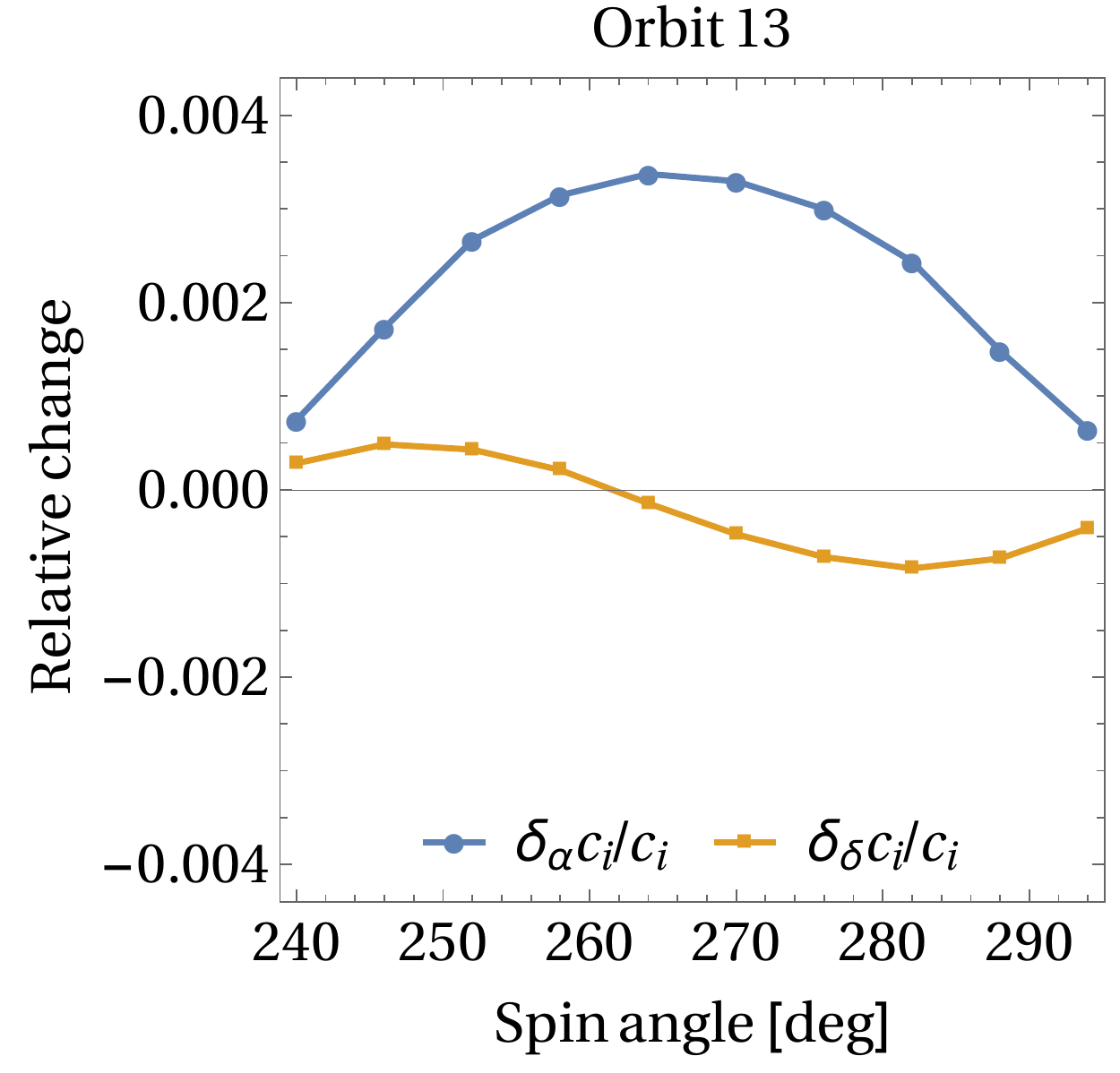}
 \includegraphics[width=.24\textwidth]{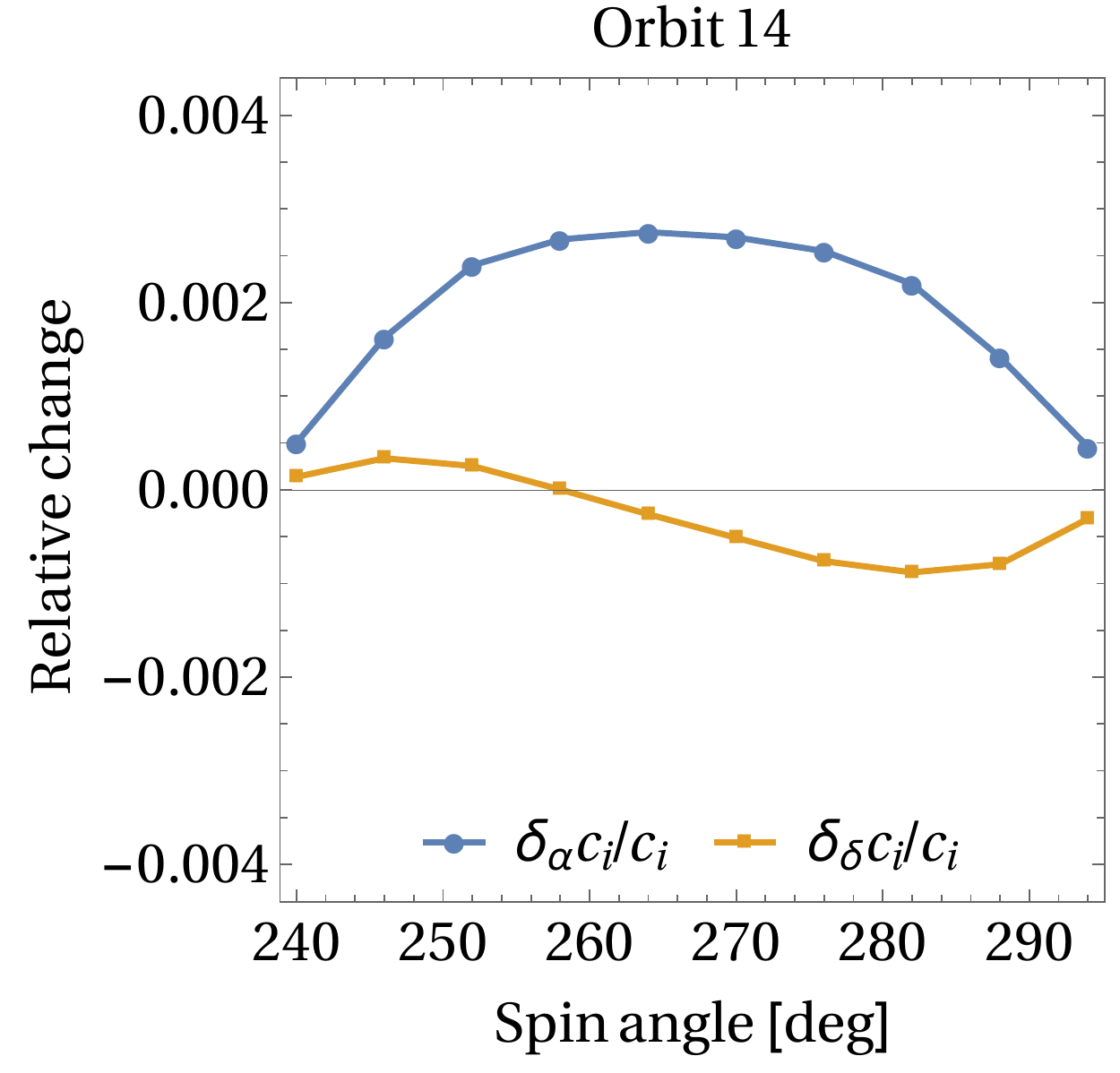}
 \includegraphics[width=.24\textwidth]{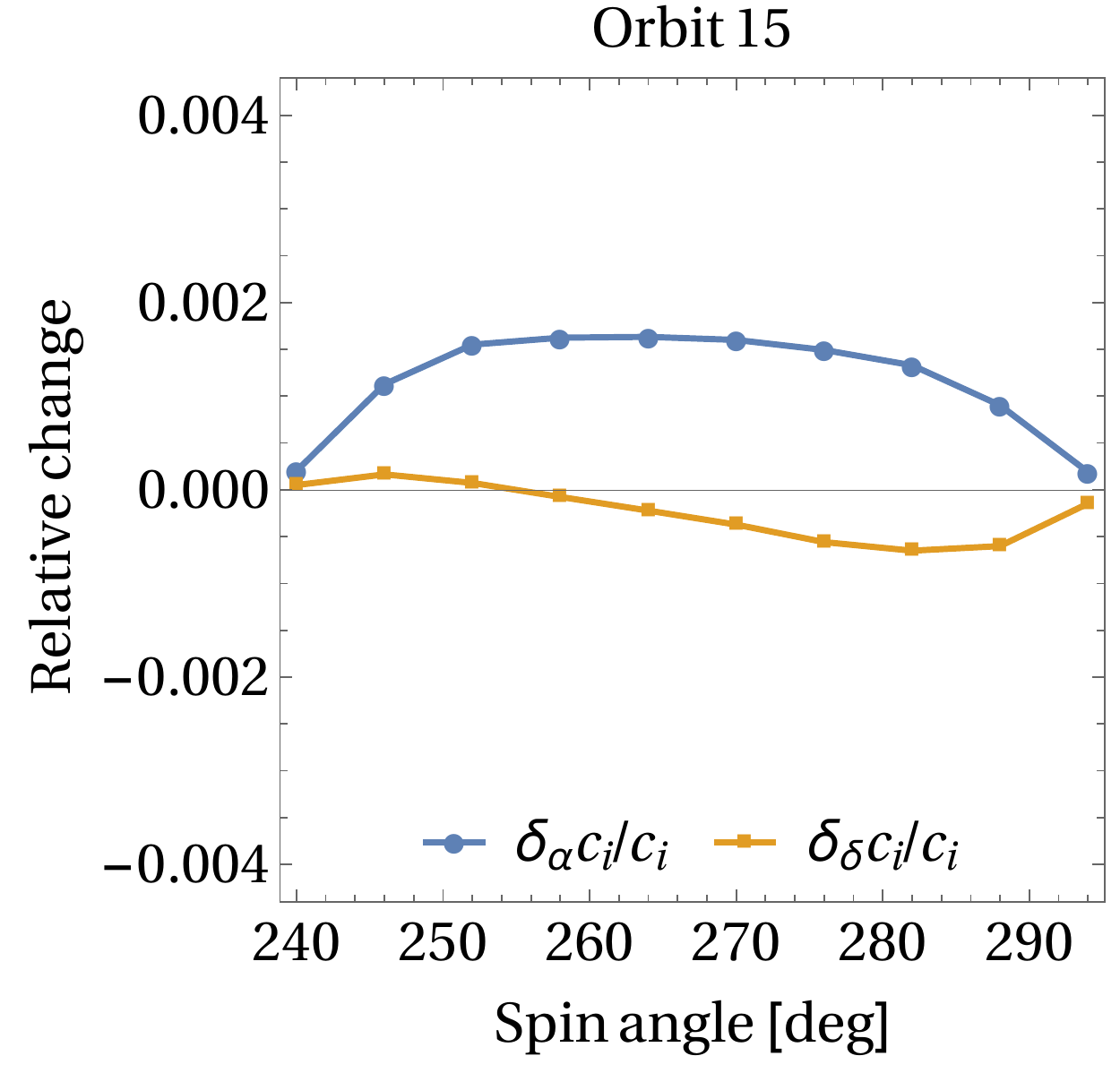}
 \includegraphics[width=.24\textwidth]{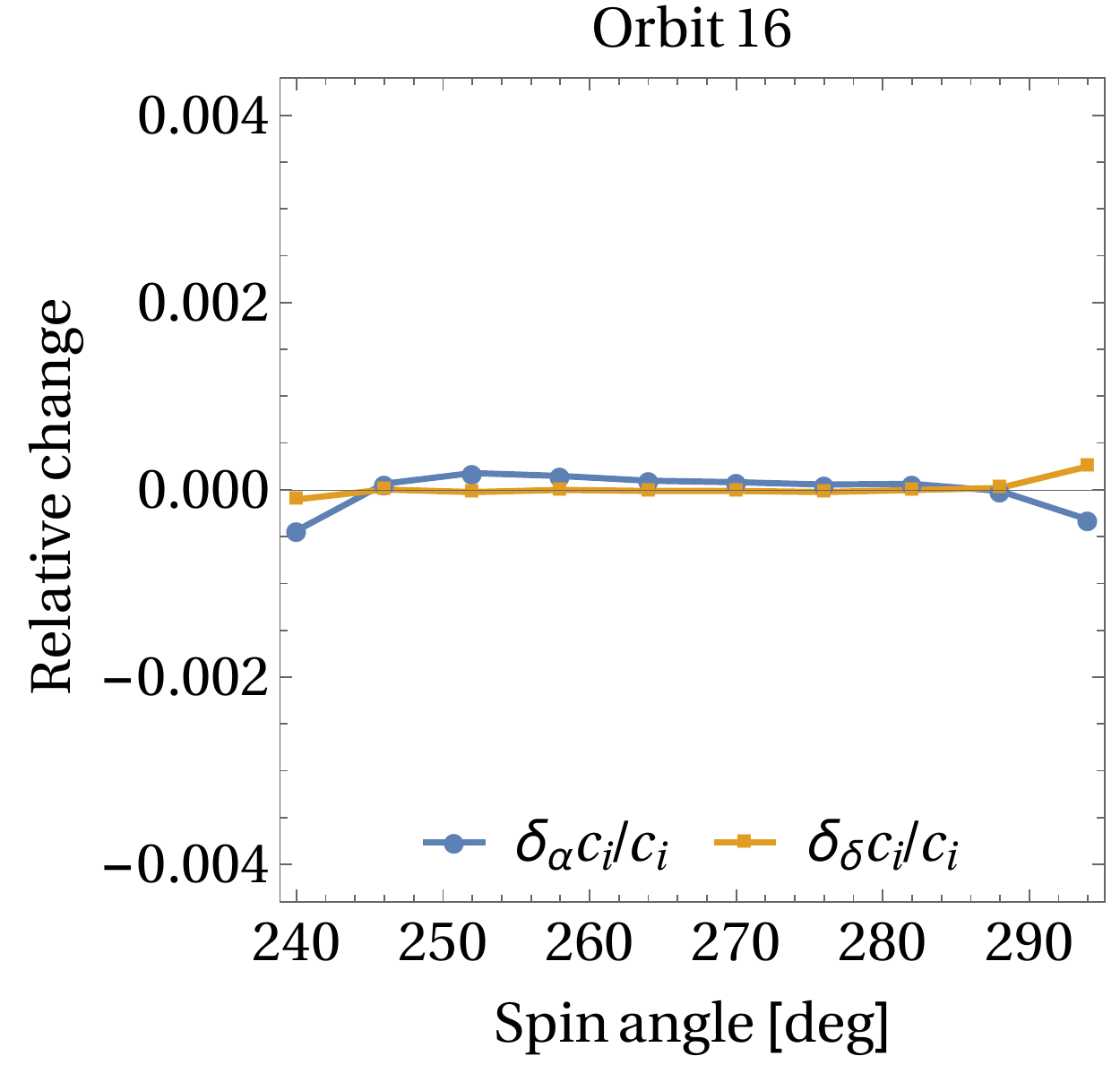}
 
  \hspace{1cm}
 
 \includegraphics[width=.24\textwidth]{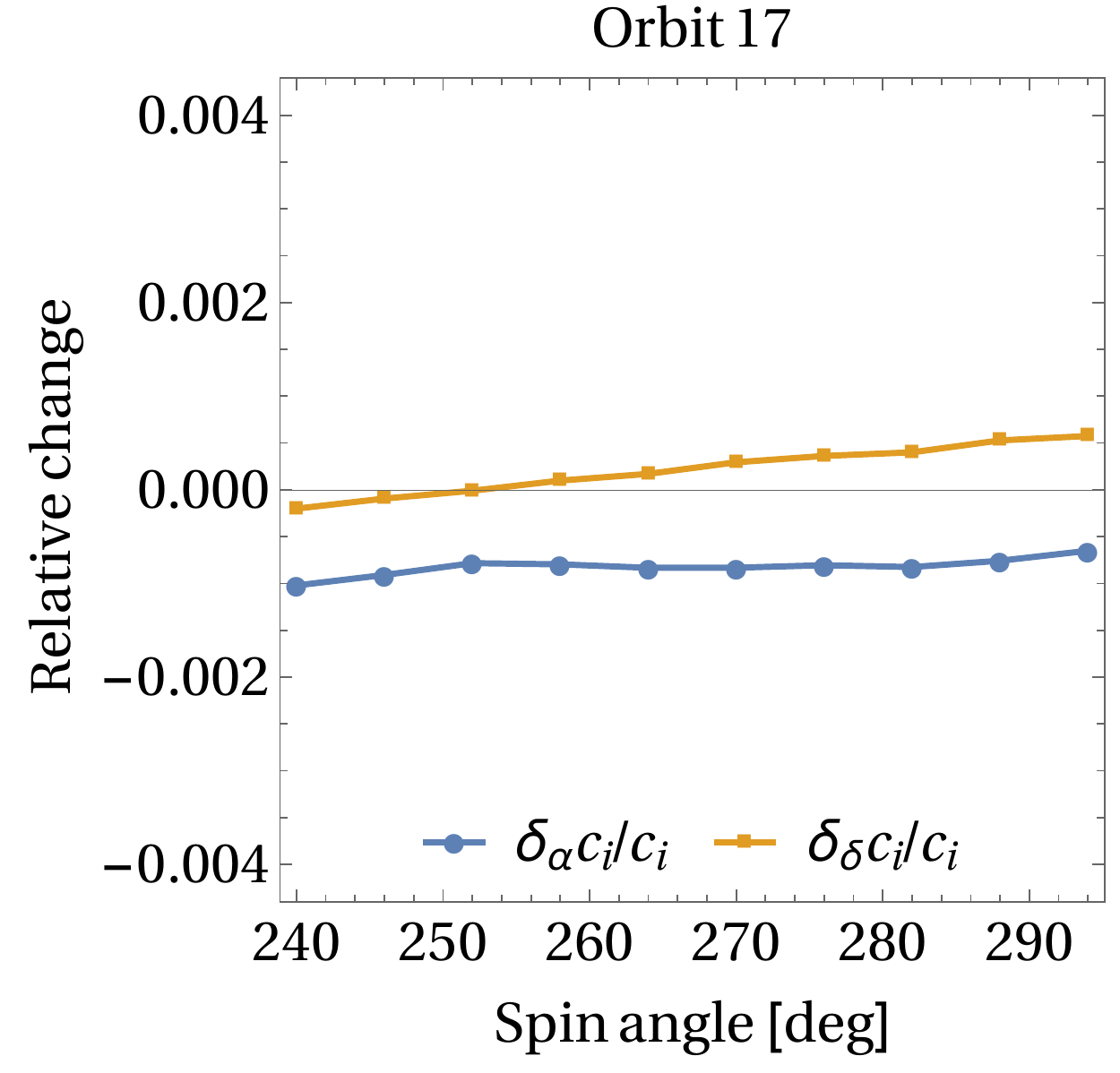}
 \includegraphics[width=.24\textwidth]{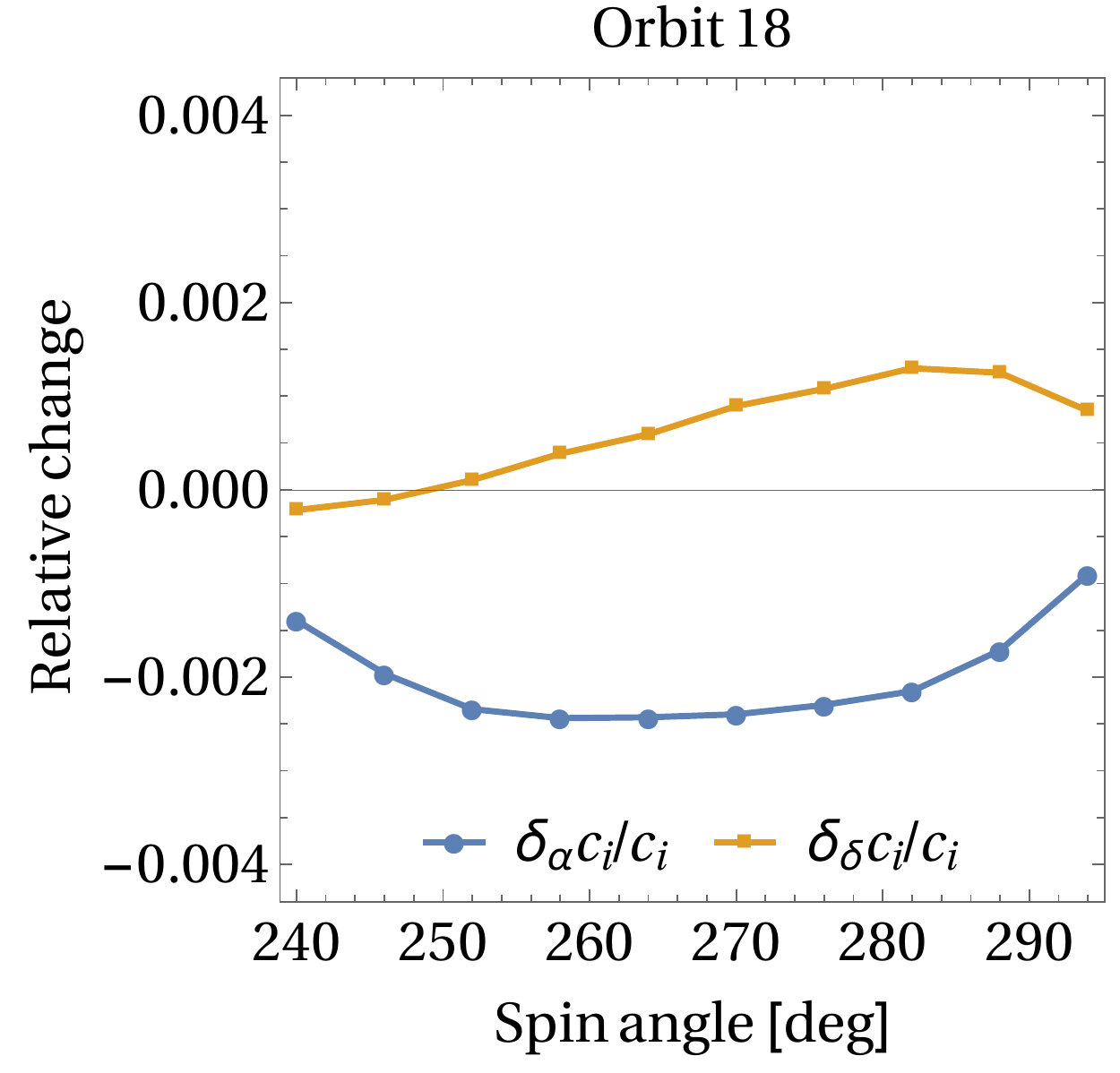}
 \includegraphics[width=.24\textwidth]{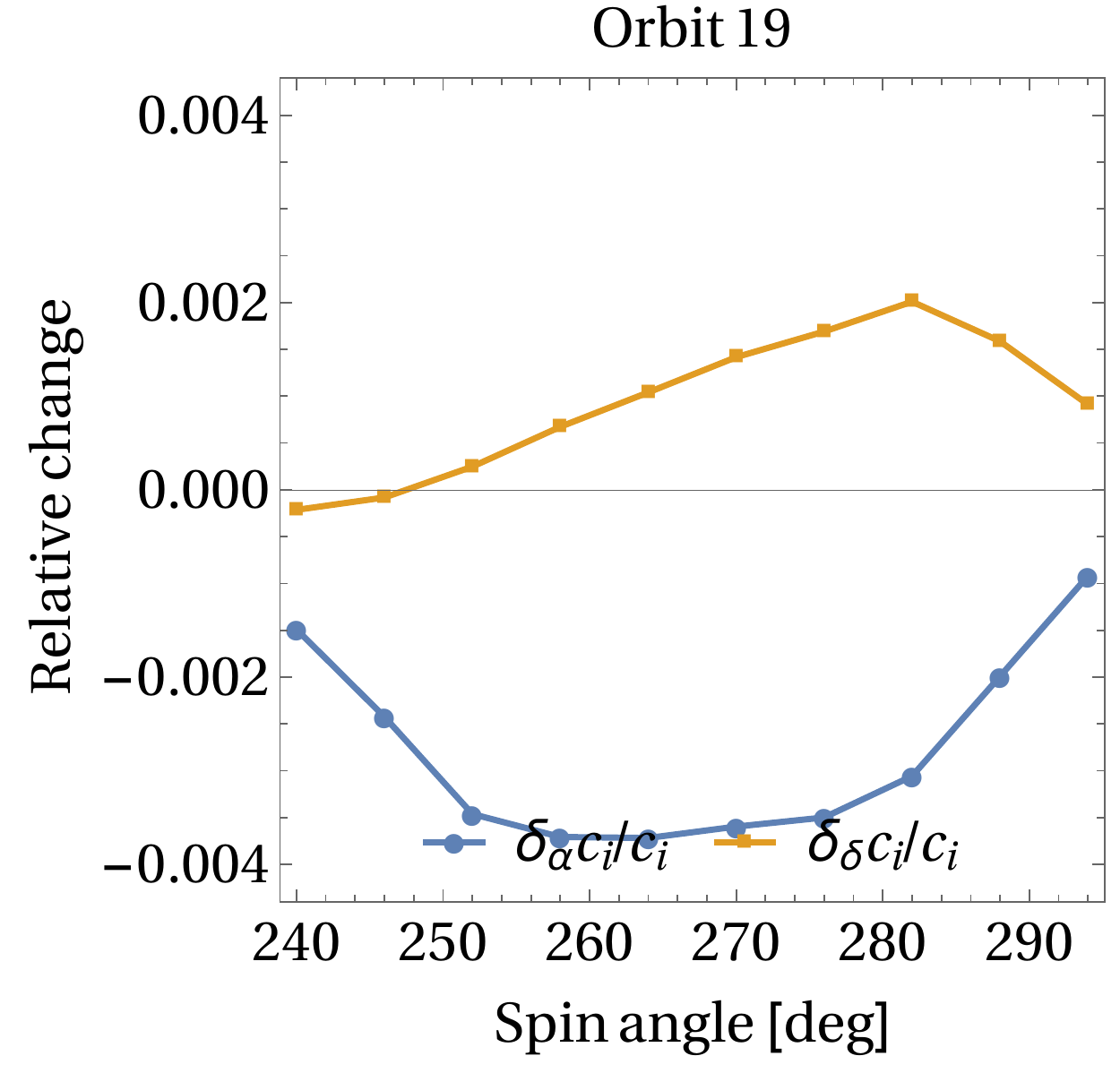}
 
 \caption{Relative changes of the count rates in orbits 13--19 for a spin-axis shifted by 0.01$^\circ$ in the directions of declination (orange) and right ascension (blue) for the ISN He parameters: $\lambda=255.4^\circ$, $\phi=5.0^\circ$, $T=7\,900$ K, $v=26.1$ km s$^{-1}$.}
 \label{unc:orient}
\end{figure*}

Equation~\eqref{eq:orientation} explicitly depends on the results of simulations, so one might suspect that this uncertainty should be calculated separately for each parameter set used in the fitting. This would have resulted in a much higher numerical complexity of the problem. However, this is a second-order effect, proportional to the second derivative of the simulated signal $s_i$ with respect to the spin-axis shift and the inflow parameters. We verified by simulation that these uncertainties do not change substantially if we investigate the proximity of the so-called correlation line of parameters \citep{bzowski_etal:12a,mccomas_etal:12b,mccomas_etal:15a,mobius_etal:12a}. Consequently, we perform these simulations once for the best currently known parameter set given by \citet{mccomas_etal:15a}. For completeness, we also add a contribution to the simulated signal from the WB \citep{kubiak_etal:14a}. In Figure~\ref{unc:orient}, we present the relative changes of the signal given by Equation~\eqref{eq:orientation}.  

The changes we obtained do not exceed 0.5\% and are smaller in the peak orbits. They are thus smaller than the relative statistical uncertainties of the number of counts in the Poisson process (Section~\ref{uncertainties:counts}). However, for the spin-axis pointing, the resulting uncertainties for count rates in bins from an individual orbit cannot not be considered as uncorrelated. For example, in orbit 13, the count rate in the bin at spin-angle 264$^\circ$ could be larger by $\sim$0.3\%, but only if other bins are also proportionally larger, as illustrated in Figure~\ref{unc:orient}. In Section~\ref{uncertainties:covariance}, we present how this uncertainty should be included in the covariance matrix. This uncertainty correlates the count rates between bins from one orbit, but does not correlate the results from different orbits because in each orbit the spin-axis pointing is determined separately and the uncertainties in the spin-axis pointing are not correlated between the orbits.

\subsection{IBEX-Lo Boresight}
\label{uncertainties:boresight}
\emph{IBEX} was designed to rotate at $\sim$4~rpm around the Z-axis in the spacecraft coordinate system. The \emph{IBEX}-Lo boresight should point toward the negative Y-axis. Thus, the angle between the spin-axis and the boresight should be equal to 90$^\circ$, and so the boresight should follow a great circle in the plane perpendicular to the spin-axis. In reality, however, the angle between the spin-axis and the boresight could differ from 90$^\circ$ by a small elevation angle $\theta$ due to the inevitable assembly imperfections (for more details see \citealt{hlond_etal:12a}). The boresight could also be shifted in spin-angle by a small angle $\eta$, i.e., at spin-angle 0$^\circ$, the boresight could be offset from the point on the great circle that is closest to the ecliptic pole. 

For an ideal spacecraft, the values for the angles $\theta$ and $\eta$ should be equal to 0. In reality, many effects could influence these values. One effect is the actual orientation of the spin-axis in the spacecraft coordinate system. Star Tracker data show that the Z-axis traces a small circle on the sky with a radius of $\sim$0.08$^\circ$ to $0.11^\circ$. \emph{IBEX}-Lo is equipped with the Star Sensor, mounted on the \emph{IBEX}-Lo baseplate. \citet{hlond_etal:12a} performed a comparison of the data from the Star Sensor and Star Tracker, and showed that the differences between boresight position derived from each of them stays within $\sim$0.1$^\circ$ in elevation ($\theta$) and spin-angle ($\eta$). Additionally, the cross-analysis of the observations performed in the high- and low-resolution modes of \emph{IBEX}-Lo \citep{mobius_etal:15b} determined that the effective boresights in both resolution modes agree within $\sim$0.15$^\circ$ in spin-angle. The precision of this measurement was not sufficient to identify an offset that statistically differs from 0. Based on this insight, we assume that $\theta$ and $\eta$ are equal to 0 and the uncertainties for these angles are $\delta\theta=0.15^\circ$ and $\delta\eta=0.15^\circ$.

The non-zero value of $\theta$ would imply that the boresight does not follow a great circle, but a somewhat smaller circle on the sky, whereas an angle $\eta \neq 0$ would result in a small shift in the count rate as a function of spin-angle. To assess the  potential magnitude of the first effect, we used simulations with a tilt of the collimator to the spin-axis $\pm$0.1$^\circ$, as previously for the case of the uncertainty in spin-axis pointing (Section~\ref{uncertainties:orientation}). The related uncertainty in count rate could be calculated from Equation~\eqref{eq:orientation} for $\zeta=\theta$. In this case, the character of the angle is different than it was previously. As previously for the uncertainties of the spin-axis pointing, the uncertainty that is obtained $\delta_\theta c_i/c_i$ does not depend strongly on the choice of the simulation parameters.

In the case of angle $\eta \neq 0$, we can use actual data because the shift is in the spin-angle direction. Thus, we use count rates measured in the neighboring bins:
\begin{equation}
 \frac{\delta_\eta c_i}{c_i}=\frac{1}{c_i}\frac{c_{i+1}-c_{i-1}}{12^{\circ}}\delta\eta\, ,
\end{equation}
where $c_{i+1}$ and $c_{i-1}$ are the measured count rates in the subsequent and the former bins, respectively, for the same orbit. 

\begin{figure*}[ht!]
 \includegraphics[width=.24\textwidth]{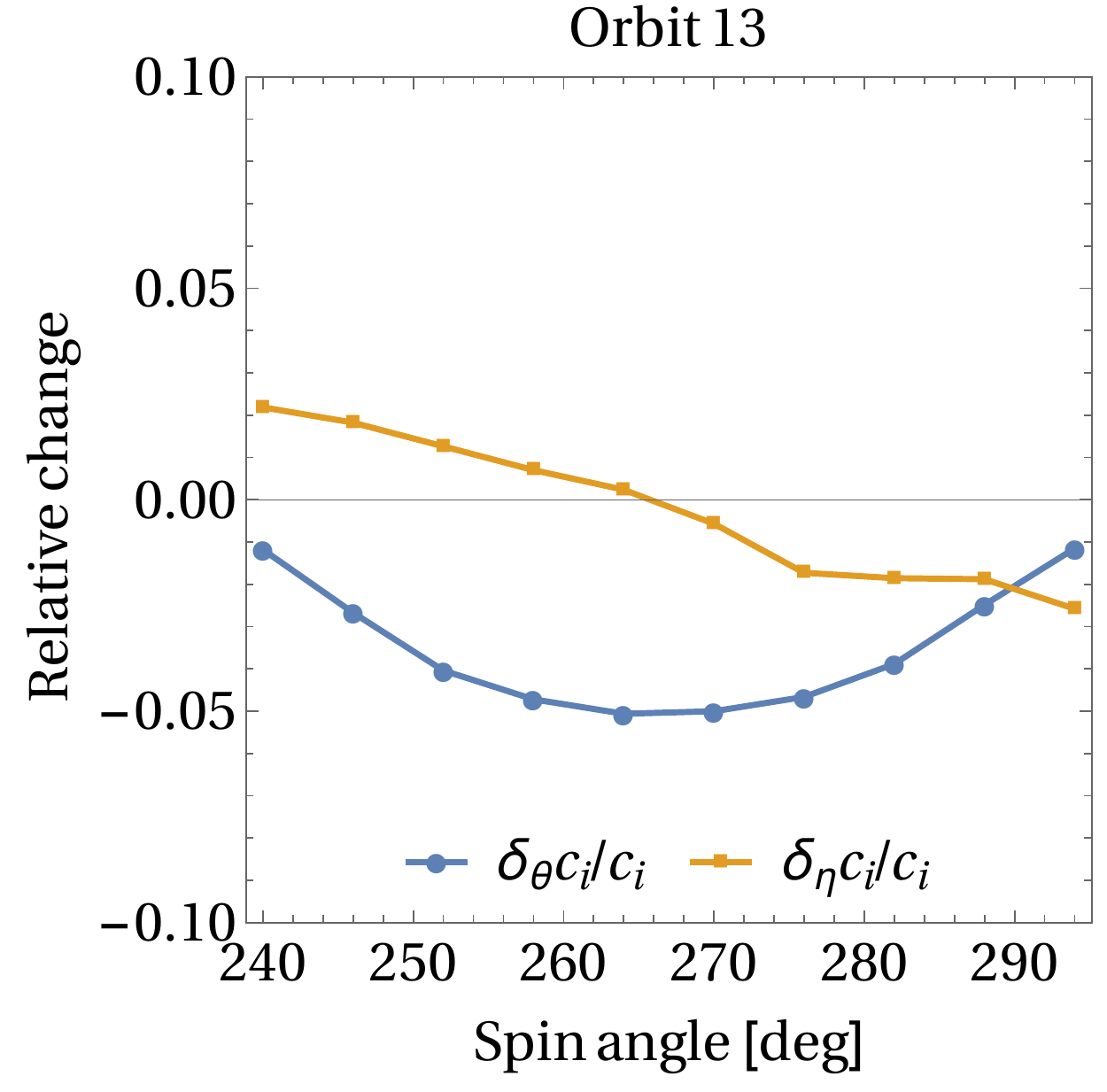}
 \includegraphics[width=.24\textwidth]{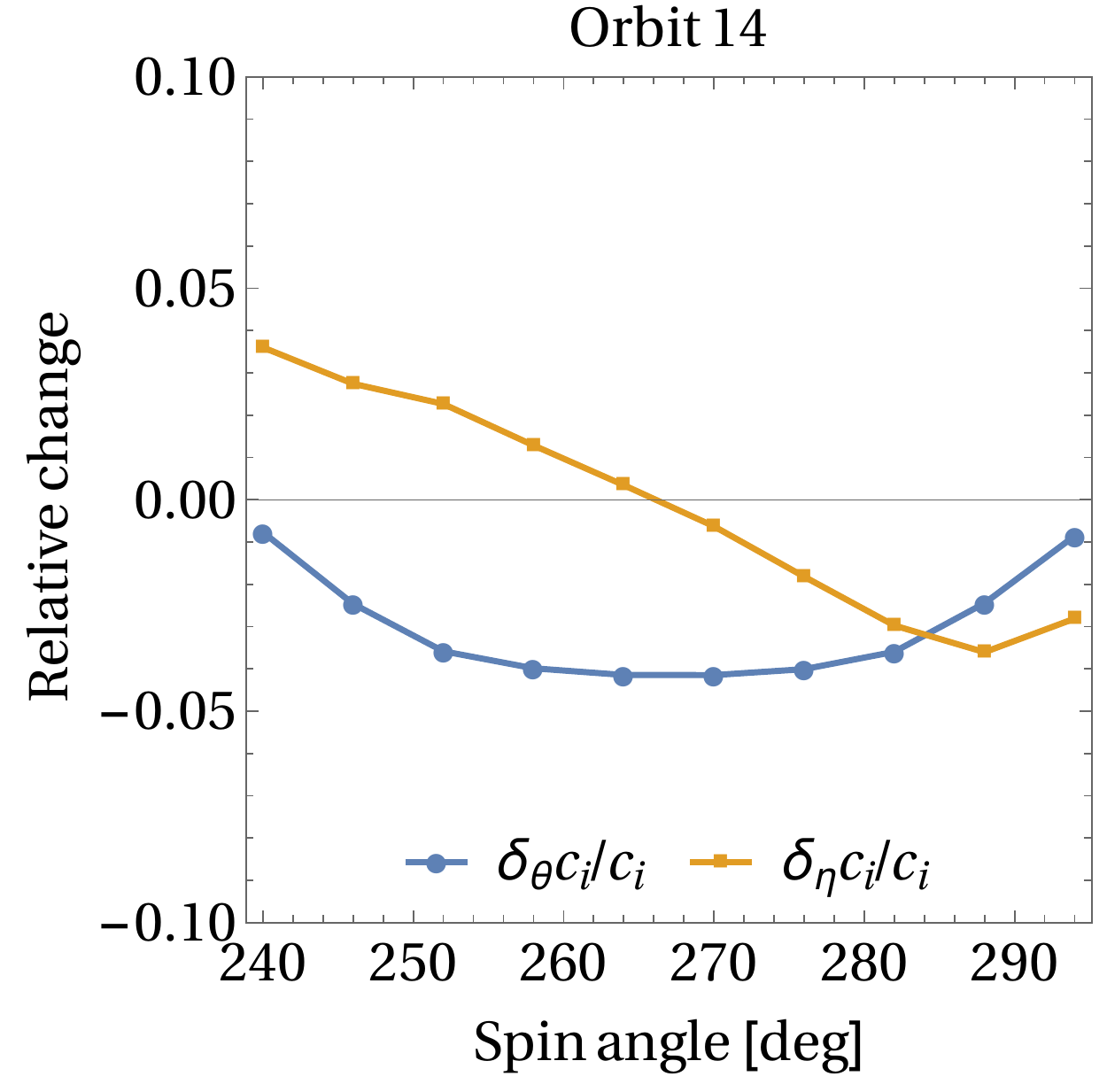}
 \includegraphics[width=.24\textwidth]{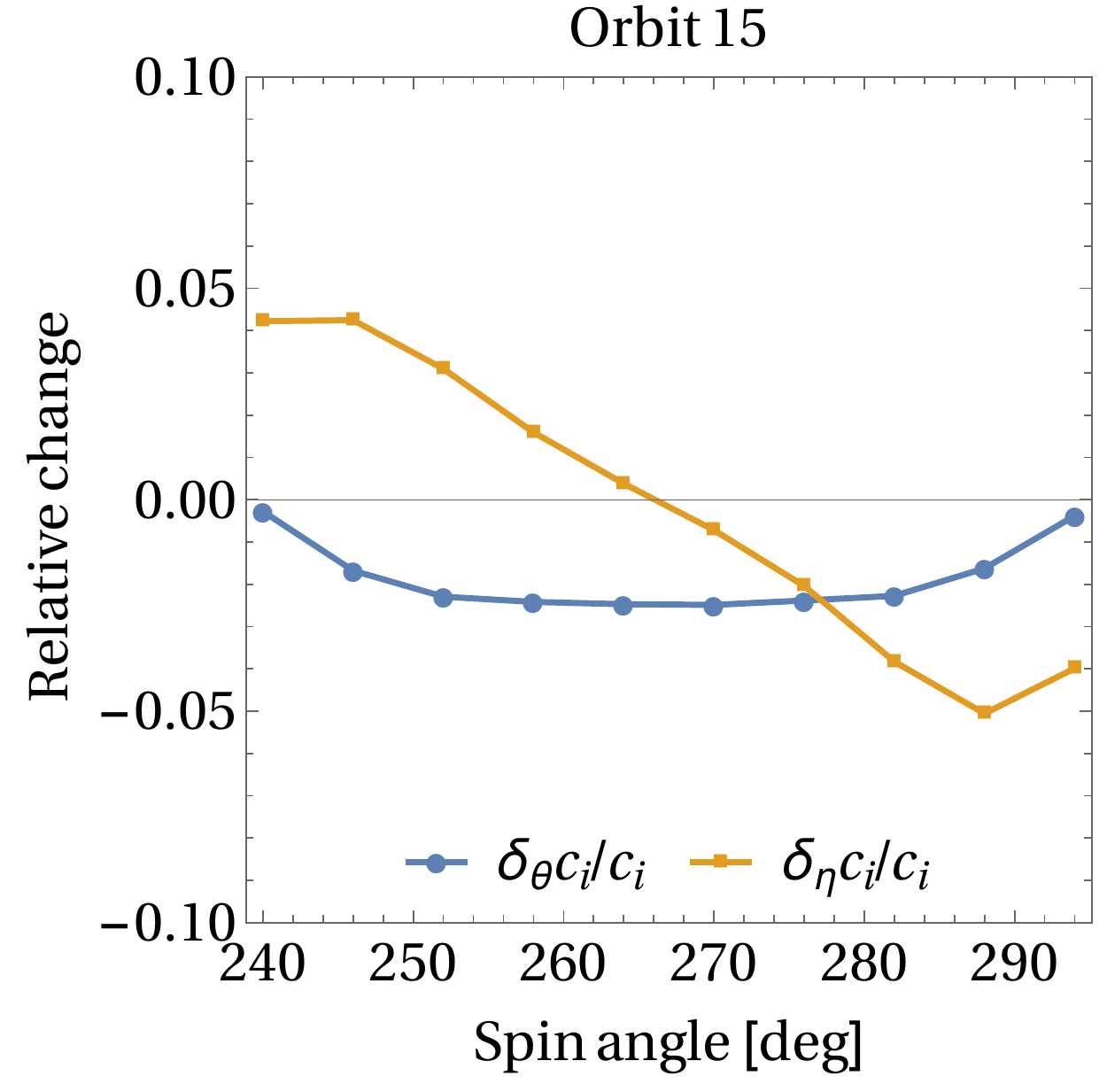}
 \includegraphics[width=.24\textwidth]{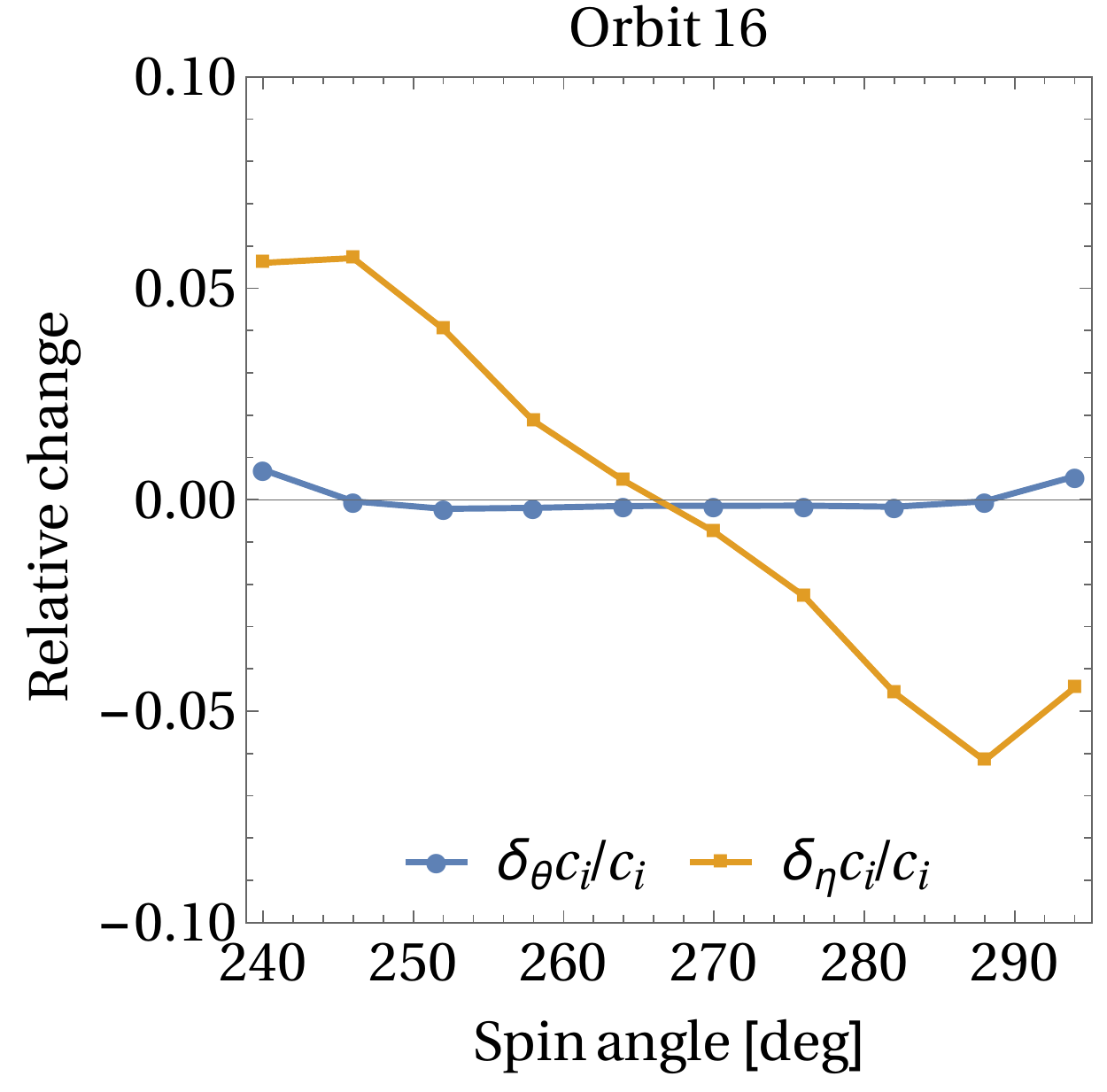}
 
  \hspace{1cm}
 
 \includegraphics[width=.24\textwidth]{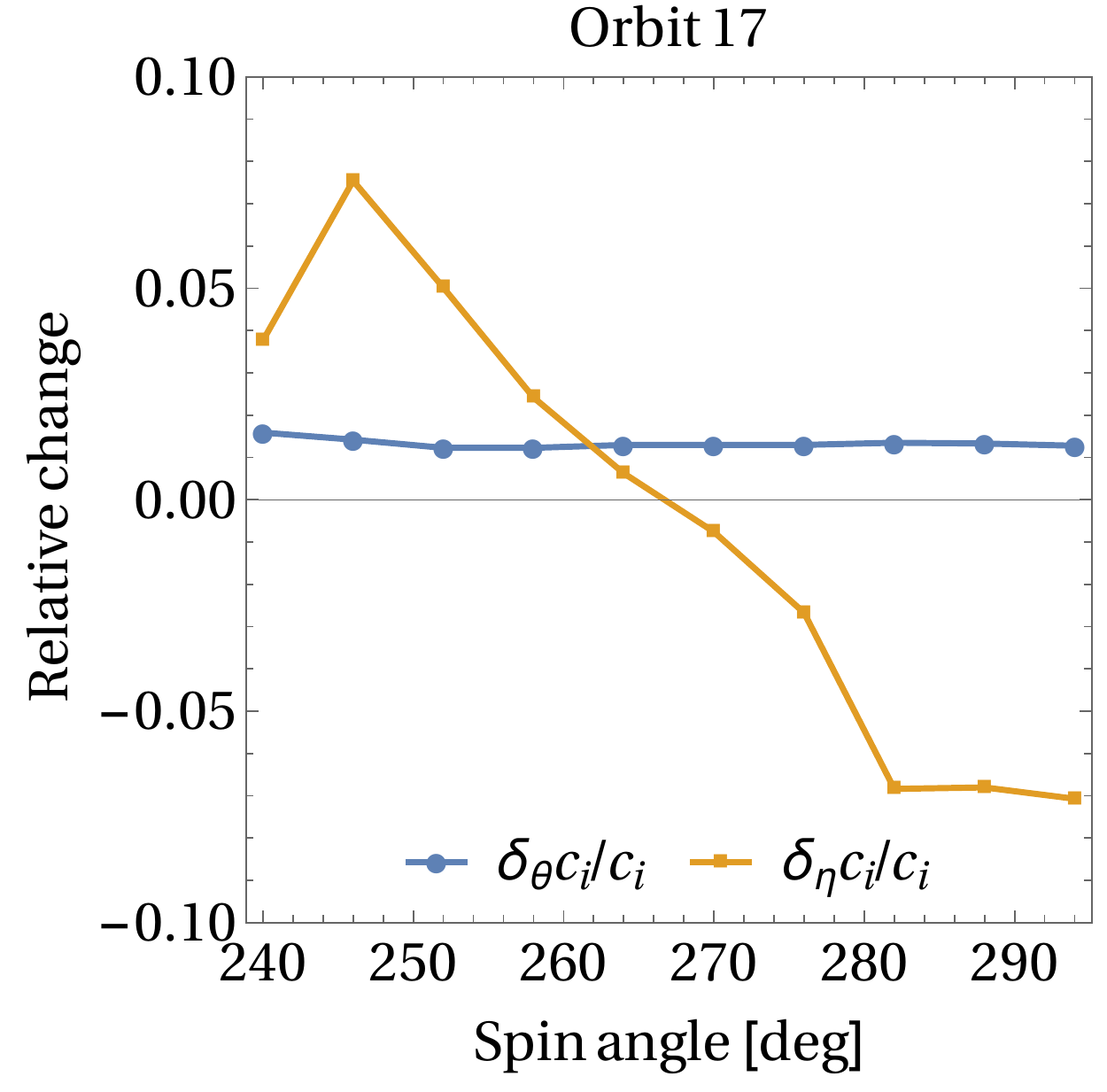}
 \includegraphics[width=.24\textwidth]{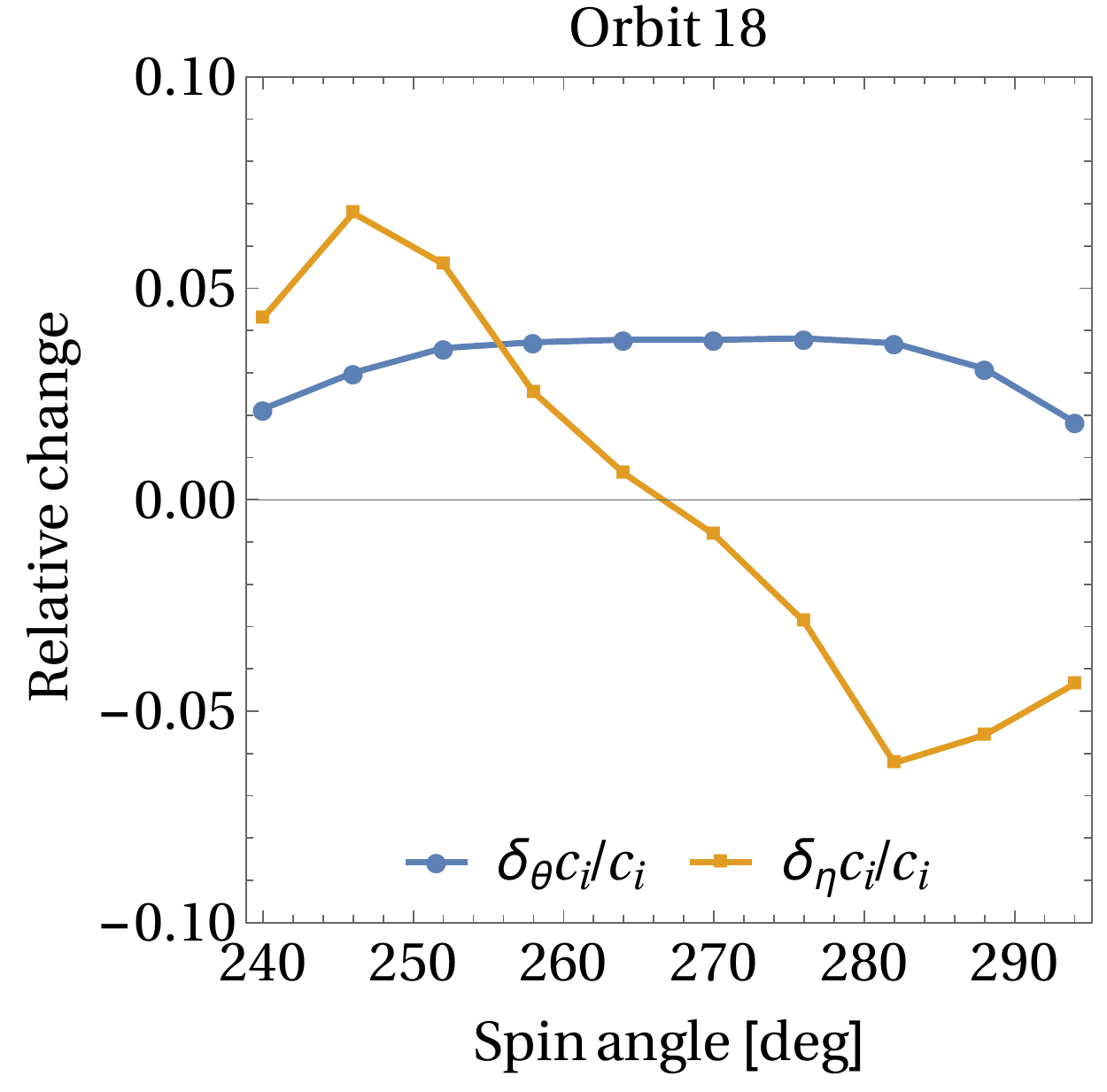}
 \includegraphics[width=.24\textwidth]{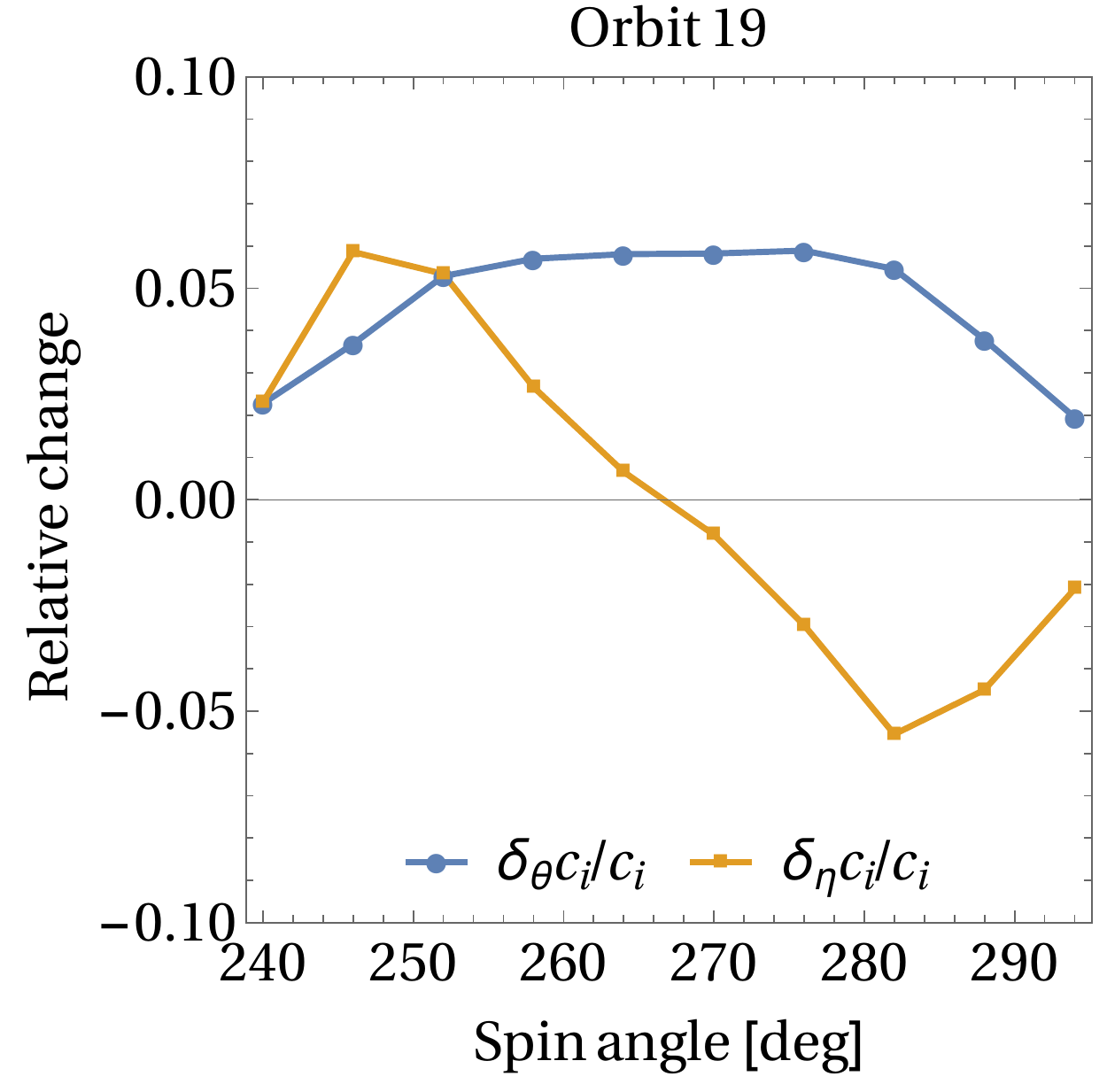}
 
 \caption{Relative changes of the count rates on orbits 13--19 for the angle between the boresight and the spin-axis shifted by $\delta\theta=0.15^\circ$ and the spin-angle shifted by $\delta\eta=0.15^\circ$ (orange). The used simulations were calculated with the same parameters as listed in Figure~\ref{unc:orient}.}
 \label{unc:boresight}
\end{figure*}

Figure~\ref{unc:boresight} shows the relative uncertainties of the count rates calculated for $\delta\theta=0.15^\circ$ and $\delta\eta=0.15^\circ$. The obtained values are relatively large and could even exceed 5\%, i.e., they are of the order of the uncertainties of the count number described in Section~\ref{uncertainties:counts}. Consequently, they must be appropriately included in the data uncertainty system. 

The angles $\theta$ and $\eta$ are constant in time and do not change between orbits. This uncertainty is systematic, i.e., it is related to the design of the instrument. Consequently, the uncertainties they introduce are correlated for all of the bins, even those from different orbits. This correlation means that the measured count rates could differ because of this uncertainty, as given by the results in Figure~\ref{unc:boresight}, e.g., the measured count rates could be smaller in orbits 13--15 only if the measured count rates in orbits 17--19 are higher. Consequently, the relatively high value of the changes in the expected count rate due to this uncertainty does not reduce the data quality significantly if properly addressed in the adopted uncertainty system.

\subsection{Other Sources}
\label{uncertainties:other}
The uncertainties described in Sections~\ref{uncertainties:counts}--\ref{uncertainties:boresight} cover all of the important, currently known sources of uncertainties. Some other sources, described in this section, are neglected in this analysis because their estimated contribution to the overall covariance matrix is small.

All of the quantities in Equation~\eqref{eq:countrate} are identified as sources of uncertainties in the analysis, except for the \emph{duration of the observations $t_i$}. This duration is fortunately known with high precision, namely, the average duration is of the order of 2 hr and the start and stop times are known with precision better than a second. This uncertainty is completely negligible in compared to those previously described.

The \emph{binning scheme} is often identified as an important source of uncertainty in many experiments. Fortunately, in the case of the 6$^\circ$ binning in the \emph{IBEX}-Lo data, it is not important. This is because the bin edges are determined with high accuracy and are affected mostly by the 1 ms granularity of the spin-period measurements by the ACS. This is equivalent to $\sim$0.025$^\circ$ uncertainty in the spin-angle. The shift in the spin-angle related to this granularity is different for each spin and the expected mean shift is 0. Thus, the related uncertainty depends only on the second- and higher-order derivatives of the count rates as a function of spin-angle, which can be neglected for the angular interval of 0.025$^\circ$. In other words, the effect could be totally neglected if the signal was changing linearly with spin-angle, which is a very reasonable assumption for small spin-angle intervals of $\sim 0.025^\circ$. The accuracy of the spin-period determination in some parts of an orbit may be much lower due to the previously described spin-pulse/spin-period synchronization problems (Section~\ref{strategy:orientation}), but the time intervals affected by this issue are excluded from the analysis \citep{mobius_etal:12a, leonard_etal:15a}.

During switching between the 6$^\circ$ bins, which takes about 3.2~ms, events are not transferred to the CEU from the interface buffer. This potentially can increase the event loss probability at the beginning of each bin, and indeed this effect was observed in the data. However, the potential variation in the number of counts in each histogram bin is limited to those counts that occur at the half of the spin-angle swept during the switching time, or equivalent to 0.04$^\circ$. Such a small effect can be neglected.

{\emph{Spin-axis stability:} from our analysis of the spacecraft orientation (Section~\ref{strategy:orientation}), we found that the spin-axis is stable during each orbit. We do not find any evidence for a significant change of the spin-axis pointing during any single orbit. We assume that the spacecraft spin-axis is stable in time at least to the level of the uncertainty of the determination of the orientation (Section~\ref{uncertainties:orientation}). 

\emph{Accuracy of the collimator transmission function:} the pre-launch calibration of the \emph{IBEX}-Lo collimator does not show any deviation from theoretical predictions \citep{fuselier_etal:09b}. Thus, in the WTPM we use the theoretical transmission function of the \emph{IBEX}-Lo collimator \citep{sokol_etal:15b}. The actual deviation from the assumed transmission is another source of statistical uncertainty. A reasonable assessment of this effect is not currently available. \citet{sokol_etal:15b} found that this could potentially affect some data points, where the observed atoms fall at the edge of the \emph{IBEX}-Lo collimator FOV. However, these data points have a low number of counts, and thus should not bias the result. We neglect this uncertainty in the current analysis.

\emph{Sensitivity threshold at low energies:} \citet{kubiak_etal:14a} and \citet{sokol_etal:15a} show that the signal modeled using the WTPM generally depends on the assumed energy threshold for detection of He atoms. Due to the lack of information on the actual limit, this would be another model related uncertainty. This is a subject of current research \citep[see][]{galli_etal:15a}. Fortunately, \citet{sokol_etal:15a} found that the signal for the data points used in our analysis is not sensitive to a variation of the assumed energy threshold over a very broad range, which certainly includes the correct threshold. Consequently, we could neglect this in the current analysis. However, in the future, this could be taken into account when we analyze the portion of the data for which a dependence on the energy threshold is important \citep[e.g., the WB,][]{kubiak_etal:14a}.

\subsection{Covariance Matrix of the Measurements}
\label{uncertainties:covariance}
The measured count rates $c_i$ form a data vector, and in the above sections we described a number of effects that are taken into account in the overall uncertainty system. Frequently in physical measurements, uncertainties are uncorrelated and then each data point has its own related uncertainty, represented by its standard deviation. The covariance matrix is then not needed (strictly speaking, it takes the form of a diagonal matrix). Since, in our analysis, data points depend on some parameters, like, e.g., spin-axis pointing, that are common for some subset of them or (as in the case of the mounting imperfections) for all of them, we need to propagate their uncertainties in the uncertainty system and form a covariance matrix that describes the uncertainties and related correlations. This system allows for discrepancies between the measured and simulated count rates that follow not only the magnitude, but also the acceptable pattern of differences. The approach using the correlation matrix is a standard technique in data analysis for situations similar to ours \citep[see, e.g.,][]{pdg:2014}.
By definition, elements of the covariance matrix $\mat{V}$ are given by
\begin{equation}
 \mat{V}_{i,j}=\mathcal{E}\left[(c_i-\mathcal{E}[c_i]) (c_j-\mathcal{E}[c_j])\right]\, , \label{eq:cov:expected}
\end{equation}
where $\mathcal{E}[...]$ denotes the expected value, i.e., the integration over uncertainty parameters with the appropriate probability distribution function. We assume that parameters describing the appropriate uncertainties are normally distributed and independent, and thus we could expand the covariance matrix into the sum:
\begin{equation}
 \mat{V}=\mat{V}^\text{p}+\mat{V}^\text{t}+\mat{V}^\text{b}+\mat{V}^\text{w}+\mat{V}^\text{o}+\mat{V}^\text{f}\, , \label{eq:covsum}
\end{equation}
where the consecutive terms relate to the Poisson uncertainty ($\mat{V}^\text{p}$), the TC uncertainty ($\mat{V}^\text{t}$), the background uncertainty ($\mat{V}^\text{b}$), the WB uncertainty ($\mat{V}^\text{w}$), the spacecraft orientation uncertainty ($\mat{V}^\text{o}$), and the boresight uncertainty ($\mat{V}^\text{f}$). 

A Gaussian distribution is the most natural choice for the distribution of the uncertainty parameters because only the second moment of the distribution is non-zero, as we assume for the uncertainty parameters in this analysis. Also, for each of the small uncertainties, we could expand the count rates only to the first derivative of $c_i$ in terms of $\epsilon$ describing each uncertainty:
\begin{equation}
 c_i(\epsilon)=c_i(0)+\frac{\partial c_i}{\partial \epsilon} \epsilon\, .
\end{equation}

With the abovementioned assumption, the covariance matrix from Equation~\eqref{eq:cov:expected} could be rewritten with respect to the parameter $\epsilon$:
\begin{equation}
 \mat{V}^\epsilon_{i,j}=\frac{\partial c_i}{\partial \epsilon}\frac{\partial c_j}{\partial \epsilon}\left(\delta \epsilon\right)^2\, .
\end{equation}
Summation over all parameters $\epsilon$ is equivalent to the commonly known formula for the folding of uncertainties.

The count rate $c_i$ for a data point $i$ depends on the number of counts $d_i$ in this bin, but not on  counts $d_j$ in a bin $j\neq i$, thus the appropriate covariance matrix is diagonal:
\begin{equation}
 \mat{V}^\text{p}_{i,j}=\frac{\partial c_i}{\partial d_i}\frac{\partial c_j}{\partial d_i}\left(\delta d_i\right)^2=\left(\frac{\gamma_i}{t_i}\right)^2\sqrt{d_i}^2\delta_{i,j}=\frac{\gamma_i^2 d_i}{t_i^2}\delta_{i,j}\, ,
\end{equation}
where $\delta_{i,j}$ is the Kronecker delta. 

We assume that the TC factors $\gamma_i$ are independent of each other, and thus the covariance matrix is also diagonal:
\begin{equation}
 \mat{V}^\text{t}_{i,j}=\frac{\partial c_i}{\partial \gamma_i}\frac{\partial c_j}{\partial \gamma_i}\left(\delta \gamma_i\right)^2=
 \frac{d_i^2\left(\delta\gamma_i\right)^2}{t_i^2}\delta_{i,j}\, ,
\end{equation}
where $\delta\gamma_i$ is given by Equation~\eqref{eq:dlambda}. 

In this analysis, we assume that the background is constant in time and spin-angle, with a single uncertainty $\delta b$ describing the overall uncertainty \citep{galli_etal:14a}. The covariance matrix takes the form
\begin{equation}
 \mat{V}^\text{b}_{i,j}=(\delta b)^2\, .
\end{equation}
The covariance matrix for the background is proportional to the matrix of ones. Since the background is described by only one parameter, then the covariance matrix is of rank 1 in this case.

The covariance matrix related to the uncertainties of the WB parameters is given as
\begin{equation}
 \mat{V}^\text{w}_{i,j}=\delta_n c_i \delta_n c_j+\delta_\lambda c_i \delta_\lambda c_j+\delta_\beta c_i \delta_\beta c_j+\delta_T c_i \delta_T c_j+\delta_v c_i \delta_v c_j\, ,
\end{equation}
where the appropriate terms are given by Equation~\eqref{eq:wbunc}, the parameters refer to the parameters of the WB. In the case of the diagonal covariance matrix, assumed for the WB parameters, the covariance given by the above equation does not include mixing terms. The correlation can be included in the future if necessary.

For the orientation uncertainty, we have already calculated the change of count rates as a function of shift from actual pointing. The covariance matrix is given by
\begin{equation}
 \mat{V}^\text{o}_{i,j}=\left(\delta_\alpha c_i \delta_\alpha c_j+\delta_\delta c_i \delta_\delta c_j\right) \delta_{\mathcal{O}(i),\mathcal{O}(j)}\, .
\end{equation}
Here, the Kronecker delta takes the numbers of orbits in which bins $i$ and $j$ are measured. This represents the fact that the shift from the actual pointing is constant over a single orbit. Consequently, the matrix $\mat{V}^\text{o}$ is block-diagonal, with each block representing the covariance between the counts rates on individual orbits. 

The boresight uncertainties correlate all of the data points:
\begin{equation}
 \mat{V}^\text{f}_{i,j}=\left(\delta_\theta c_i \delta_\theta c_j+\delta_\eta c_i \delta_\eta c_j\right)\, .
\end{equation}
Here, all of the data points are correlated, and thus all elements of the matrix generally have non-zero values. 

For the application of the covariance matrix in the least-squares method, inversion of the matrix is needed. This is possible if the matrix is of the maximal rank. All of the component matrices in the sum given by Equation~\eqref{eq:covsum} are positive-semidefinite, and at least the Poisson uncertainty covariance matrix $\mat{V}^\text{p}$ is positive-definite because it only has diagonal positive coefficients. Thus, the total covariance matrix is positive-definite, and hence of maximal rank. Inversion of this matrix can be performed using well-known numerical schemes.

\section{DATA RELEASE}
\label{datarelease}
This paper is a part of the Special Issue of the \emph{Astrophysical Journal Supplement Series} \citep{mccomas_etal:15b} that is accompanied by \emph{IBEX} Data Release 9. In this data release, the \emph{IBEX} team publish the spin-axis pointings for all \emph{IBEX} orbits until 252b (Section~\ref{strategy:orientation}), the \emph{IBEX} ephemeris data, the ``good time'' intervals \citep{mobius_etal:12a,leonard_etal:15a}, and the count rates $c_i$ averaged over the good times (Section~\ref{uncertainties}), along with the full covariance matrix of these rates (Section~\ref{uncertainties:covariance}). Additionally, we provide the observed number of counts in each bin $d_i$, the TC factors $\gamma_i$, the durations of observations $t_i$, and the subtracted values of the WB count rates $w_i$. We provide these quantities for the spin-angle range 240$^\circ$ to 294$^\circ$ on the key ISN orbits from seasons 2009--2014. Note that the TC is needed for data in seasons 2009--2012, but we could only provide the correction factors for 2009 and 2010 due to the special mode of the instrument in the latter two years. In addition to the data products mentioned here, the data release also includes other data products, presented by \citet{schwadron_etal:15a} and \citet{sokol_etal:15b}.

\section{APPLICATION IN THE WTPM}
\label{application}
In the previous analysis using WTPM, the least-squares method was used to find the parameters of ISN He: inflow direction, speed, and temperature \citep{bzowski_etal:12a}. The same method was used by \citet{kubiak_etal:14a} to determine the WB properties. We also use this method in the current analysis; however, here we need to use a more general form for the least-squares estimator \citep{pdg:2014}:
\begin{equation}
 \chi^2(\vec{\pi})=(\vec{c}-a\vec{F}(\vec{\pi}))^\text{T} \mat{V}^{-1} (\vec{c}-a\vec{F}(\vec{\pi}))\, , \label{eq:chi2}
\end{equation}
where $\vec{c}$ and $\vec{F}(\vec{\pi})$ are vectors of the measured count rates and the expected time-, spin-angle-, and collimator-averaged flux, respectively, $a$ is the proportionality factor depending on the unknown instrumental efficiencies, and $\vec{\pi}=(\lambda,\,\beta,\,T,\,v)^\text{T}$ denote the parameters of the ISN He gas sought: the direction of inflow in ecliptic coordinates ($\lambda$, $\beta$), speed $v$, and temperature $T$. The WTPM calculates the flux in physical units. Details of the flux calculation, its integrating over the collimator transmission function, and averaging over spin-angle bins and good times are described by \citet{sokol_etal:15b}. In the analysis, we do not rely on an absolute calibration of the instrument. Instead, we obtain the scaling factor $a$ in Equation~\ref{eq:chi2} as described by \citet{sokol_etal:15b}: we calculate it from the condition $\partial \chi^2(\vec{\pi})/\partial a=0$. Using this method, this quantity could be determined analytically for any simulated flux. 

In the least-squares method we look for the global minimum of the $\chi^2\left(\vec{\pi}\right)$ estimator:
\begin{equation}
 \frac{\partial \chi^2 (\vec{\pi})}{\partial\lambda}=0\, ,  
 \frac{\partial \chi^2 (\vec{\pi})}{\partial\beta}=0\, ,  
 \frac{\partial \chi^2 (\vec{\pi})}{\partial T}=0\, ,  
 \frac{\partial \chi^2 (\vec{\pi})}{\partial v}=0\, .
\end{equation}
This condition is not sufficient to be the global minimum. However, we have searched for the minimum over a relatively wide part of the possible parameter space and we have not found any additional stationary point in this analysis. Note that in the original analysis of the ISN He data by \citet{bzowski_etal:12a}, a secondary minimum in $\chi^2(\lambda)$ was found. This secondary minimum is absent in the present analysis and the $\chi^2$ values are substantially lower, as we show further in the text, which we attribute to the refined uncertainty treatment that we describe in the current paper and the refinements in the modeling presented by \citet{sokol_etal:15b}. 

For the determination of the minimum we performed simulations of the expected flux for a large number of different parameter sets in the proximity of the 4D correlation tube found by \citet{bzowski_etal:12a}, \citet{mccomas_etal:12b,mccomas_etal:15a}, and \citet{mobius_etal:12a}. For this purpose, we used a regular grid of parameters. In ecliptic longitude, we sampled the range $252^\circ \leq \lambda \leq 264^\circ$ with a step of $\Delta \lambda=1^\circ$. For each ecliptic longitude, we chose the latter parameters close to the correlation line on the integer multiplicity of the grid steps. In ecliptic latitude, we used a step of $\Delta\beta=0.1^\circ$, in temperature $\Delta T=250 \text{ K}$, and in speed $\Delta v=0.5 \text{ km s}^{-1}$. After that, we probed the points by $n_\beta$ steps in latitude, $n_T$ in temperature, and $n_v$ in speed, for which $\sqrt{n_\beta^2+n_T^2+n_v^2}\leq 4$. The selected points cover the vicinity of the correlation line with safe margins. This gives 257 points on a regular grid for each ecliptic longitude. 
This regular grid was then supplemented by additional points on a denser grid with $\Delta T=125\text{ K}$ and $\Delta v=0.25 \text{ km s}^{-1}$, selected in a manner similar to that described above. We ended up with 4297 probed points for all of the ecliptic longitudes on a regular grid, which guarantees that we are not missing any hypothetical secondary minima of $\chi^2$. Note that the sizes of the grid steps are small enough to show the $\chi^2$ variation in the parameter space both along and across the correlation line.

We calculate the expected fluxes of ISN He in each bin taken into account in the analysis (see Section~\ref{strategy:dataselection}) for each of the 4297 parameter sets on the regular grid. The simulations are performed using nWTPM, i.e., the full version of WTPM, with all details taken into account \citep[see ][]{sokol_etal:15b}. 
In the proximity of the minimum $\chi^2(\vec{\pi})$, the $\chi^2$ estimator could be expanded as a second-order polynomial of the inflow parameters:
\begin{equation}
 \chi^2(\vec{\pi})\approx\chi^2(\vec{\pi}_0)+(\vec{\pi}-\vec{\pi}_0)^\text{T}\mat{A}(\vec{\pi}-\vec{\pi}_0)\, , \label{eq:chi2exp}
\end{equation}
where $\mat{A}$ is a $4\times4$ positive-definite, symmetric matrix of the polynomial coefficients, and $\vec{\pi}_0$ is the parameter set for which the estimator is the smallest. Using the least-square method with known values of the estimator for a set of grid points in the proximity of the minimum, we could find the 15 independent unknowns in Equation~\eqref{eq:chi2exp}: the value at the minimum $\chi^2(\vec{\pi}_0)$, the ISN parameter vector at the minimum $\vec{\pi}_0$ (4 values), and the matrix $\mat{A}$ (due to symmetry, only 10 coefficients are independent). 

The approximation given by the quadratic expansion is valid only for a limited part of the parameter space. However, we checked that the values provided by this approximation do not differ by more than 20\% from those calculated on the regular grid in the region where $\chi^2(\vec{\pi})-\chi^2(\vec{\pi}_0)\leq 16.25$, i.e., in the 3$\sigma$ region for four parameters. This justifies the approximation used in our calculations.

In general, the estimator at the minimum ($\chi^2(\vec{\pi}_0)$) should be distributed as the $\chi^2$ distribution with the number of degrees of freedom ($\nu$) equal to the number of data points minus the number of model parameters. The expected value for this estimator at minimum is equal to the number of degrees of freedom, with the variance equal to this number doubled. In our analysis, the number of model parameters is five: four describe the physical properties of ISN He, and the last one is the proportionality factor $a$ scaling the simulations to the data.

The covariance matrix $\mat{\Sigma}$ for the parameter set $\vec{\pi}$ is calculated from the matrix of the second derivatives of the estimator at minimum \citep{pdg:2014}:
\begin{equation}
 \mat{\Sigma}=\left(\frac{1}{2}\left.\frac{\partial^2 \chi^2(\vec{\pi})}{\partial \vec{\pi}^2}\right|_{\vec{\pi}=\vec{\pi}_0}\right)^{-1}= \mat{A}^{-1}\, , \label{eq:rescov}
\end{equation}
where we use the fact that the expansion given by Equation~\eqref{eq:chi2exp} is exact at $\vec{\pi}=\vec{\pi}_0$. The uncertainties of the parameters can then be obtained as the square root of the diagonal elements of the covariance matrix: $\sigma_{\lambda}=\sqrt{\Sigma_{1,1}}$, $\sigma_{\beta}=\sqrt{\Sigma_{2,2}}$, $\sigma_T=\sqrt{\Sigma_{3,3}}$, $\sigma_{v}=\sqrt{\Sigma_{4,4}}$.

The correlation matrix $\mat{C}$ is calculated from the following equation:
\begin{equation}
 \mat{C}=\mat{S}^{-1}\mat{\Sigma}\mat{S}^{-1}\, , \label{eq:rescor}
\end{equation}
where $\mat{S}=\text{diag}(\sigma_\lambda,\sigma_\beta,\sigma_T,\sigma_v)$ is a diagonal matrix containing the square roots of the diagonal elements of matrix $\mat{\Sigma}$. Matrix $\mat{C}$ describes the correlations between the parameters of ISN He. Due to the observation strategy, a strong correlation is expected \citep{mccomas_etal:12b}.

Let $\mat{D}=(\vec{\zeta}_1,\, \vec{\zeta}_2,\, \vec{\zeta}_3,\, \vec{\zeta}_4)$ be the normalized and ordered set of eigenvectors of matrix $\mat{C}$ with the eigenvalues $z_1,\, z_2,\, z_3,\, z_4$, respectively, with $z_i>z_j$ for $i<j$. Then, $\mat{D}^\text{T}\mat{C}\mat{D}$ is the diagonal matrix of the eigenvalues. The vector $\vec{\zeta}_1$ determines the direction of the highest correlation in the parameter space, that is, the correlation line. Namely, the correlated parameters lie on the line given by
\begin{equation}
 \vec{\pi}_0+t\mat{S}\vec{\zeta}_1\qquad t\in(-k z_1,k z_1)\, , \label{eq:corrline}
\end{equation}
for the confidence interval $k\sigma$. It is convenient to separate the uncertainties along and across the correlation line. For this purpose, we define the covariance matrices along and across the correlation line as
\begin{align}
 \mat{\Sigma}_\text{along}&=\mat{S}\mat{D}
 \left(\begin{matrix}
  z_1 & 0 & 0 & 0\\
  0 & 0 & 0 & 0\\
  0 & 0 & 0 & 0\\
  0 & 0 & 0 & 0\\
 \end{matrix}\right)
 \mat{D}^\text{T}\mat{S}\, ,\\
 \mat{\Sigma}_\text{across}&=\mat{S}\mat{D} 
 \left(\begin{matrix}
  0 & 0 & 0 & 0\\
  0 & z_2 & 0 & 0\\
  0 & 0 & z_3 & 0\\
  0 & 0 & 0 & z_4\\
 \end{matrix}\right)
 \mat{D}^\text{T}\mat{S}\, .
\end{align}
We define the uncertainties along and across the correlation line as the square roots of the appropriate diagonal elements of these matrices. 

Throughout the text, we will discuss the uncertainties for different confidence intervals. As commonly chosen, we adopted the confidence intervals $\alpha=68.27\%$, $\alpha=95.45\%$, and $\alpha=99.73\%$, which we refer to as the 1$\sigma$, 2$\sigma$, and 3$\sigma$ intervals, respectively. The allowed region in the parameter space is determined by the $\chi^2$ estimator value \citep{avni:1976,lampton_etal:1976}:
\begin{equation}
 \Delta \chi^2(\vec{\pi})\equiv\chi^2(\vec{\pi})-\chi^2(\vec{\pi}_0)\leq \text{CDF}^{-1}(\chi^2\text{-dist}(m),\alpha)\, .
\end{equation}
where $\text{CDF}^{-1}(\chi^2\text{-dist}(m),\alpha)$ is the inverse cumulative distribution function at the confidence level $\alpha$ for the $\chi^2$ distribution with $m$ degrees of freedom. In this case, $m$ describes the number of interesting parameters in the analysis. Thus, the allowed regions in the 4D parameter space of $\vec{\pi}$ are bound by the surfaces of $\Delta \chi^2(\vec{\pi})=4.72,\, 9.72,\, 16.25$ for the 1$\sigma$, 2$\sigma$, and 3$\sigma$ levels, respectively. However, if we are interested only in an $m$-dimensional subspace, then $\Delta \chi^2(\vec{\pi})$ should be minimized over the ($4-m$)-dimensional quotient space. For example, in the case of the ecliptic longitude, $\Delta \chi^2(\lambda)\equiv \min_{\beta,T,v} \Delta\chi^2(\lambda,\beta,T,v)$ is limited by the values 1, 4, and 9 for the selected confidence intervals. 

\section{RESULTS}
\label{results}
In this paper, we focus on the development of the uncertainty system and understanding the effects of various uncertainties on the results of ISN He parameter fitting. To that end, we analyze the first season of ISN He observations in spring 2009 using a similar data selection as in the original analysis, with the exceptions discussed earlier. In the previous sections, we have developed a complex system of uncertainties that takes into account all of the instrumental effects we could quantify. Fitting the ISN He flow parameters, using the method given in Section~\ref{application}, to the data from the 2009 observation season, with all uncertainties described in Section~\ref{uncertainties} included, yields the following set of parameters:
\begin{equation}
 (\lambda_0,\beta_0,T_0,v_0)=(257.73^\circ,\, 5.04^\circ,\, 6655\text{ K},\, 24.47\text{ km s}^{-1})\, .
\end{equation}
The uncertainties and correlations of this parameter set are described by the following covariance matrix:
\begin{equation}
 \mat{\Sigma}=\left(\begin{matrix}
 0.942 & -0.0599 & -489 & -0.755 \\
 -0.0599 & 0.0225 & 33.2 & 0.0532 \\
 -489 & 33.2 & 267500 & 412 \\
 -0.755 & 0.0532 & 412 & 0.650 \\                 
 \end{matrix}
 \right)\, .
\end{equation}
The square roots of the diagonal terms in the matrix are the $1\sigma$ uncertainties of the corresponding parameters, thus $\sigma_{\lambda}=0.97^\circ$, $\sigma_{\beta}=0.15^\circ$, $\sigma_T=517\text{ K}$, and $\sigma_v=0.81\text{ km s}^{-1}$. 

After the normalization of each row and column of the covariance matrix (see Equation~\eqref{eq:rescor}) we get the following correlation matrix:
\begin{equation}
 \mat{C}=\left(\begin{matrix}
 1 & -0.411 & -0.974 & -0.965 \\
 -0.411 & 1 & 0.427 & 0.440 \\
 -0.974 & 0.427 & 1 & 0.988 \\
 -0.965 & 0.440 & 0.988 & 1 \\           
 \end{matrix}
 \right)\, .
\end{equation}
This matrix shows a very high correlation between $T$ and $v$. This correlation is related to the fact that \emph{IBEX} measures the ratio of the thermal speed to the bulk speed through the measurement of the Mach cone \citep{mobius_etal:12a}. Also, a high anticorrelation exists between $\lambda$ and $v$. The reason for this is the way in which \emph{IBEX} actually measures interstellar atoms and is explained by the existence of the correlation line \citep{bzowski_etal:12a,mccomas_etal:12b,mobius_etal:12a}. The inflow latitude $\beta$ is much less correlated with the other parameters. 

The correlation line in the linear approximation is given by the eigenvector of matrix $\mat{C}$ with the highest eigenvalue. The largest eigenvalue for this matrix is equal to 3.20, thus a relatively large part of the uncertainties is distributed along the line given by Equation~\eqref{eq:corrline}:
\begin{align}
 \vec{\pi}=&(257.73,\, 5.04,\, 6655,\, 24.47)+\nonumber\\&+(0.528t,\, -0.048t,\, -284t,\, -0.442t)\, ,
\end{align}
where $t\in(-3.2k,\,3.2k)$ and $k\sigma$ determines the confidence interval. The covariance matrices along and across this correlation line are the following:
\begin{align}
 \mat{\Sigma}_\text{along}&=
 \left(\begin{matrix}
 0.891 & -0.0807 & -479 & -0.747 \\
 -0.0807 & 0.0073 & 43.4 & 0.0676 \\
 -479 & 43.4 & 258109 & 402 \\
 -0.747 & 0.0676 & 402 & 0.626 \\
 \end{matrix}\right)
 \, ,\\
 \mat{\Sigma}_\text{across}&=
 \left(\begin{matrix}
 0.051 & 0.0207 & -9 & -0.008 \\
 0.0207 & 0.0152 & -10.2 & -0.0144 \\
 -9 & -10.2 & 9372 & 9 \\
 -0.008 & -0.0144 & 9 & 0.023 \\
 \end{matrix}\right)
 \, .
\end{align}
Calculation of the correlation matrix based only on the part of covariance along the correlation line gives a matrix containing only $\pm1$ elements. This is a consequence of the construction of this matrix. For the covariance across the correlation line, most of the correlation between $\lambda$, $T$, and $v$ is subtracted, but it is still not removed entirely.

In Figure~\ref{res:ellipses}, we present the allowed parameter regions in two-dimensional (2D) cuts in the parameter space. For these plots, we used the approximation given by Equation~\eqref{eq:chi2exp}. The results discussed here are denoted by red lines. In Figure~\ref{res:lamwide}, we present $\Delta\chi^2(\lambda)$ illustrating the allowed ranges of inflow longitudes in the discussed case. We show the actual dependence of $\Delta \chi^2$ on $\lambda$, obtained from $\chi^2$ minimization for each $\lambda$ in the regular grid. 

\begin{figure*}[ht!]
 \includegraphics[width=.32\textwidth]{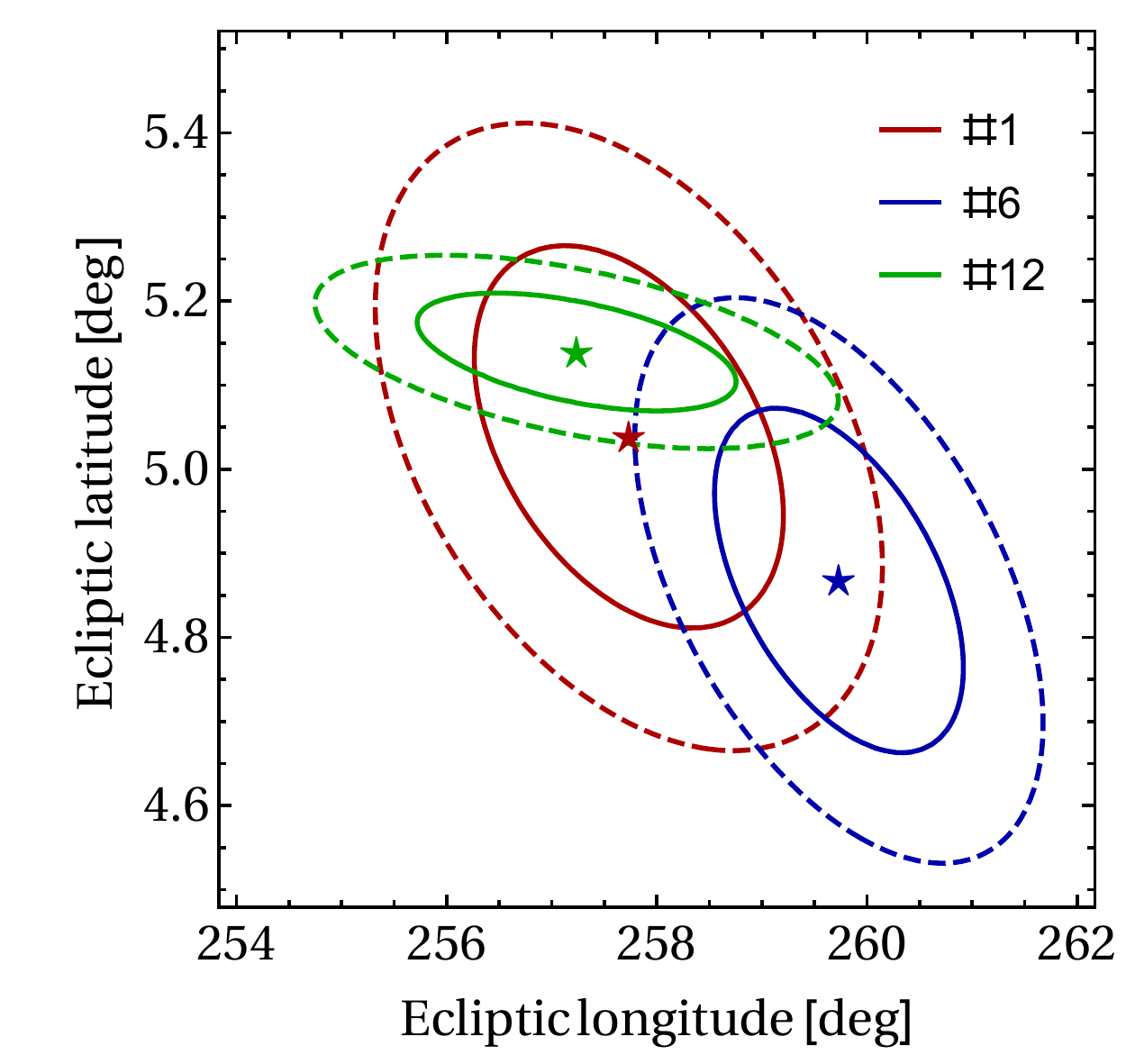}
 \includegraphics[width=.32\textwidth]{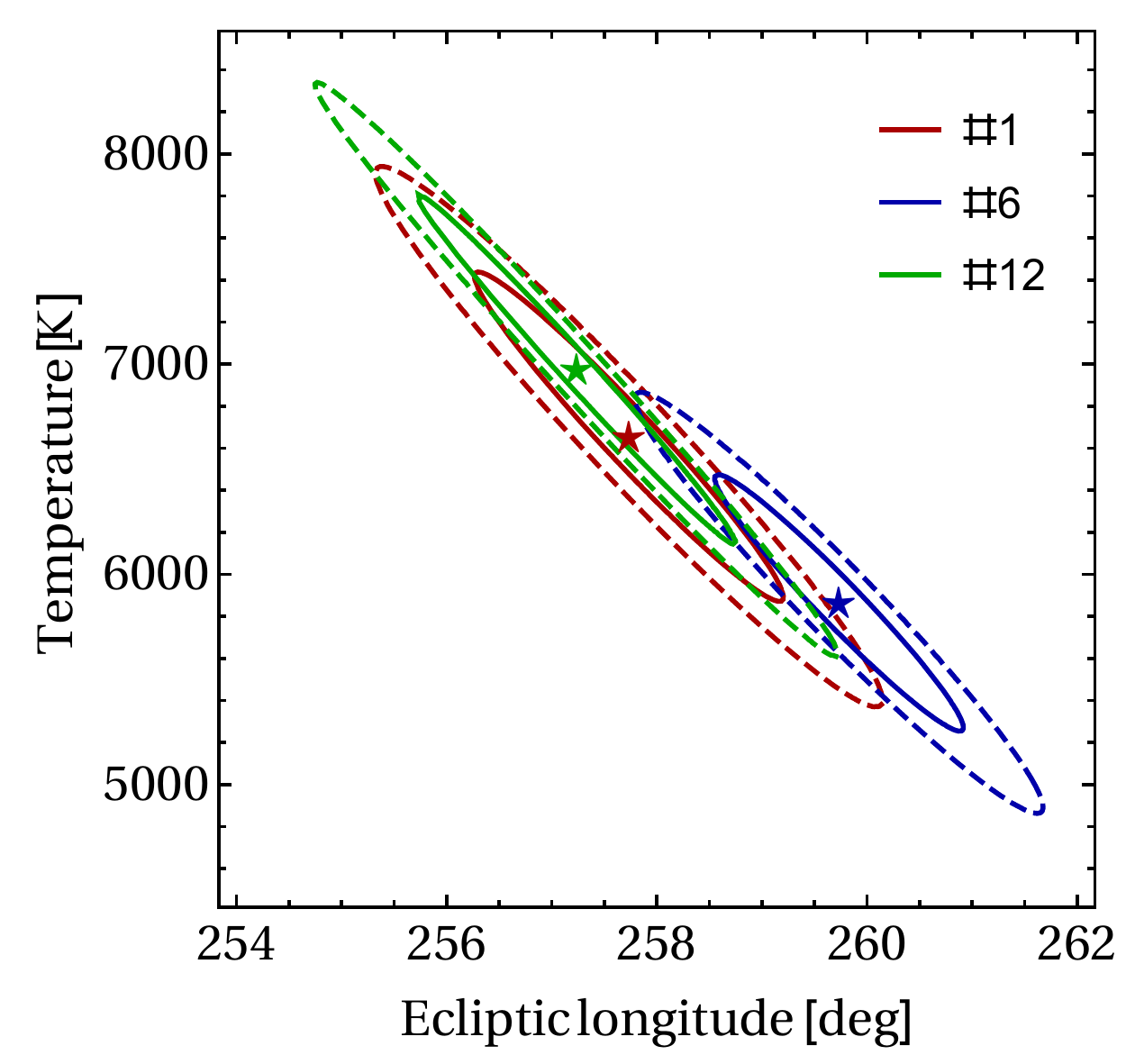}
 \includegraphics[width=.32\textwidth]{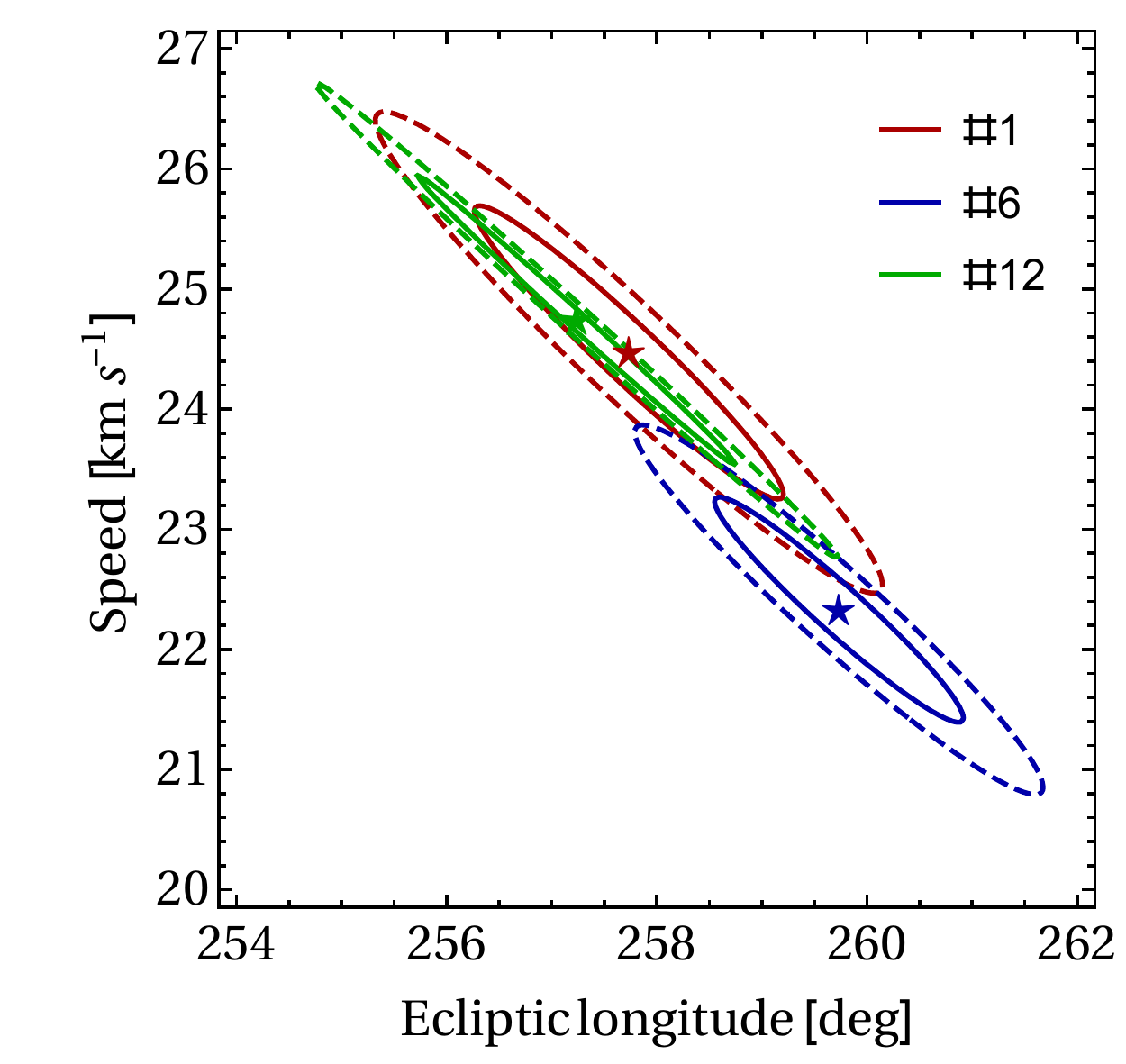}
 
 \includegraphics[width=.32\textwidth]{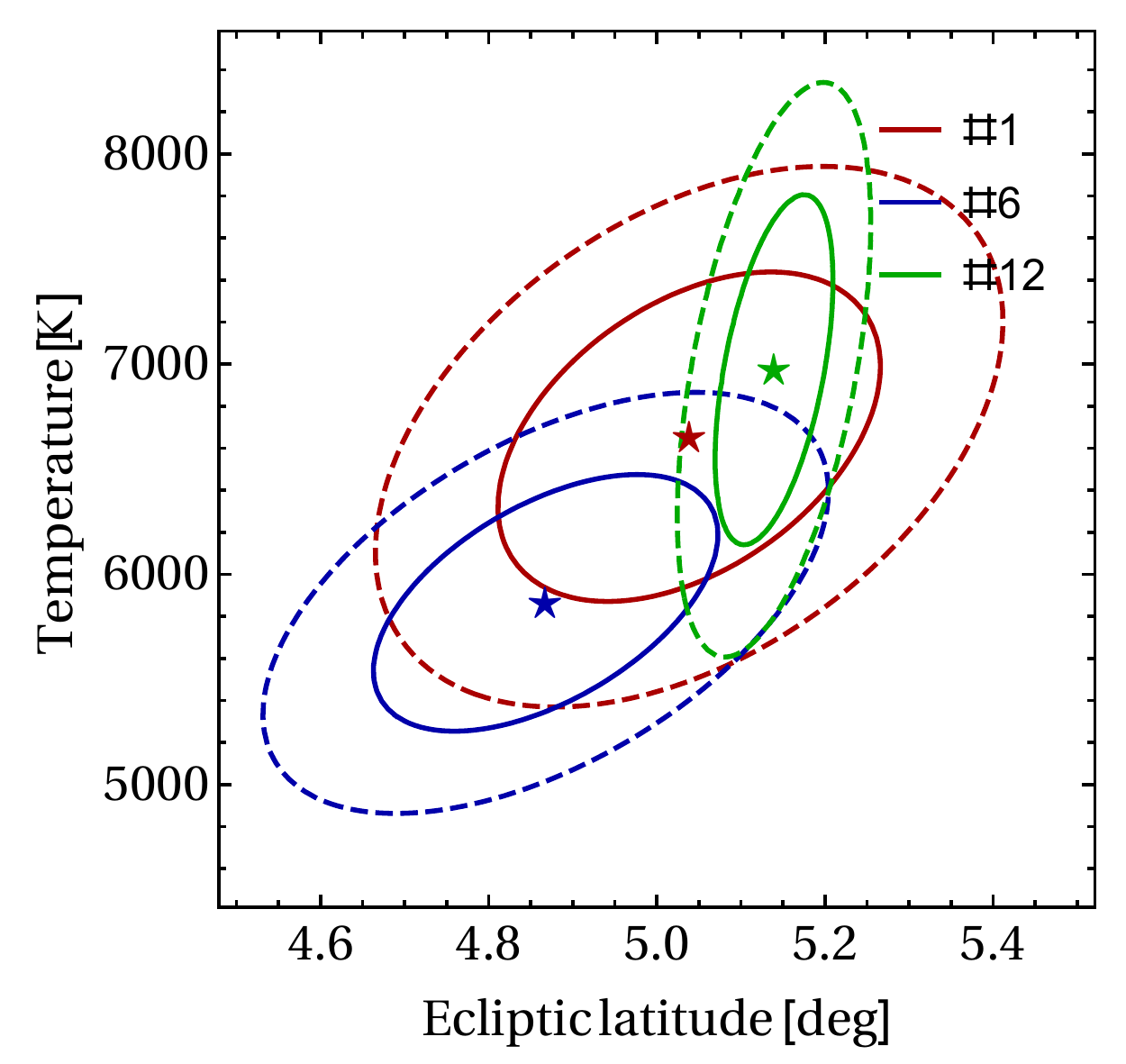}
 \includegraphics[width=.32\textwidth]{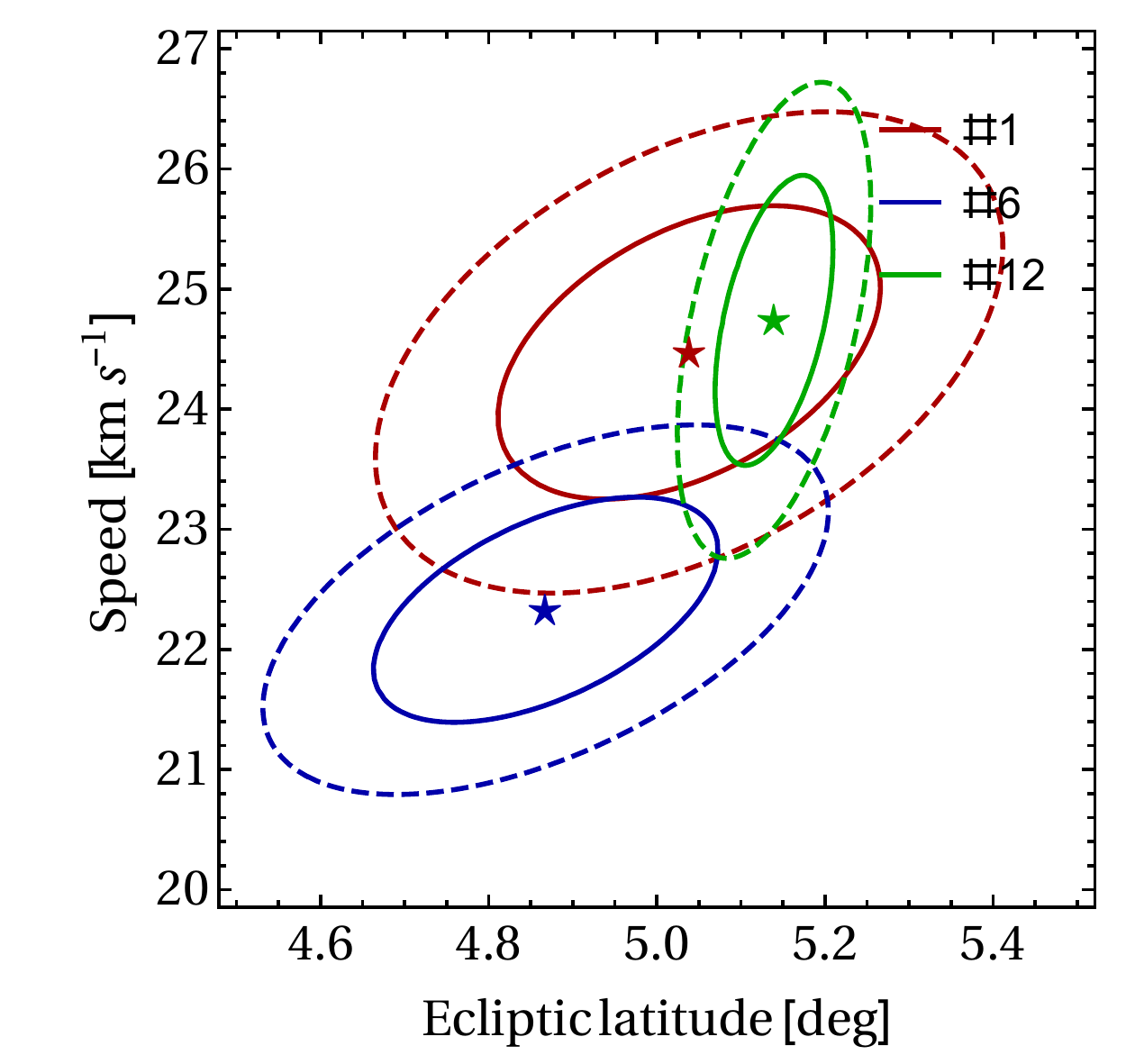}
 \includegraphics[width=.32\textwidth]{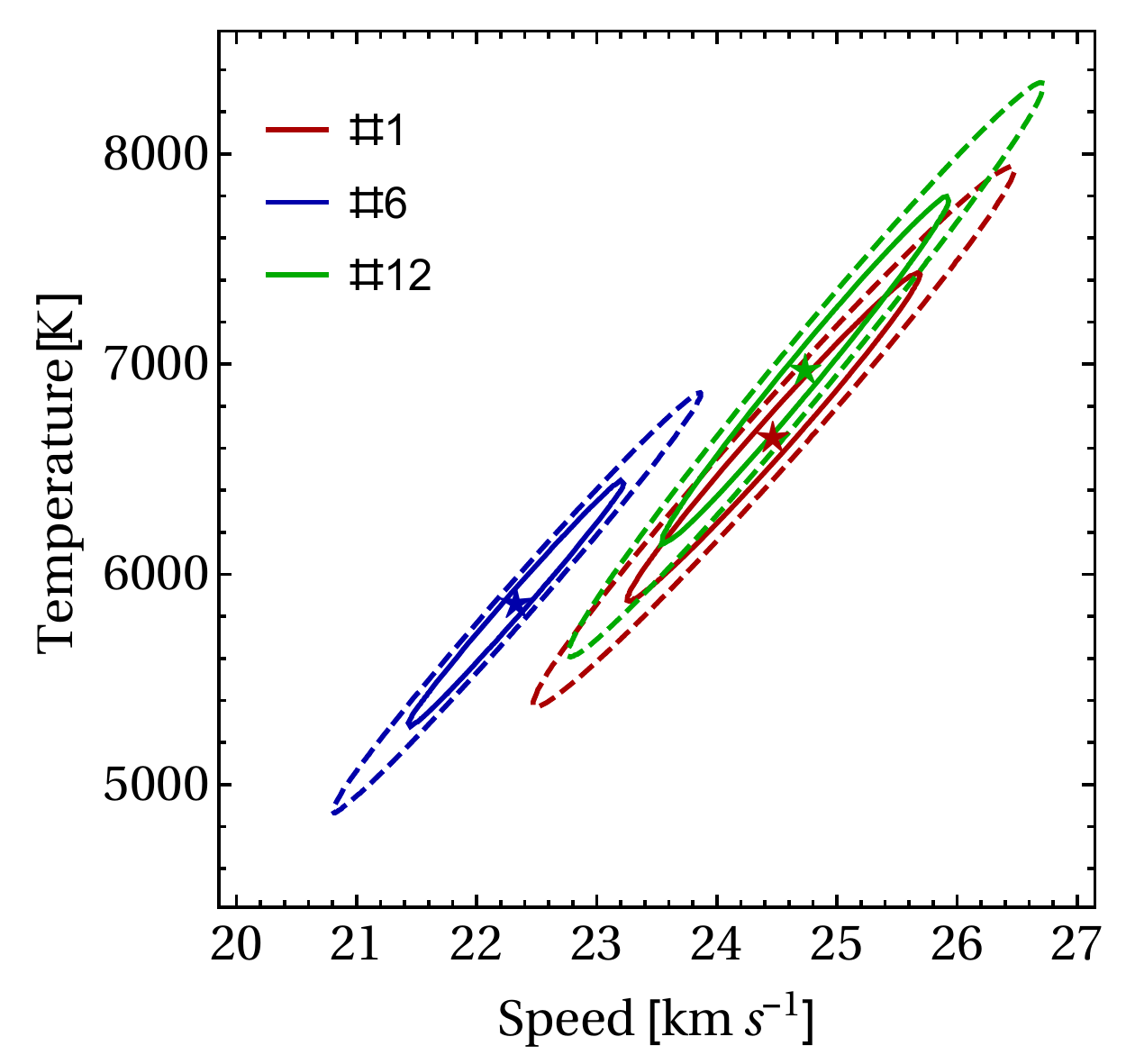}
 
 \caption{Ellipses of the allowed regions of the 2D cuts in the parameter space for selected cases from Table \ref{tab:resunc}: \#1 (red), \#6 (blue) and \#12 (green). Case \#1 is the model with the Warm Breeze and all uncertainties included, \#6 is the case without the Warm Breeze, but with all uncertainties included, \#12 is the case with the Warm Breeze included, for the data \emph{not} corrected for the throughput effect. The solid and dashed lines mark the regions allowed for the $1\sigma$ and $2\sigma$ confidence level, respectively.}
 \label{res:ellipses}
\end{figure*}

\begin{figure}
\centering
 \includegraphics[width=.45\textwidth]{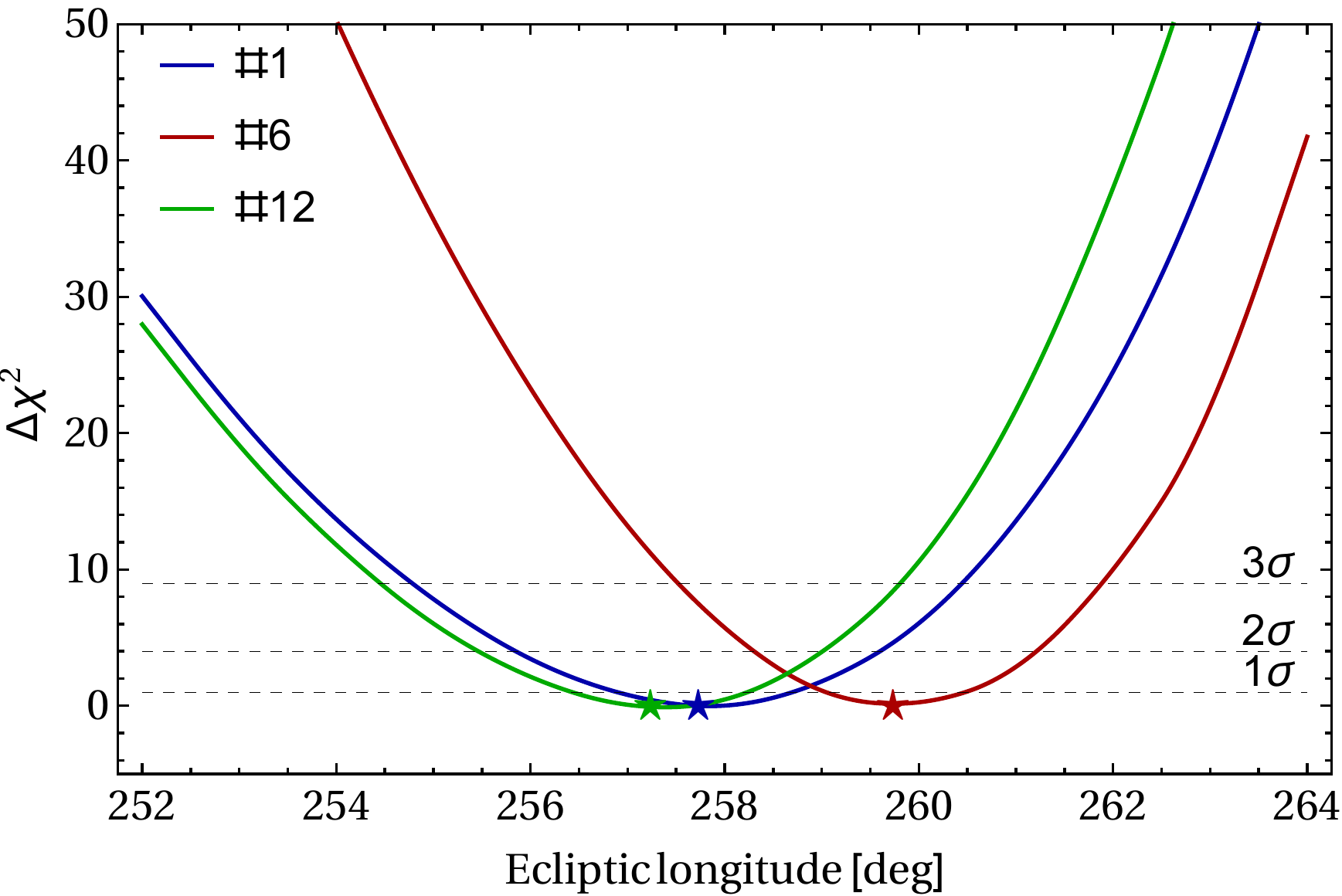}
 \caption{Minimized $\Delta\chi^2(\lambda)$ showing the allowed ranges of ecliptic longitudes at the $1\sigma$, $2\sigma$, and $3\sigma$ level for $\Delta\chi^2=1,\,4,$ and 9 respectively. The color coding is the same as in Figure~\ref{res:ellipses} for the cases from Table~\ref{tab:resunc}.}
 \label{res:lamwide}
\end{figure}

A comparison of the data (set \#1) with the best fit, with all uncertainties included, is presented in Figure~\ref{res:datasimu}. The error bars represent the square root of the diagonal terms in matrix $\mat{V}$. We present the fit residuals (observed minus simulated count rates) and their ratios to the uncertainties and values in each data point. Note that the residuals at orbit 17 are the largest, however, due to the large uncertainties, they do not stick out strongly compared with the uncertainties in other data points. Note also that the residuals are not distributed symmetrically due to the presence of the off-diagonal terms in the data covariance matrix. A notable feature is the excess of the data over simulations in all orbits earlier than 18, which may suggest that an additional signal, not accounted for in the model, was present. The magnitude of this excess signal is on the order of 0.25 counts/s in orbits 13 through 16, which is on a level of $\sim$1\% of the peak value.

\begin{figure*}
\centering
 \includegraphics[width=.8\textwidth]{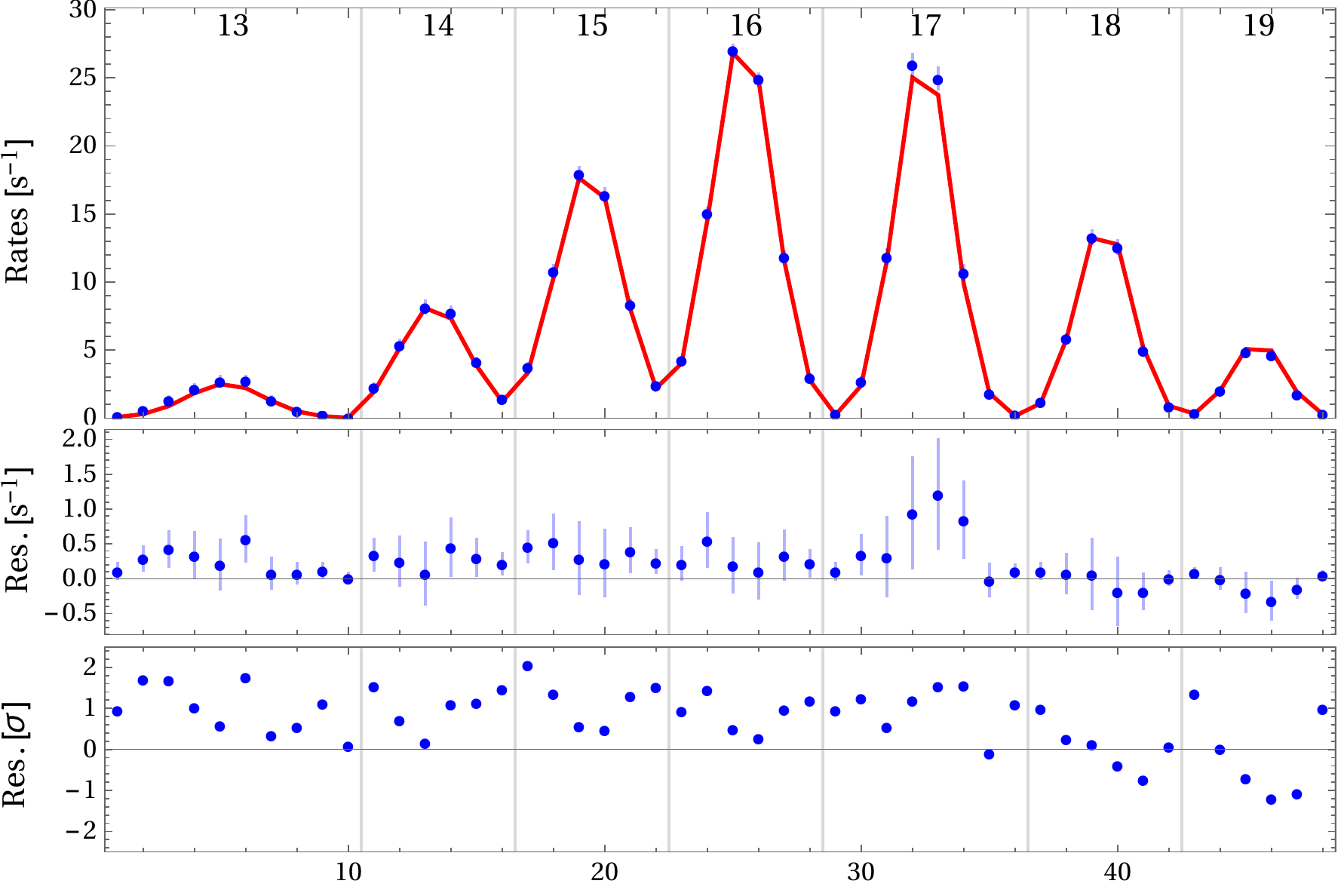}
 \caption{Upper panel: comparison of the data used in the fitting with the simulations for the ISN parameters obtained for set \#1, i.e., $(\lambda_0,\beta_0,T_0,v_0)=(257.73^\circ,\, 5.04^\circ,\, 6655\text{ K},\, 24.47\text{ km s}^{-1})$. Middle panel: the fit residuals. Lower panel: the fit residuals divided by their uncertainties.}
 \label{res:datasimu}
\end{figure*}

\subsection{Uncertainty Sources}
In this analysis, we discuss the impact of various uncertainties on the results of fitting the ISN He flow parameters using WTPM for the data from the 2009 ISN season. To that end, we construct covariance matrices which differ by the types of included sources, given as the terms in Equation~\eqref{eq:covsum}. In addition, we check the impact of subtracting the WB \citep{kubiak_etal:14a} and the constant background \citep[BG;][]{galli_etal:14a}. Also, the impact of the TC is shown. In Table~\ref{tab:resunc}, we present the best fits for the ISN He parameters obtained including different sources of uncertainties. Set \#1 presents the results with all uncertainties, the background, and the WB taken into account as discussed in detail above. Further rows in the table eliminate one or more effects from the list. 

\begin{deluxetable}{rccccccccccrrrrrrr}
\rotate
 \tablewidth{0pt}
 \tablecolumns{18}
 \tabletypesize{\scriptsize}
 \tablecaption{ISN He parameters from fits with different uncertainty sources included \label{tab:resunc}}
 \tablehead{
    \colhead{\#}					&
    \colhead{Selection}				&
    \colhead{$\mat{V}^{p}$}		&    
    \colhead{TC}					&
    \colhead{$\mat{V}^{t}$}		&    
    \colhead{BG}					&
    \colhead{$\mat{V}^{b}$}		&
    \colhead{WB}					&
    \colhead{$\mat{V}^{w}$}		&
    \colhead{$\mat{V}^{o}$}		&
    \colhead{$\mat{V}^{f}$}		&
    \colhead{$\chi^2$}				&
    \colhead{$\nu$}					&
    \colhead{$\frac{\chi^2-\nu}{\sqrt{2\nu}}$}	&
    \colhead{$\lambda\, [^\circ]$}	&
    \colhead{$\beta\, [^\circ]$}	&
    \colhead{$T\, [\text{K}]$}		&
    \colhead{$v\, [\text{km s}^{-1}]$}		
 }
 \startdata
1 & G & \cm & \cm & \cm & \cm & \cm & \cm & \cm & \cm & \cm & 64.8 & 44  & 2.2 & $257.7\pm1.0$ & $5.04\pm0.15$ & $6650\pm520$ & $24.5\pm0.8$ \\
2 & G & \cm & \cm & \cm & \cm & \cm & \cm & \cm & \cm &     & 65.9 & 44  & 2.3 & $257.7\pm1.1$ & $5.12\pm0.05$ & $6590\pm610$ & $24.3\pm0.8$ \\
3 & G & \cm & \cm & \cm & \cm & \cm & \cm & \cm &     & \cm & 66.0 & 44  & 2.3 & $257.7\pm1.0$ & $5.04\pm0.15$ & $6690\pm530$ & $24.5\pm0.8$ \\
4 & G & \cm & \cm & \cm & \cm & \cm & \cm & \cm &     &     & 67.2 & 44  & 2.5 & $257.7\pm1.0$ & $5.12\pm0.05$ & $6610\pm580$ & $24.4\pm0.8$ \\
5 & G & \cm & \cm & \cm & \cm & \cm & \cm &     & \cm & \cm & 86.6 & 44  & 4.5 & $258.5\pm0.9$ & $4.98\pm0.15$ & $6400\pm470$ & $23.7\pm0.7$ \\
6 & G & \cm & \cm & \cm & \cm & \cm &     &     & \cm & \cm & 216.0 & 44  & 18.3 & $259.7\pm0.8$ & $4.87\pm0.14$ & $5860\pm400$ & $22.3\pm0.6$ \\
7 & G & \cm & \cm & \cm & \cm &     & \cm & \cm & \cm & \cm & 64.8 & 44  & 2.2 & $257.7\pm1.0$ & $5.04\pm0.15$ & $6650\pm520$ & $24.5\pm0.8$ \\
8 & G & \cm & \cm & \cm &     &     & \cm & \cm & \cm & \cm & 65.4 & 44  & 2.3 & $257.6\pm1.0$ & $5.05\pm0.15$ & $6700\pm520$ & $24.5\pm0.8$ \\
9 & G & \cm & \cm &     & \cm & \cm & \cm & \cm & \cm & \cm & 67.3 & 44  & 2.5 & $257.8\pm0.9$ & $5.03\pm0.15$ & $6620\pm500$ & $24.4\pm0.8$ \\
10 & G & \cm &     &     & \cm & \cm & \cm & \cm & \cm & \cm & 63.5 & 44  & 2.1 & $257.4\pm1.0$ & $5.06\pm0.15$ & $6960\pm530$ & $24.7\pm0.8$ \\
11 & G & \cm &     &     &     &     &     &     &     &     & 235.5 & 44  & 20.4 & $258.7\pm0.7$ & $5.17\pm0.04$ & $6620\pm440$ & $23.3\pm0.6$ \\
12 & G & \cm &     &     & \cm & \cm & \cm & \cm &     &     & 66.0 & 44  & 2.3 & $257.2\pm1.0$ & $5.14\pm0.05$ & $6970\pm550$ & $24.7\pm0.8$ \\
13 & G & \cm & \cm & \cm &     &     &     &     & \cm & \cm & 227.0 & 44  & 19.5 & $259.7\pm0.8$ & $4.87\pm0.13$ & $5880\pm440$ & $22.3\pm0.7$ \\
\hline
14 & $>2\%$ & \cm & \cm & \cm & \cm & \cm & \cm & \cm & \cm & \cm & 54.1 & 39  & 1.1 & $258.4\pm1.0$ & $4.97\pm0.16$ & $6290\pm500$ & $24.1\pm0.8$ \\
15 & $>10\%$ & \cm & \cm & \cm & \cm & \cm & \cm & \cm & \cm & \cm & 40.9 & 23  & 1.7 & $258.4\pm1.7$ & $5.01\pm0.16$ & $6340\pm990$ & $24.0\pm1.4$ \\
16 & 240--294 & \cm & \cm & \cm & \cm & \cm & \cm & \cm & \cm & \cm & 125.7 & 72  & 3.9 & $258.4\pm0.9$ & $5.07\pm0.14$ & $6330\pm490$ & $23.9\pm0.8$ \\
17 & G $\neq13$ & \cm & \cm & \cm & \cm & \cm & \cm & \cm & \cm & \cm & 43.4 & 33  & 0.6 & $258.7\pm1.0$ & $4.98\pm0.15$ & $6070\pm530$ & $23.7\pm0.9$ \\
18 & G $\neq14$ & \cm & \cm & \cm & \cm & \cm & \cm & \cm & \cm & \cm & 53.7 & 38  & 1.2 & $257.0\pm1.1$ & $5.16\pm0.15$ & $7030\pm590$ & $24.9\pm0.9$ \\
19 & G $\neq15$ & \cm & \cm & \cm & \cm & \cm & \cm & \cm & \cm & \cm & 57.6 & 38  & 1.6 & $256.2\pm1.3$ & $5.13\pm0.15$ & $7410\pm770$ & $25.7\pm1.1$ \\
20 & G $\neq16$ & \cm & \cm & \cm & \cm & \cm & \cm & \cm & \cm & \cm & 60.6 & 38  & 1.9 & $257.4\pm1.1$ & $5.05\pm0.16$ & $6830\pm590$ & $24.8\pm0.9$ \\
21 & G $\neq17$ & \cm & \cm & \cm & \cm & \cm & \cm & \cm & \cm & \cm & 53.4 & 36  & 1.4 & $257.9\pm0.9$ & $5.05\pm0.16$ & $6620\pm490$ & $24.4\pm0.8$ \\
22 & G $\neq18$ & \cm & \cm & \cm & \cm & \cm & \cm & \cm & \cm & \cm & 58.3 & 38  & 1.7 & $258.3\pm1.1$ & $5.01\pm0.16$ & $6400\pm590$ & $24.0\pm0.9$ \\
23 & G $\neq19$ & \cm & \cm & \cm & \cm & \cm & \cm & \cm & \cm & \cm & 37.9 & 38  & -0.6 & $259.5\pm0.8$ & $4.89\pm0.15$ & $5920\pm390$ & $23.2\pm0.6$ \\
\enddata
\tablecomments{Selection: G---shape based criterion (see Section~\ref{strategy:dataselection}), $>B\%$---points with counts rate higher than $B\%$ of maximum, 240--294---wide selection around the peak, $\neq o$---orbit $o$ excluded, $\mat{V}^\text{p}$---Poisson uncertainties, TC---throughput corrected, $\mat{V}^\text{t}$---throughput correction uncertainties, BG---background removed, $\mat{V}^\text{b}$---background uncertainties, WB---Warm Breeze removed, $\mat{V}^\text{w}$---Warm Breeze uncertainties, $\mat{V}^\text{o}$---pointing uncertainties, $\mat{V}^\text{f}$---boresight uncertainties, $\chi^2$---the estimator for the best fit, $\nu$---number of degree of freedom, ($\lambda,\,\beta,\,T,\,v)$---best-fit parameters. }
\end{deluxetable}

In this table, we mark the effects included in a given case, list the $\chi^2$ value obtained from minimization for this case, and the best parameters obtained from the minimization. We also show how the $\chi^2$ value obtained differs from the expected value, equal to the number of degrees of freedom $\nu$, in terms of the expected standard deviation of its distribution. In an ideal case, this value should be in the range $(-1,\,1)$ for a $1\sigma$ confidence interval. An actually obtained value $>2$ suggests that the proposed model likely is not able to correctly describe the observations, which may be either because the uncertainty system is incomplete or because the adopted physical model is incorrect/incomplete. 

The geometrical uncertainties related to the boresight position ($\mat{V}^f$) and spacecraft orientation ($\mat{V}^o$) reduce the $\chi^2$ value by a few (see sets \#2--\#4 in the table). Inclusion of the boresight position uncertainty changes the fitted latitude of the inflow by $0.08^\circ$, which is a large change compared to the uncertainty. The uncertainty in this case grows significantly as well. The other parameters are almost untouched by these two sources of uncertainty.

In sets \#5 and \#6, we show how the results react to the inclusion of the WB. If the WB is not subtracted from the \emph{IBEX} signal, then $\chi^2$ increases by more than a factor of three, and the parameters are shifted toward larger longitudes with respect to the longitude fitted with the WB subtracted. Specifically, they are very close to the values originally quoted by \citet{bzowski_etal:12a} and \citet{mobius_etal:12a}, with an inflow longitude of $\sim$259$^\circ$. Similar results were obtained when we only took the Poisson uncertainties into account (set \#11). The inclusion of the WB uncertainties also reduces the $\chi^2$ value substantially. Fitting with the WB subtracted without including its uncertainty (set \#5) results in a minimum $\chi^2$ that is larger by $\sim20$ compared to the baseline set \#1. Thus, this is a very important part of the overall uncertainty budget. 

Subtraction of the constant background (sets \#7 and \#8) does not change the inflow parameters, but the effect of the subtraction is unexpectedly large for the magnitude of minimum $\chi^2$: it is reduced by 0.6. Note that the background level $0.0089$ cts s$^{-1}$ is more than two orders of magnitude smaller than the count rate in the data point with the lowest rate. 

Including the TC for this particular data selection made the $\chi^2$ value larger. This, however, is mainly caused by the contribution from orbit 17, for which the correction factor is the largest (see Figure~\ref{thr_factor}) and where the measured rates are larger than expected from the best-fit parameters. This may have to do with the fact that the good times for this orbit were very short and occurred at the  boundaries of the HASO times, i.e., when the spacecraft was relatively close to the terrestrial environment \citep[see][Figure~1]{bzowski_etal:15a}. This also affects the obtained parameters and the parameter correlation line. The change in the temperature is mostly related to the change in the direction of the correlation line, but the fitted temperatures are also lower by $\sim$100~K if we include the correction.

In set \#12, we list the parameters obtained when we only include the WB and the background without including the other effects. In that case, the inflow longitude changes to 257.2$^\circ$, while with only the new uncertainties taken into account (set \#13) without subtracting the WB and the background, the inflow longitude is 259.7$^\circ$ and the temperature decreases to $\sim$5900~K. 

\subsection{Data Range Selection}
\label{results:selection}
The \emph{IBEX}-Lo detector distinguishes helium atoms from hydrogen only on a statistical basis. Consequently, the observed signal in energy steps 1--3 is potentially a mixture of helium and hydrogen. In general, hydrogen should be observed at later times during the ISN season \citep{saul_etal:12a,saul_etal:13a,kubiak_etal:13a,rodriguez_etal:13a, schwadron_etal:13a}. Potentially, an additional signal could have a large impact on the fitted parameters (as can be seen above for the WB). It is important to choose only those data points for which the ISN He signal is dominant. Our criteria for data selection are presented in Section~\ref{strategy:dataselection}. In the following, we show the sensitivity of the fitted parameters to data selection (see sets \#14 through \#23 in Table~\ref{tab:resunc}).

The first test repeated the fits with data selected using arbitrary limits for the percentage of the lowest admitted count rates with respect to the maximum observed count rate. Rows \#14 and \#15 in Table~\ref{tab:resunc} show the parameters fitted to the data sets for which the observed count rates are larger than 2\% and 10\% of the maximal signal in the season, respectively. For these selections, we obtain $\chi^2$ values closer to the number of degree of freedom (39 and 23, respectively). However, the obtained result is significantly shifted along the correlation line to a longitude of 258.4$^\circ$.  

A similar longitude was obtained for a wider selection of data around the peak, i.e., using a spin-angle range between 240$^\circ$ and 294$^\circ$ (set \#16). However, the resulting $\chi^2=125.7$ is much larger than the expected value, equal to the 72 degrees of freedom in this data set. This could suggest that even after subtraction of the WB and the background, some source of background or additional signal is still left. Note that the parameters obtained for this data selection agree better than $1\sigma$ with the results of the original data selection (case \#1).

In sets \#17--\#23 we examined how sensitive our results are to excluding data points from individual orbits. The obtained inflow longitude is most strongly shifted to the lower values when we drop orbit 15. When we leave out earlier or later orbits, the inflow longitude is generally larger. The largest longitude is obtained when we drop orbit 13 or 19. Also, dropping the same orbits results in the most significant reduction of $\chi^2$. This may be understandable because orbit 13 still contains a non-vanishing contribution from the WB (even though we tried to account for it in the fitting), and orbit 19 is likely to feature the largest contribution from ISN H, which is not accounted for at all in our simulations or the uncertainty system. 

\citet{bzowski_etal:15a}, in their Figure 1, present the distribution of good times on the key ISN orbits. The valid time intervals of the observations in each orbit are different. Thus, the data points with the highest count rates are not necessarily connected to the highest statistics. For example, in orbit 17, we have only $\sim$8 hr of observations, and thus the statistics is low.
From these facts, it is hard to conclude if the change of best-fit parameters due to the elimination of individual orbits is related to the drop of the orbits with worse or better statistics, or with the additional signal not subtracted from the data. However, the systematic dependence suggests the second possibility. 

\subsection{Other studies of the 2009 data}
Another analysis of \emph{IBEX}-Lo data was performed by \citet{schwadron_etal:15a}. In their method, the observed count-rate peak positions from orbits 14--19 were fit to the model with the inflow longitude treated as a free parameter, and the other parameters were adopted based on the correlation line given by \citet{mccomas_etal:12b}. The existence of this correlation is a direct result of the analytical theory developed by \citet{lee_etal:12a}. The fitting was carried out on data with a fine time resolution during the orbits, unlike in our method where the count rates analyzed are averaged over good times for individual orbits. For the 2009 observation season, they obtained $\lambda=256.6^\circ\pm2.7^\circ$, $\beta=5.1^\circ\pm0.3^\circ$, $v=24.8\pm2.1$~km~s$^{-1}$, and $T=7400\pm200$~K.

The small difference between the inflow longitude obtained by \citet{schwadron_etal:15a} for the 2009 season and the longitude $257.7^{\circ} \pm 1.0^{\circ}$ we obtain in the first line of Table~\ref{tab:resunc} is within the Schwadron et al. error bars. For an identical set of orbits (model \# 17 in Table~\ref{tab:resunc}), we obtained $\lambda = 258.7^{\circ} \pm 1.0^{\circ}$, which is still inside the Schwadron et al. error bars. The differences between the other parameters result from the correlation between the parameters. The small difference in longitude may be due to the difference in the used form of the data, which are not completely equivalent. This is shown by \citet{bzowski_etal:15a}, where an analysis based on the Gauss fit parameters is compared with the direct fit of the observed signal to the expected fluxes. Also, in our method, we do not assume that parameters must adhere to any correlation line, and all four parameters are fit independently. The parameter correlation emerges as a natural result of the fitting.

Note that the result for the 2009 data presented in \citet{bzowski_etal:15a} is also different from the result we obtained here because they use spin-angle bins $254^\circ$--$282^\circ$ from orbits 14--19. Here, this region is supplemented by one additional orbit, orbit 13. Despite the low count rate, the counting statistics from this orbit is rather large due to relatively long ``good time'' intervals. Note that the result of $\lambda=258.6^\circ\pm1.2^\circ$, $\beta=4.99^\circ\pm0.16^\circ$, $v=23.8\pm1.0$~km~s$^{-1}$, and $T=6140\pm590$~K, as obtained by \citet{bzowski_etal:15a}, are very close to the results of this analysis for set \#17, i.e., with orbit 13 dropped.
These differences lend support to the conclusion drawn by \citet{mobius_etal:15b} on the importance of the WB in the search for the parameters of ISN He, and by \citet{bzowski_etal:15a} that the WB needs revisiting with the use of a broader data set than originally used by \citet{kubiak_etal:14a}. Also, the observation by \citet{schwadron_etal:15a} that some of the time-resolved data points they used seem not to fit the model, while other adjacent points from the same orbit do fit well to the found optimum solution may potentially be important. This may be indicative of an additional intermittent, presumably local, component in some portions of the data.

\section{CONCLUSIONS}

We improved both the modeling and the data preparation systems in the analysis of ISN He observations from \emph{IBEX}-Lo. Modeling improvements are discussed in the accompanying paper by \citet{sokol_etal:15b}. In this paper, we analyzed the observational aspects of \emph{IBEX}-Lo measurements of ISN He and refined the system of searching the ISN He flow parameters. To that end, we reviewed the spin-axis determination and corrected the data for the throughput reduction due to the ambient electron flux. We developed a consistent system of uncertainties for the data, which includes correlations between data points due to a number of effects. We also analyzed the influence of various uncertainties on the results and set the field for actual data analysis, which is presented by \citet{bzowski_etal:15a} in this volume. 

We carried out a comparison study of the influence of various elements of the uncertainty system on the obtained ISN He inflow parameters, taking as the test case the observations from the 2009 ISN He observation season, analyzed previously by \citet{bzowski_etal:12a}. These data were behind the prior suggestion that the ISN He inflow vector differs by $\sim$4$^{\circ}$ in direction and $\sim$3~km~s$^{-1}$ in speed, however, this was based on an unsatisfactorily high value of minimized $\chi^2$. 

Our analysis showed that the ISN He flow parameters obtained with the uncertainty system developed in this study do not significantly differ from the values obtained by \citet{bzowski_etal:12a}. The upwind direction of ISN He emerging from this study based on 2009 data is $\lambda = 257.7^{\circ} \pm 1.0^{\circ}$, $\beta = 5.04^{\circ} \pm 0.15^{\circ}$ in ecliptic coordinates, speed $ v = 24.5 \pm 0.8$~km~s$^{-1}$, and temperature $T = 6650 \pm 520$~K. This solution, however, is found for the minimized $\chi^2_{\text{min}}$ value which exceeds the expected value by $2.2 \sigma$ and cannot be  regarded as satisfactory. We believe the reason for this is either (1) a still incomplete uncertainty system (i.e., a hidden influence from some still unrecognized measurement effect), or (2) an underestimated background, or (3) an inadequate assumed physical model, which would imply that the model of the ISN He with the WB 150 AU in front of the heliosphere described by the superposition of two homogeneous Maxwell--Boltzmann distribution functions with two different parameter sets is not fully valid.

The uncertainties presented in this paper result from the fitting of the assumed model to different data sets with the full uncertainty system taken into account. Since, based on this analysis, we suspect that the adopted model is not able to fully describe all of the sources of the signal, the parameter uncertainties we obtained should not be treated as the appropriate confidence intervals. In fact, they cannot be determined based on an analysis of a single season. Further discussion of the estimation of uncertainties is given by \citet{bzowski_etal:15a}.

From an analysis of various contributors to the $\chi^2$ we found that the single-Maxwellian model for the ISN He gas population in front of the heliosphere does not describe the physical reality well. This conclusion was already hypothetically proposed by \citet{bzowski_etal:12a} and supported by the discovery of the WB by \citet{kubiak_etal:14a}. Without the WB, the longitude of the inflow vector would be still larger, $259.7^{\circ} \pm 0.8^{\circ}$, with the remaining parameters following the correlation line approximately shown by \citet{bzowski_etal:15a}, which is very close to the result reported in this paper. Now we can quantify the contribution from the WB: when it is included in the analysis, the minimum $\chi^2$ value is reduced drastically, from 216 to 64.8 (note that we mention the absolute $\chi^2$, not the reduced $\chi^2$, which is the absolute $\chi^2$ divided by the number of degrees of freedom). Surprisingly, such a large reduction in $\chi^2$ is given by a population with a very low contribution to the signal observed in the data points that were taken to the analysis. It also exposes the sensitivity of the results to small extra components.

We have shown that data selection has a relatively low impact on the fitted parameters, but it does have a notable influence on the absolute magnitude of the minimum $\chi^2$. Taking the data from wider ranges of spin-angles in each orbit tends to worsen the quality of the fit, which lends support to the hypothesis that the signal contains unidentified contributions other than ISN He and the WB or that the WB parameters need to be refinement. Additional evidence supporting this conclusion is the fact that rejecting a larger number of low count-rate data points (in other words, leaving only the data points with intensities not less than 0.02 of the maximum count rate) reduces $\chi^2$ so that it deviates from the expected value only by $1.1 \sigma$, which can be regarded as a satisfactory fit. Interestingly, a larger influence on the solution has the selection of orbits taken to the analysis. Solutions with individual orbits dropped from the analyzed sample vary within $\pm 1.5^\circ$ in the inflow longitude, with the remaining parameters changing accordingly. 

The interface buffer losses, misidentified by \citet{lallement_bertaux:14a} as a dead time effect, turned out to have a negligible effect on the fitted parameters, even though the absolute magnitude of the correction is relatively high: $\sim$10\%. This conclusion is in full agreement with the independent analysis by \citet{mobius_etal:15b} in this volume. Including the uncertainty of the boresight orientation in relation to the spin-axis statistically significantly changes the fitted inflow latitude and its uncertainty. The change is by 0.08$^\circ$.
Other effects taken into account in the uncertainty system have a very low impact both on the fitted parameters and on the $\chi^2$ value. Therefore, we conclude that either all of them should be included in the analysis, or all of them should be excluded. We recommend the first option, since they are now well identified and including them does not complicate the analysis very much, and due to the presence of the WB and its uncertainties, one must anyway institute an uncertainty system similar to that presented in this study. One additional uncertainty is related to the uncertainty of the distance from which the atoms that \emph{IBEX} observes actually come. In the interpretation model that we adopted, this distance is assumed to be equal to 150 AU, which is approximately the mean free path against collisions for ISN He atoms in the LIC, but in reality this distance has never been modeled. The effect of changing this atom tracking distance on the simulated signal is presented by \citet{sokol_etal:15b}, and its potential influence on the derived parameters is discussed by \citet{mccomas_etal:15b}.

The refined observation parameters and the uncertainty system, as well as the conclusions we obtained here, are used by \citet{bzowski_etal:15a} in their analysis of data from 2009--2014. 

\acknowledgments
The research portion carried out at SRC PAS was supported by the Polish National Science Center grant 2012-06-M-ST9-00455. Work by US authors was supported by the \emph{IBEX} mission as a part of NASA's Explorer Program.

\appendix

\section{Correction for the throughput reduction in the interface buffer}
\label{throughput}
The \emph{IBEX}-Lo detector transfers measured events to the Central Electronic Unit (CEU) via an interface buffer in order to adapt the stochastic sequence of the observed events to their intake into the CEU with a certain processing time for each event. To mitigate the smoothing effect, the interface buffer has been designed as a double buffer which temporarily stores up to two events. Only the third consecutive event is lost if it occurs within one processing time window of the CEU. Losses, and thus any reduction of the event rate, are negligible as long as the typical time between successive events is much shorter than the processing time, but they become noticeable as the typical time between successive events approaches this value.

In this appendix, we derive an analytical model of the interface buffer (Section~\ref{throughput:derivation}) to statistically correct for the losses caused by limited throughput. In Section~\ref{throughput:before}, we describe the application of the model to data before orbit 168. Starting with orbit 169, \emph{IBEX}-Lo was commanded into a new operational mode that requires a TOF2 coincidence, which reduces the total rate to well below 100 events per second. This eliminates the throughput reduction almost entirely.

\subsection{Derivation}
\label{throughput:derivation}
In this section, we derive an analytical model of a First In First Out (FIFO) double buffer designed to compensate for the natural irregularity of the time intervals between successive events. Let $p_2(t)$, $p_1(t)$, and $p_0(t)$ be the probabilities that at time $t$ two, one, and zero events are stored in the buffer. These three states are all possible buffer states, and thus $p_0(t)+p_1(t)+p_2(t)=1$ at any time. We use this to substitute $p_1(t)=1-p_2(t)-p_0(t)$. The buffer processing time, i.e., the time needed to transmit a single event, is $t_\text{b}$. The probabilities of the buffer state at time $t+t_\text{b}$ can be expressed as
\begin{align}
 p_2(t_0+t_\text{b})&=\eta_{2\to2}p_2(t_0)+\eta_{1\to2}\big(1-p_2(t_0)-p_0(t_0)\big)+\eta_{0\to2}p_0(t_0)\, , \label{eqp2}\\
 p_0(t_0+t_\text{b})&=\eta_{2\to0}p_2(t_0)+\eta_{1\to0}\big(1-p_2(t_0)-p_0(t_0)\big)+\eta_{0\to0}p_0(t_0)\, , \label{eqp0}
\end{align}
where $\eta_{a\to b}$ are the probabilities that during an interval $\Delta t=t_\text{b}$ the buffer changes the number of stored events from $a$ to $b$. For the FIFO buffer, the probabilities for buffer state changes are as follow:
\begin{align}
 \eta_{2\to2}&=\int_0^{t_\text{b}} \mathcal{P}_\text{time,2}(y)\left[1-\mathcal{P}_\text{Poiss}(\lambda (t_\text{b}-y),0)\right]\text{d}y\, , \label{eta22}\\
 \eta_{1\to2}&=\sum_{n=2}^{\infty} \mathcal{P}_\text{Poiss}(\lambda t_\text{b},n) \left(1-\int_0^{t_\text{b}}\mathcal{P}_\text{time,1}(y)\left(\frac{y}{t_\text{b}}\right)^n\text{d}y\right) \label{eta12}\, ,\\
 \eta_{0\to2}&=1-\mathcal{P}_\text{Poiss}(\lambda t_\text{b},0)-\mathcal{P}_\text{Poiss}(\lambda t_\text{b},1)\, , \label{eta02}\\
 \eta_{2\to0}&=0\, , \label{eta20}\\
 \eta_{1\to0}&=\mathcal{P}_\text{Poiss}(\lambda t_\text{b},0)\, , \label{eta10}\\
 \eta_{0\to0}&=\mathcal{P}_\text{Poiss}(\lambda t_\text{b},0)\, , \label{eta00}
\end{align}
where $\mathcal{P}_\text{Poiss}(\alpha,k)$ is the probability of $k$ counts in the Poisson process with parameter $\alpha$, and $\lambda$ is the event rate. $\mathcal{P}_\text{time,j}(y)$ is the probability density that the event currently in process will be fully transferred at the time $t=t_0+y$ if it is known that exactly $j$ events are stored in the buffer at time $t=t_0$:
\begin{align}
 \mathcal{P}_\text{time,1}(y)&=\frac{1}{\mathcal{N}_1}\mathcal{P}_\text{Poiss}(\lambda (t_\text{b}-y),0)\, ,\\
 \mathcal{P}_\text{time,2}(y)&=\frac{1}{\mathcal{N}_2}\left[1-\mathcal{P}_\text{Poiss}(\lambda (t_\text{b}-y),0)\right]\, ,
\end{align}
$\mathcal{N}_j$ are the normalization constants obtained from the condition $\int_0^{t_\text{b}} \mathcal{P}_\text{time,j}(y)\text{d}y=1$ for $j=1,2$. The probability density that the transmission of the currently processed event will be finished at time $t=t_0+y$, if exactly one event is stored in the buffer $\mathcal{P}_\text{time,1}(y)$, is directly proportional to the probability that another event has not been added to the buffer during the interval $\Delta t$ between the start of the transmission ($t=t_0+y-t_\text{b}$) and now ($t=t_0$): $\Delta t=t_0-(t_0+y-t_\text{b})=t_\text{b}-y$, in other words, the probability of zero counts in the Poisson process with parameter $\lambda\Delta t=\lambda(t_\text{b}-y)$. 
Conversely, if it is known that at time $t=t_0$ exactly two events are stored in the buffer, then the probability density for the time of the end of transmission of the first processed event $\mathcal{P}_\text{time,2}(y)$ is proportional to the probability that a non-zero number of events have been added to the buffer from the start of the transmission, i.e., the complementary event to that previously described. 

The probabilities given by equations \eqref{eta22}--\eqref{eta00} can be inferred based on the following reasoning. The probability $\eta_{2\to2}$ means that after the first processed event was transmitted, at least one additional event has been added. Thus, one needs to integrate the previously described probability density $\mathcal{P}_\text{time,2}(y)$ with the complement probability to the probability of zero events during the interval $(t_0+y,t_0+t_\text{b})$. 

The transition from one event in the buffer to two $\eta_{1\to2}$ implies that at least two events were observed during the interval and at least one of them after the transmission of the first stored event is completed. The probability could be expanded into a series\footnote{This is an infinite series, but fortunately it quickly converges in the case of the \emph{IBEX}-Lo buffer, for which we have $\lambda t_\text{b} \ll 1$. In the numerical approach, we take the first 20 terms in the series.} of probabilities of different numbers ($n$) of events (folded with the complementary event) to account for all of the events possibly accumulated in a time shorter than the time needed to complete the transmission of one stored event. Due to the statistical independence of the events, one needs to fold the fraction of $y/t_\text{b}$ $n$ times. 

The probability $\eta_{0\to2}$ means that at least two new events are added during the interval; this is the complementary event to the sum of zero and one events. If at the time $t=t_0$ two events are stored in the buffer, then both of them cannot be transmitted during the time $t_\text{b}$, and thus $\eta_{2\to0}=0$. The transition from the state with one ($\eta_{1\to0}$) and zero ($\eta_{0\to0}$) events in the buffer to the empty buffer occurs if new events are not added to the buffer during this interval.

Assuming that the event rate does not change with time, i.e., the stationary Poisson process, we have $p_j(t+t_\text{b})=p_j(t)$, and we can solve equations \eqref{eqp2} \& \eqref{eqp0} and express the probabilities $p_2$, $p_1$, and $p_0$ as a function of the probabilities $\eta_{i\to j}$, defined in Equations \eqref{eta22} through \eqref{eta00}. In the left panel of Figure~\ref{bufprob}, we present these probabilities as a function of $\lambda$ for the two values of the buffer transmission time $t_\text{b}$ in the \emph{IBEX}-Lo interface buffer. Hereafter, we treat these probabilities $p_j(\lambda,t_\text{b})$ as a function of $\lambda$ and $t_\text{b}$, which were assumed to be constant in the above reasoning. The probability that the event is added to the buffer, i.e., that it is transferred to the CEU, is equal to the probability that the buffer is not full: $p_\text{tr}(\lambda,t_\text{b})=1-p_2(\lambda,t_\text{b})$. In the right panel of Figure~\ref{bufprob}, we compare the transmission probabilities for the cases with and without the double buffer included in the interface.

\begin{figure}[ht!]
\plottwo{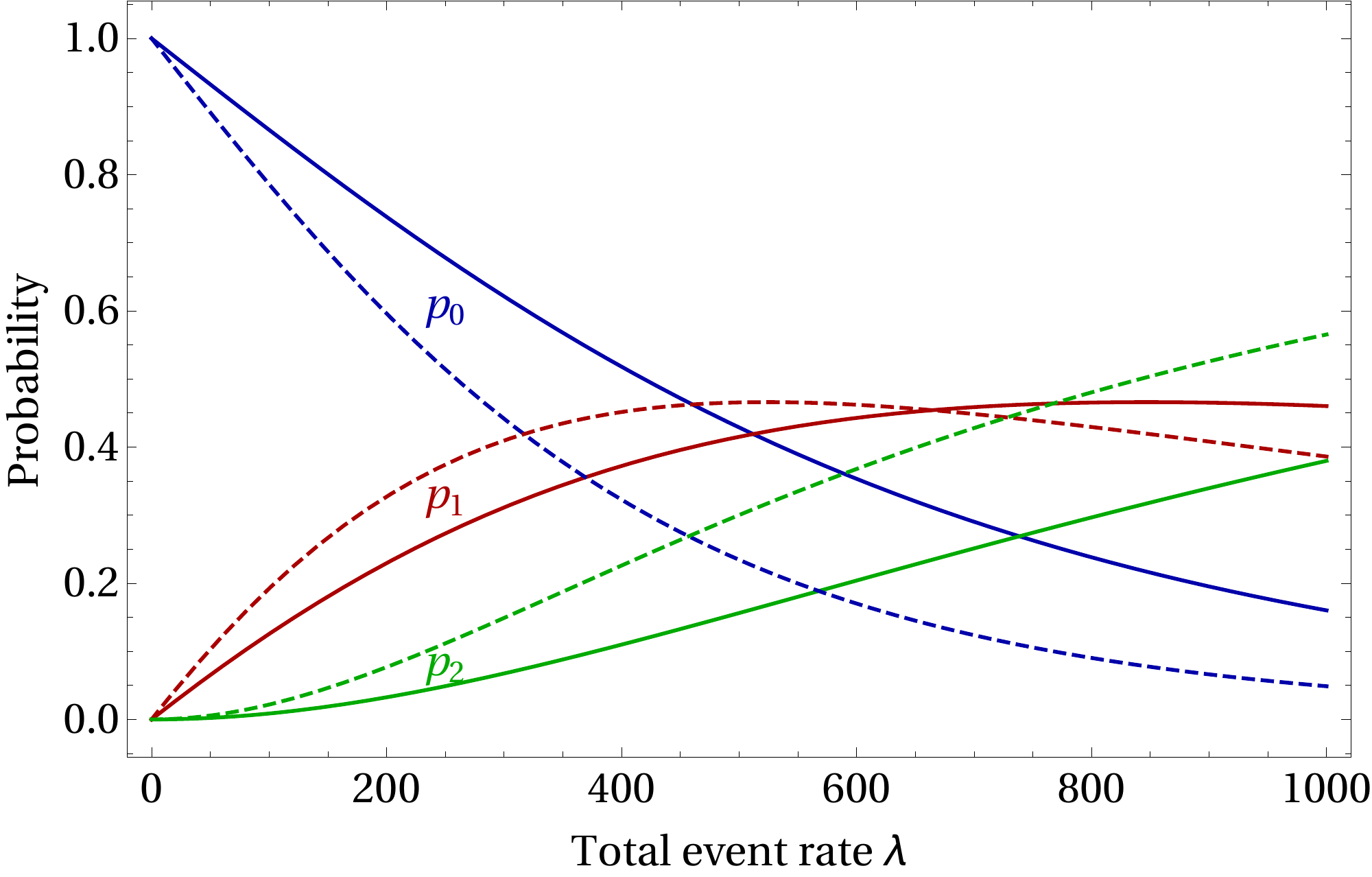}{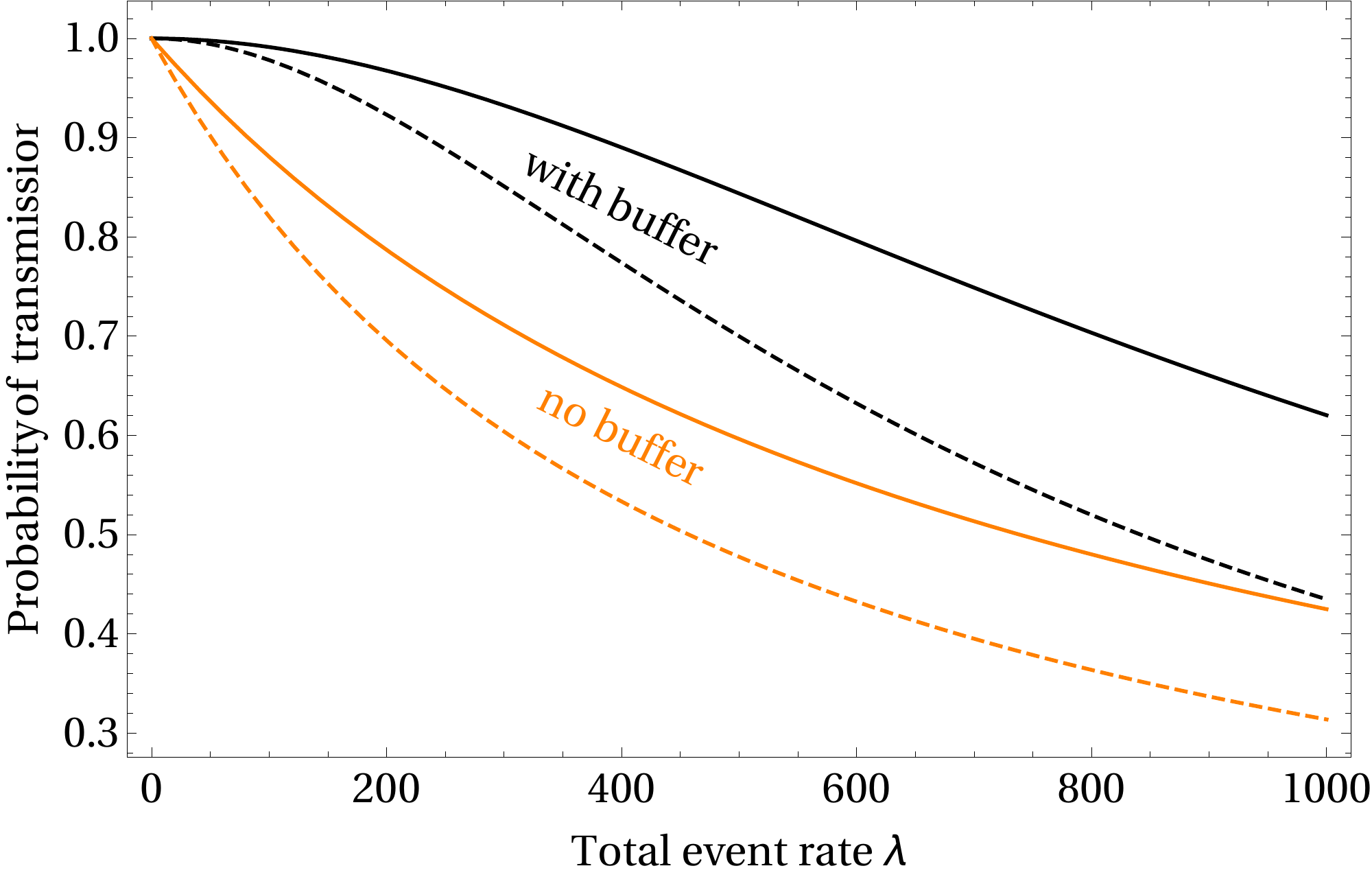}
 \caption{Probabilities of the buffer states (left panel) and probability of event transmission (right panel) as a function of the total event rate ($\lambda$). In the right panel, we present probability of event transmission for an interface without a buffer (orange). Solid lines are plotted for $t_\text{b}=t_\text{oHK}$ and dashed for $t_\text{b}=t_\text{wHK}$. Typical values of the total event rates measured by \emph{IBEX}-Lo are 200--500 s$^{-1}$.}
 \label{bufprob}
\end{figure}

In addition, at the beginning of each 6$^\circ$ bin, 20 House Keeping and Rate packets must be transferred to the CEU. The transmission of each packet takes $t_\text{HK}=0.834\text{ ms}$, whereas $t_\text{event}=1.354\text{ ms}$ is needed for the transmission of a single event. Each packet is transferred together with the currently processed event, or consecutively, if the interface buffer is empty. During the part of the 6$^\circ$ bin when these packets are transferred, the buffer transmission time is $t_\text{wHK}=t_\text{event}+t_\text{HK}$, whereas outside this part it is $t_\text{oHK}=t_\text{event}$. Transmission of the House Keeping and Rate packet right after the event takes an additional time $t_\text{HK}$.

The time needed to transfer all of the House Keeping and Rate packets varies depending on the measured event rate. If the buffer is empty during the transfer, it takes $20\times t_\text{HK}=16.68\text{ ms}$, whereas if the buffer is not empty for the whole time, it takes $20\times t_\text{wHK}=43.76\text{ ms}$. For an appropriate estimate of the transfer time for the House Keeping and Rate packets, we perform a linear interpolation of the actually transmitted event rate between the following two values:
\begin{equation}
 t_\text{transfer}(\lambda)\approx16.68\text{ ms}+\frac{p_\text{tr}(\lambda,t_\text{wHK})\lambda}{\tilde{\lambda}_\text{max}}(43.76-16.68\text{ ms})\, ,
\end{equation}
where $\tilde{\lambda}_\text{max}=t_\text{wHK}^{-1}$ is the maximum possible transmission rate during the transfer of House Keeping and Rate packets.  

The data are accumulated in 6$^\circ$ bins. The accumulation time for a bin is equal to $t_\text{bin}=\frac{6}{360}t_\text{spin}$, where $t_\text{spin}\approx 14.4\text{ s}$ is the time of one spacecraft rotation. To correct the data for the limited throughput, we need to estimate the ratio of the total event rate to the rate of transmitted events as a weighted average:
\begin{equation}
 \gamma(\lambda)=\frac{\lambda}{\tilde{\lambda}}=\left(\frac{t_\text{transfer}(\lambda)}{t_\text{bin}}p_\text{tr}(\lambda,t_\text{wHK})+\frac{t_\text{bin}-t_\text{transfer}(\lambda)}{t_\text{bin}}p_\text{tr}(\lambda,t_\text{oHK}) \right)^{-1}\, . \label{eqthcorr}
\end{equation}

Equation \eqref{eqthcorr} gives the correction factor for limited throughput as a function of the total event rate ($\lambda$). In this derivation, we assume that the total event rate is constant over time. This is not in the case in \emph{IBEX} observations, as the spacecraft rotates and the total event rate varies. However, the timescales of the changes in the total event rate and the time of scanning through one spin-angle bin are much longer than the transmission time $t_\text{b}$, and thus we are able to use the stationarity condition as a reasonable assumption.

\subsection{Application of the model to the \emph{IBEX} ISN flow observations}
\label{throughput:before}
In this section, we will apply the model for the event reduction across the interface buffer to the ISN flow observations with as much information about the total event rate that arrives at the buffer as is available in the data. The TOF system of the \emph{IBEX}-Lo sensor can measure up to three TOF values for each incoming particle: TOF0 between the first C-foil and the microchannel plate (MCP) detector, TOF1 between the second C-foil and the MCP, and TOF2 between the two foils. In addition, it records a delay line value TOF3, which indicates the position of incidence on the stop MCP detector which is subdivided into four sectors \citep[for details see][]{fuselier_etal:09b}. Up to orbit 168, each event with at least one TOF measured (TOF0--TOF3), was transmitted to the interface buffer. The majority were events with only TOF3 present, most of them caused by electrons.

In addition to the full event information, the \emph{IBEX}-Lo TOF logic generates the monitor rates for each of the individual TOF channels, integrated over 60$^\circ$ wide spin-angle sectors, denoted as S0--S5, at a cadence of 64 spins. The total event rate $\lambda$ transmitted to the interface buffer is given by the sum of the TOF2 ($\lambda_\text{TOF2}$) and TOF3 ($\lambda_\text{TOF3}$) monitor rates, with the 'triple' event rate ($\lambda_\text{triple}$) subtracted. The start and stop signals used in TOF2 and TOF3 are mutually exclusive, except for 'triple' events, which involve all detector parts, and thus they are recorded by both the TOF2 and TOF3 monitor rates:
\begin{equation}
 \lambda=\lambda_\text{TOF3}+\lambda_\text{TOF2}-\lambda_\text{triple}\, .
\end{equation}
These rates are not reduced by the losses at the interface buffer. Based on this information, we estimate the total event rate $\lambda$ for each 6$^\circ$ bin separately.

We realize that the contribution of the measured ISN atoms to these rates can be determined separately for each 6$^\circ$ bin based on the histogram events. We sum these events over the appropriate 60$^\circ$ sectors and subtract them from the total event rate for now to obtain the background for each sector S$j$, which can be expressed as
\begin{equation}
 \lambda_\text{bg}(\text{S}j)=\lambda_\text{TOF3$|$bg}(\text{S}j)+\lambda_\text{TOF2$|$bg}(\text{S}j)=\lambda_\text{TOF3}(\text{S}j)+\lambda_\text{TOF2}(\text{S}j)-\sum_{i\in \text{S}j}(\alpha+1)\lambda_\text{triples}(i)\, .
\end{equation}
The summation is over ten 6$^\circ$ bins $i$ in sector S$j$, where $\alpha=2.56\pm0.29$ denotes the ratio of the rate of all possible atom events to the rate of triple atom events. The additional 1 in the brackets accounts for the presence of triple events in both TOF2 and TOF3. In Figure~\ref{thr_rates}, we present two examples of the angular distributions of these rates within the 60$^\circ$ sectors. Because the monitor TOF rates only represent averages over the wide sector, the actual rates in each 6$^\circ$ bin cannot be restored unambiguously. Retaining constant values for each 60$^\circ$ sector would lead to discontinuities at the edges. Therefore, we use a quadratic spline $\mathcal{S}_\text{bg}$ with knots at the sector edges, defined so that the integral of the spline over the respective sector is equal to the sector background rate $\lambda_\text{bg}(\text{S}j)$. Only one unique spline fulfills these conditions. Subsequently, we assess the uncertainties of this approximation. In the last step we reconstruct the total event rate in each 6$^\circ$ bin $i$ by adding the previously subtracted counts:
\begin{equation}
 \bar{\lambda}(i)=\mathcal{S}_\text{bg}(i)+\alpha\lambda_\text{triples}(i)\, .
\end{equation}
$\mathcal{S}_\text{bg}(i)$ is the value of the spline for each bin $i$. 

\begin{figure*}[ht!]
 \plottwo{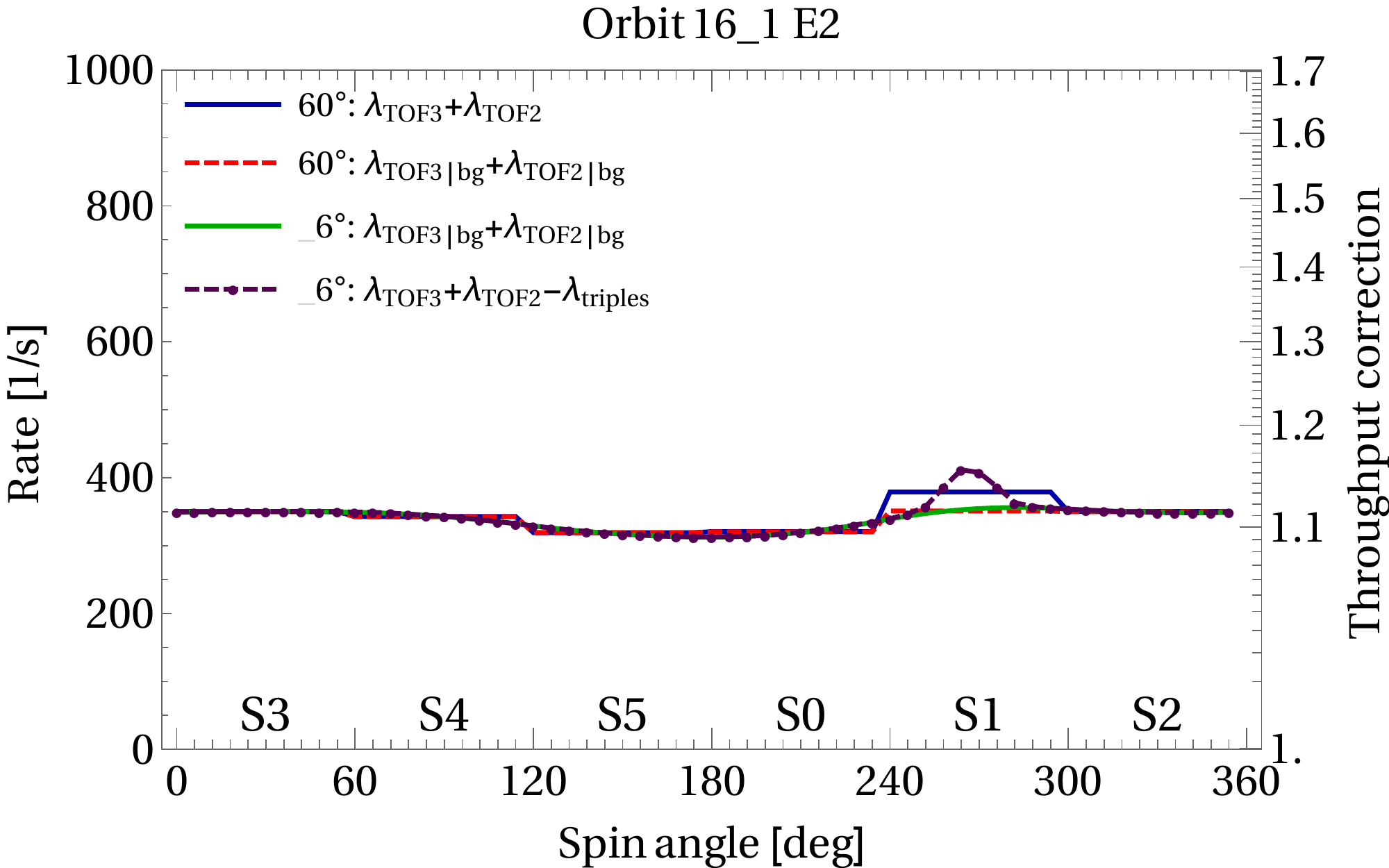}{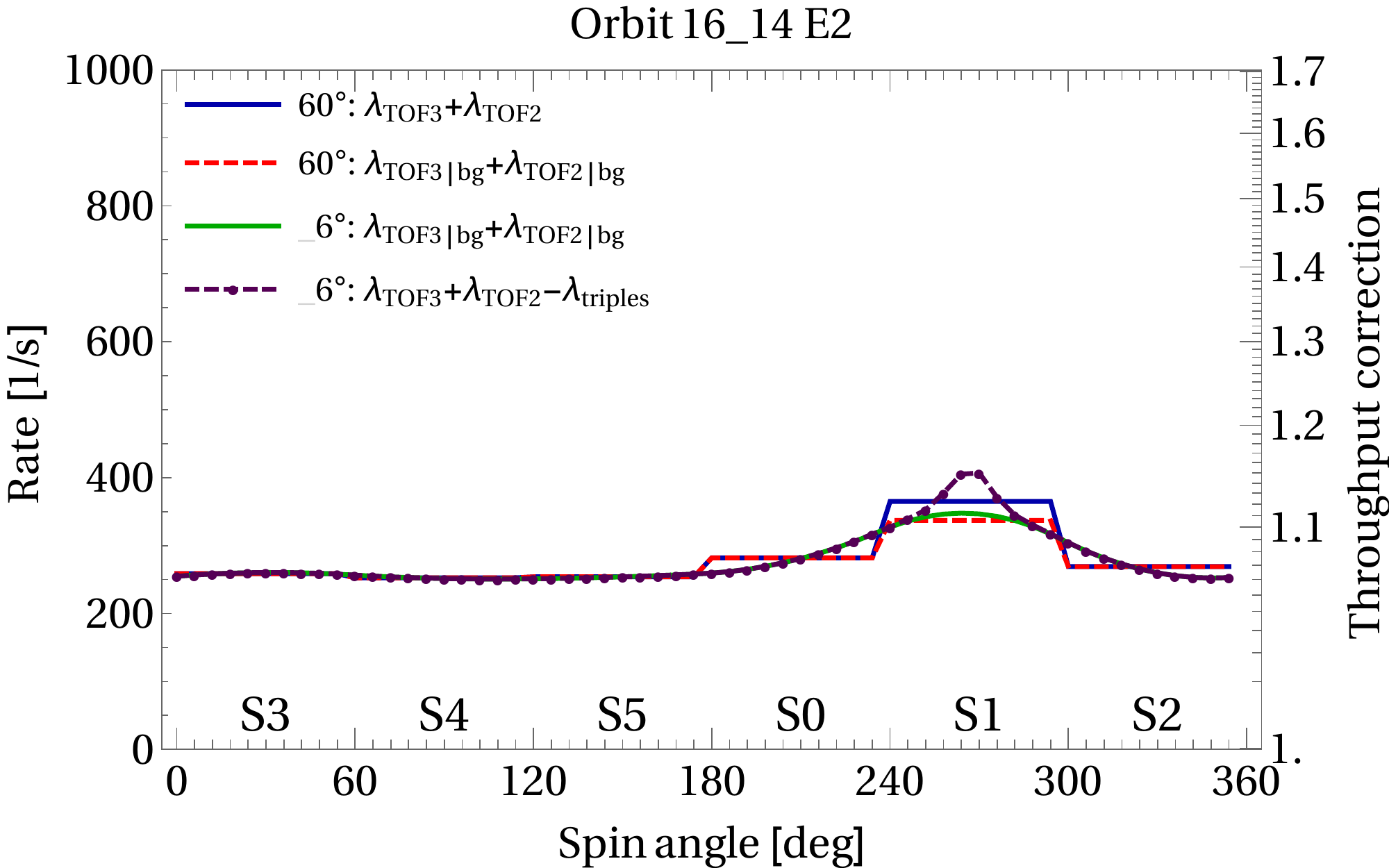}
 \caption{Illustration of the steps of the procesure used to estimate the total event rate for two selected time blocks on orbit 16: sum of the TOF3 and TOF2 event rates for the wide 60$^\circ$ sectors (solid blue line); the sum with a subtracted contribution from the measured atoms (dashed red); determined spline $\mathcal{S}_\text{bg}$ (solid green); the spline with added atom contribution (dashed purple). The scale on the right represents the appropriate correction factor. The two panels illustrate the event rates and the correction factors for two different time blocks during the observations. Note the substantial difference between the values of the correction factor for these two blocks. }
 \label{thr_rates}
\end{figure*}

The total event rate varies substantially with time during the observations, as does the correction factor. Therefore, the procedure to obtain estimates of the total event rate is applied separately for each 512-spin time block, enumerated with $q$. The number of counts $d_{i,q}$ in each bin $i$ for this time block $q$ is multiplied by the correction factor given by Equation \eqref{eqthcorr} for $\lambda=\bar{\lambda}_q(i)$. The effective TC factor for a given bin $i$ is the ratio of the corrected-to-uncorrected cummulative number of counts:
\begin{equation}
  \gamma_i  = \frac{\sum_{q=0}^{n}\gamma_{i,q} d_{i,q}}{\sum_{q=0}^{n} d_{i,q}}\, ,\label{eq:lambda}
\end{equation}
where $\gamma_{i,q}=\gamma\big(\bar{\lambda}_q(i)\big)$ is the TC factor for the time block $q$. 

In Figure~\ref{thr_factor} we show effective TC factors for those bins with a spin-angle in the range (240$^\circ$, 294$^\circ$) for orbits 13--19. The maximum in the correction factor around spin-angle 270$^\circ$ in individual orbits is caused by the events due to the ISN flow. A typical value for the correction factor in this range is $\sim$1.1, while the maximum value reaches $\sim$1.15. These values represent systematic differences in the data from orbit to orbit and over spin-angle for the fitting of the He ISN parameters. This systematic variation is statistically significant at least for the data points with highest statistics.

Given the statistical nature of the correction and the limited resolution of the data on which the correction is based, it is important to also assess the resulting uncertainties. In the described procedure, we quantify two types of uncertainties. The first one is related to the unknown shape of the  background rate $\lambda_\text{bg}$ over each 6$^\circ$ bin. The other one is related to the statistical fluctuations of the total event rate during each 512-spin time block.

\begin{figure}[ht!]
\centering
 \includegraphics[width=.45\textwidth]{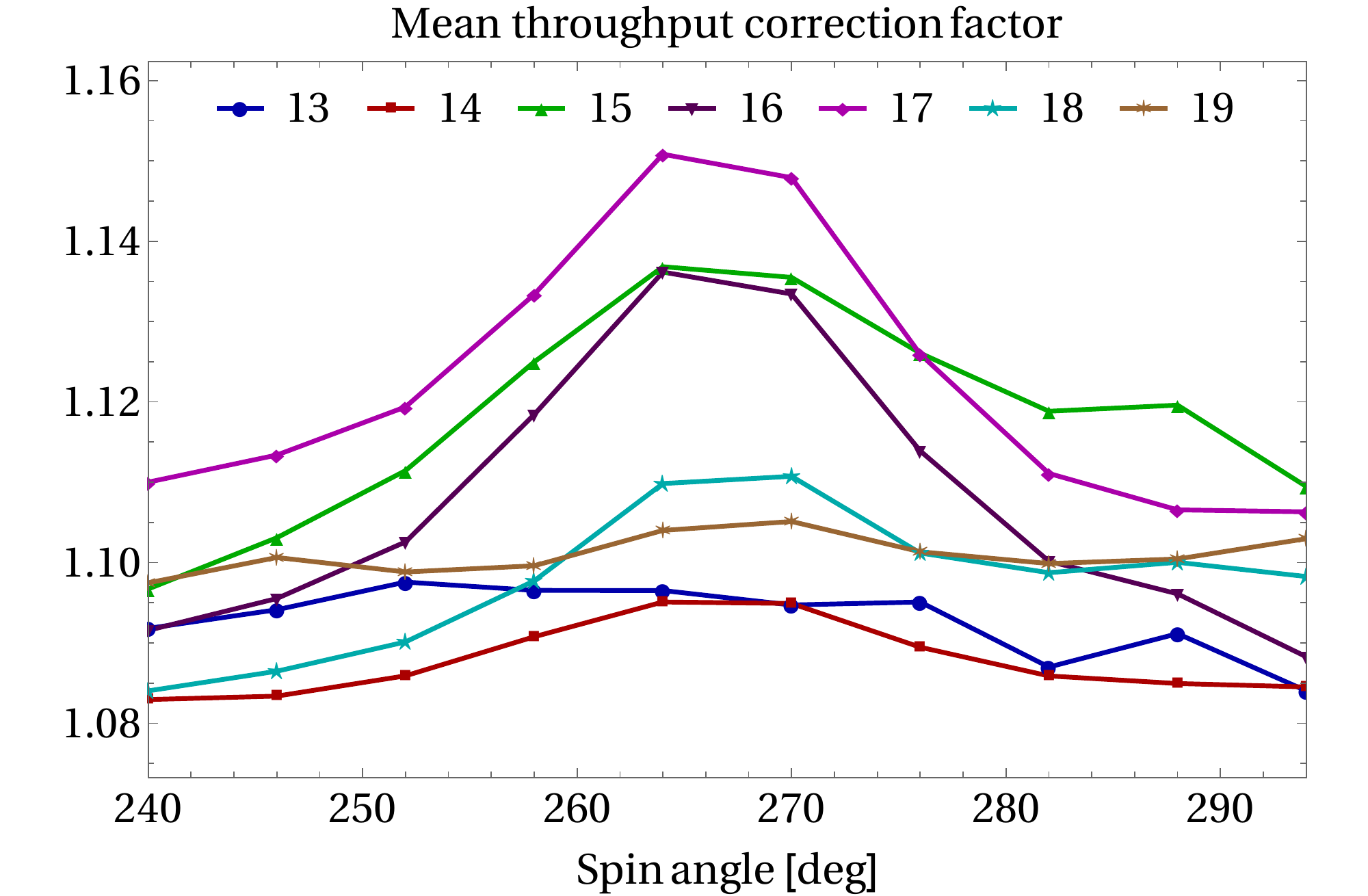}
 \caption{Mean value of the correction factor compensating for the throttling of the interface buffer for the data points around the ISN flow for the orbits used in the analysis.}
 \label{thr_factor}
\end{figure}

Determining the structure of the background rate $\lambda_\text{bg}$ with a 6$^\circ$ resolution is not possible because the sensor does not store the monitor rates at that resolution. We approximate the uncertainty related to this lack of information based on the previously described spline, and for each bin we take the value
\begin{equation}
 \delta_\text{A}\bar{\lambda}(i)=\left[\frac{1}{10}\int_{i-5}^{i+5}\big(\mathcal{S}_\text{bg}(x)-\mathcal{S}_\text{bg}(i)\big)^2\text{d}x\right]^{1/2}\, .
\end{equation}
We connect the uncertainty for a single bin with the variation of the spline over all 10 bins in the 60$^\circ$ sector. In other words, we assume that if the variation over the sectors is larger, then the uncertainty for a single bin is larger. Variation in the neighboring bins could have different directions and we use this procedure to asses the absolute value of this variation. Thus, despite the integral for each single bin depending on the value in the neighboring bins, we treat the uncertainties for the different bins as independent. Still, we take the correlation over time into account. We assume that a certain sector pattern in the background rate $\lambda_\text{bg}$ implies a similar pattern in the 6$^\circ$ bins. Quantitatively, we define this pattern as the following vector:
\begin{equation}
\vec{\xi}=\big(\lambda_\text{bg}(\text{S}0),\lambda_\text{bg}(\text{S}1), \lambda_\text{bg}(\text{S}2), \lambda_\text{bg}(\text{S}3),\lambda_\text{bg}(\text{S}4),\lambda_\text{bg}(\text{S}5) \big)-\frac{1}{6}\sum_{j=0}^5 \lambda_\text{bg}(\text{S}j)(1,1,1,1,1,1)\, .
\end{equation}
Then, the correlation coefficient between two 512-spin time blocks $q$ and $p$ is given by
\begin{equation}
 a_{q,p}=\frac{\vec{\xi}_q\cdot\vec{\xi}_p}{|\vec{\xi}_q||\vec{\xi}_p|}\, .
\end{equation}

Each 512-spin time block consists of 8 separate 64-spin blocks, for which the rates $\lambda_\text{TOF3}$ and $\lambda_\text{TOF2}$ are transmitted. As the statistical uncertainty, we take the standard deviation of the mean of the sum of these rates over the interval. We denote this quantity as $\delta_\text{B}\bar{\lambda}(i)$. The statistical variation of $\lambda_\text{triple}$ is small and can be neglected here. These uncertainties are not correlated in time or from bin to bin.

Then, the total uncertainty of the effective TC factor for each bin $i$ is
\begin{equation}
 \delta \gamma_i= \left[ \frac{1}{\sum_{q=0}^{n} d_{i,q}^2}\sum_{q=0}^{n}\sum_{p=0}^{n} d_{i,q}d_{i,p}\gamma'_{i,q}\gamma'_{i,p}\left[ \delta_\text{A}\bar{\lambda}_q(i)\delta_\text{A}\bar{\lambda}_p(i)a_{q,p}+\delta_\text{B}\bar{\lambda}_q(i)\delta_\text{B}\bar{\lambda}_p(i)\delta_{q,p} \right]\right]^{1/2}\, , \label{eq:dlambda}
\end{equation}
where $\gamma'_{i,q}=\text{d}\gamma(\bar{\lambda}_q(i))/\text{d}\lambda$, and $\delta_{q,p}$ is the Kronecker delta.

Starting from orbit 169, \emph{IBEX}-Lo was configured into such a mode that the total event rate is equal to the TOF2 rate, which amounts to at most a few tens per second. In this case, the correction factor is at most 1.002, and thus negligible compared to the statistical uncertainties of the measured count rates.

\bibliographystyle{apj}
\bibliography{paper}

\end{document}